\documentclass[a4paper,11pt]{article}
\usepackage{heppub}
\pdfoutput=1
\usepackage{graphicx}
\graphicspath{{plots/}}
\usepackage{amsmath}
\usepackage{amssymb}
\usepackage{xspace}
\usepackage{xcolor}
\usepackage{enumitem}

\newcommand{\WidthTwoSubfigs}{0.5\textwidth}

\newcommand{\ord}[1]{\mathcal{O}(#1)}

\newcommand{\df}{\mathrm{d}}
\newcommand{\img}{\mathrm{i}}

\newcommand{\eps}{\epsilon}

\newcommand{\GeV}{\,\mathrm{GeV}}

\newcommand{\nn}{\nonumber}

\newcommand{\cusp}{\mathrm{cusp}}

\newcommand{\MSbar}{$\overline{\text{MS}}$\xspace}

\DeclareMathOperator{\Tr}{Tr}

\allowdisplaybreaks[3]

\title{Higher-Order Sudakov Resummation in Coupled Gauge Theories}

\author{Georgios Billis,}
\author{Frank J.~Tackmann,}
\author{and Jim Talbert}

\affiliation{Theory Group, Deutsches Elektronen-Synchrotron (DESY), D-22607 Hamburg, Germany}

\emailAdd{georgios.billis@desy.de, frank.tackmann@desy.de, james.talbert@desy.de}

\abstract{We consider the higher-order resummation of Sudakov double logarithms
in the presence of multiple coupled gauge interactions. The associated evolution
equations depend on the coupled $\beta$ functions of two (or more) coupling
constants $\alpha_a$ and $\alpha_b$, as well as anomalous dimensions that have joint
perturbative series in $\alpha_a$ and $\alpha_b$.  We discuss possible strategies for
solving the system of evolution equations that arises. As an example, we obtain the complete
three-loop (NNLL) QCD$\otimes$QED Sudakov evolution factor. Our results also readily
apply to the joint higher-order resummation of electroweak and QCD Sudakov logarithms.

As part of our analysis we also revisit the case of a single gauge interaction
(pure QCD), and study the numerical differences and reliability of various
methods for evaluating the Sudakov evolution factor at higher orders. We find
that the approximations involved in deriving commonly used analytic
expressions for the evolution kernel can induce noticeable numerical differences
of several percent or more at low scales, exceeding the
perturbative precision at N$^3$LL and in some cases even NNLL. Therefore, one
should be cautious when using approximate analytic evolution kernels for
high-precision analyses.
}
\date{July 05, 2019}

\preprint{DESY 19-128}

\arxivnumber{1907.02971}
\journalref[JHEP]{https://doi.org/10.1007/JHEP03(2020)182}

\begin{document}

\maketitle

%%%%%%%%%%%%%%%%%%%%%%%%%%%%%%%%%%%%%%%%%%%%%%%%%%%%%%%%%%%%%%%%%%%%%%%%%%%%%%%%
\section{Introduction}
\label{sec:intro}
%%%%%%%%%%%%%%%%%%%%%%%%%%%%%%%%%%%%%%%%%%%%%%%%%%%%%%%%%%%%%%%%%%%%%%%%%%%%%%%%

In perturbative quantum field theories it is well known that observables sensitive
to physics at different scales $\mu \ll \mu_{0}$ can have their perturbative
expansions in the coupling constant $\alpha$ generically enhanced by
Sudakov logarithms of the form
%%%
\begin{align}
\label{eq:genlog}
\alpha(\mu)^n \ln^m \frac{\mu}{\mu_0}
\qquad\text{with} \qquad
m\leq 2n
\,.\end{align}
%%%
For sufficiently separated scales, the logarithms grow large
enough to dominate (and eventually deteriorate) the perturbative series.
The reliability and precision of theoretical predictions can therefore be improved
(or restored) by a reorganization of the perturbative series into a form that keeps the
highest-power logarithms to all orders in $\alpha$, a procedure called resummation.
Various different formalisms have been developed to achieve the resummation, with
strongly interacting processes being the most widely studied due to their dominant
contribution to many key collider processes. Indeed, multiple observables have been
resummed to next-to-next-to-leading logarithmic (NNLL) and even N$^{3}$LL accuracy
within QCD, in an effort to match the ever increasing experimental precision.

Despite their smaller couplings in the Standard Model (SM), corrections from the
emission of electroweak (EW) bosons can be comparable to those of QCD calculated at
NNLO (cf.\ $\alpha_{e} \sim \alpha_{s}^2$).
Furthermore, the exchange of massive virtual
EW bosons in high-energy processes can generate EW Sudakov logarithms of the form of \eq{genlog}, which can cause
sizeable EW corrections. The resummation of EW Sudakov logarithms has been studied
for many years, albeit typically at lower orders than in QCD, see
e.g.~\refscite{Ciafaloni:1998xg, Fadin:1999bq, Kuhn:1999nn, Jantzen:2005az,
Denner:2006jr, Chiu:2007dg, Chiu:2008vv, Bell:2010gi, Becher:2013zua,
Christiansen:2014kba, Krauss:2014yaa, Jadach:2016zsp, Bauer:2017isx, Manohar:2018kfx}.
As a result, achieving sufficiently precise predictions for many collider observables requires
considering them in a joint QCD$\otimes$EW environment to fully capture all relevant effects.
This is of course true when considering an extremely high energy future collider~\cite{Mangano:2016jyj},
where EW corrections can be $\mathcal{O}(1)$, but also for measurements at the LHC
reaching percent-level precision, the prime example being the high-precision
measurements of $W$ and $Z$ production~\cite{Aad:2014xaa, Aad:2015auj, Aaboud:2017svj,
Chatrchyan:2013tia, Khachatryan:2015oaa, CMS:2019yfw}.
The recent literature reflects this: The impact of EW and mixed QCD$\otimes$EW
corrections on the $W$-mass measurement have received much attention
(see e.g.~\refscite{Dittmaier:2015rxo, Alioli:2016fum, CarloniCalame:2016ouw, Jadach:2017sqv}
and references therein). QED corrections to the evolution
of parton distribution functions (PDFs) have also been obtained, see
e.g.~\refscite{Roth:2004ti, Martin:2004dh, Ball:2013hta, deFlorian:2015ujt,
deFlorian:2016gvk, Mottaghizadeh:2017vef}.
The full NNLO $\mathcal{O}(\alpha_{e}\alpha_{s})$ mixed QCD$\otimes$QED corrections
and $\mathcal{O}(\alpha_{e}^{2})$ QED corrections for on-shell $Z$ production were
calculated recently in \refcite{deFlorian:2018wcj} and for Higgs production
in bottom-quark annihilation in \refcite{H:2019nsw}.
The one-loop QED corrections to the Sudakov resummation were included in the
high-precision analysis of thrust in $e^+e^-$ collisions~\cite{Abbate:2010xh}.
The resummed $p_T$ spectrum of $Z$-boson production including QED corrections was
obtained in \refcite{Cieri:2018sfk}, capturing the pure QED logarithmic contributions
at NLL and the mixed QCD$\otimes$QED contributions at LL. Closely related, the QED
corrections to the two-loop anomalous dimensions of $p_T$-dependent distributions
were obtained in \refcite{Bacchetta:2018dcq} and of the quark form factors in
\refcite{H:2019nsw}.

In this paper, we step back and reevaluate some of the technical aspects of
Sudakov resummation when the interactions of two gauge symmetries $G_{a} \otimes G_{b}$
are involved, staying agnostic as to the precise resummation formalism utilized.
In particular, we analyze the integrand structure of the Sudakov evolution factor
%%%
\begin{align} \label{eq:generalsudakov}
U(\mu_0,\mu)
= \exp\biggl\{\int_{\mu_0}^{\mu} \frac{d\mu^{\prime}}{\mu^{\prime}} \Gamma_\cusp [\alpha_a(\mu^{\prime}),\alpha_b(\mu^{\prime})]\ln\frac{Q}{\mu^{\prime}} +\gamma [\alpha_a(\mu^{\prime}),\alpha_b(\mu^{\prime})]
\biggr\}
\,.\end{align}
%%%
An evolution factor of this form necessarily appears in all formulations of
higher-order Sudakov resummation. It implicitly depends on the $\beta$ functions
controlling the renormalization group evolution (RGE) of both gauge coupling
constants $\alpha_a(\mu)$ and $\alpha_{b}(\mu)$, which in general are a coupled
system of differential equations, as well as anomalous dimensions ($\Gamma_\cusp, \gamma$)
whose perturbative expansions are themselves joint series in $\alpha_{a,b}$.
We attempt to be as generic as possible in our discussion and therefore do not
immediately specify $G_{a,b}$.  To draw some phenomenological conclusions we will
eventually consider the example of QCD$\otimes$QED, i.e.\ $G_{a} \equiv SU(3)_{c}$
and $G_{b} \equiv U(1)_{\rm em}$, for which we obtain the complete three-loop (NNLL)
Sudakov evolution factor.

Interestingly, we also find that the approximations made in obtaining closed-form
analytic expressions for \eq{generalsudakov} that are
commonly used in the literature can lead to nonnegligible numerical differences.
When evolving to low scales, where the resummation becomes most important, the
resulting effects can reach several percent or more. In other words, the use of
different strategies for evaluating \eq{generalsudakov} can be a source of
nontrivial systematic differences between different resummation implementations
that can potentially exceed the perturbative precision one is aiming for.
As one manifestation, we find
nontrivial violations of the consistency of the evolution, which we probe by
testing the closure condition
%%%
\begin{equation} \label{eq:closure}
U(\mu,\mu_{0})\, U(\mu_0,\mu) \overset{!}{=} 1
\,.\end{equation}
%%%

Since these issues already appear for a single gauge interaction we devote a
substantial fraction of our paper to exploring them in this simpler case, using
pure QCD as test case. We consider various combinations of numerical
and approximate analytic treatments of the Sudakov factor and study their
accuracy.
Ultimately, we are forced to conclude that the commonly used approximate analytic
expressions for \eq{generalsudakov} are not sufficiently reliable when
aiming for percent-level precision.
We find that a seminumerical approach, where the $\mu$-integration
in the exponent of \eq{generalsudakov} is carried out numerically while using
an approximate analytic solution for the running of the coupling, provides a good compromise,
which also straightforwardly generalizes to multiple gauge interactions.

The structure of the paper is as follows. In \sec{beta}, we discuss the coupled
$\beta$ functions for $\alpha_{a,b}$. We first review the standard approximate
analytic solutions for $\alpha$ in the one-dimensional case, and then derive
corresponding approximate analytic solutions for the coupled system up to
three-loop (NNLO) running.
In \sec{1D}, we move to analyzing the Sudakov evolution kernels for the
one-dimensional case, discussing several methods for evaluating it and studying their
numerical performance up to N$^3$LL. In \sec{2D} we then apply the lessons learned
to the case of two coupled gauge interactions. We conclude in \sec{conclude}.
In \app{pertresults}, we provide the perturbative ingredients utilized in our
numerical comparisons, including the complete set of three-loop QCD$\otimes$QED
coefficients. In \app{betaextract}, we discuss the extraction of the complete
three-loop mixed QCD$\otimes$QED $\beta$-function coefficients.

%%%%%%%%%%%%%%%%%%%%%%%%%%%%%%%%%%%%%%%%%%%%%%%%%%%%%%%%%%%%%%%%%%%%%%%%%%%%%%%%
\section{Iterative solutions to \texorpdfstring{\boldmath $\beta$}{beta}-function RGEs}
\label{sec:beta}
%%%%%%%%%%%%%%%%%%%%%%%%%%%%%%%%%%%%%%%%%%%%%%%%%%%%%%%%%%%%%%%%%%%%%%%%%%%%%%%%

Before discussing the evaluation of the Sudakov evolution kernel, it will be important
to analyze one of its key ingredients, namely the solution of the $\beta$-function RGE
of the coupling constant. In \subsec{beta1D}, we review the
well-known case of a single gauge theory, with particular attention given to the numerical accuracy
of different approximate analytic RGE solutions.
In \subsec{beta2D} we then discuss the solution of the coupled RGE system for
the coupling constants of two gauge theories.

%===============================================================================
\subsection{Single gauge theory}
\label{subsec:beta1D}
%===============================================================================

We start from the well-known $\beta$-function RGE for the
coupling constant $\alpha(\mu)$ of a generic gauge theory,
%%%
\begin{align} \label{eq:betageneral}
\frac{\df\alpha(\mu) }{\df\ln\mu} \equiv \beta[\alpha(\mu)]
&= -2\alpha(\mu)\sum_{n=0}^{\infty}\eps^n \beta_n \Bigl[\frac{\alpha(\mu)}{4\pi}\Bigr]^{n+1}
\nn \\
&= -2\beta_0\,\frac{\alpha(\mu)^2}{4\pi}\, \biggl[1+\eps\, \frac{\alpha(\mu)}{4\pi}\, b_1
+\eps^2\,\frac{\alpha(\mu)^2}{(4\pi)^2}\, b_2 + \ord{\eps^3} \biggr]
\,.\end{align}
%%%
Here, $\mu$ is the renormalization scale, and we introduced a formal expansion parameter $\eps \equiv 1$, which we use to keep track of the evolution order. The perturbative coefficients $b_n$ in the second line
are defined by the ratio
%%%
\begin{equation}
b_{n} = \frac{\beta_n}{\beta_0}
\,.\end{equation}
%%%

We take the viewpoint that \eq{betageneral} \emph{defines} the running order of the RGE.
That is, keeping the terms up to $\ord{\eps^k}$ in \eq{betageneral} defines
the N$^k$LO or ($k+1$)-loop running of $\alpha(\mu)$.
At leading order, $\ord{\eps^0}$, \eq{betageneral} has the well-known exact analytic solution,
%%%
\begin{align} \label{eq:alphaLO}
\alpha(\mu) = \frac{\alpha(\mu_0)}{X}
\,,\qquad
X \equiv X(\mu_0,\mu) = 1 + \frac{\alpha(\mu_0)}{2\pi}\,\beta_0\,\ln \frac{\mu}{\mu_0}
\,,\end{align}
%%%
where $\alpha(\mu_0)$ is a boundary condition for the coupling constant.

As is well known, \eq{betageneral} does not admit an exact analytic solution at NLO and beyond.%
\footnote{More precisely, \eq{betageneral} can still be integrated analytically at NLO and even
NNLO. The resulting expressions, however, cannot be analytically solved for $\alpha(\mu)$ in terms of $\alpha(\mu_0)$ anymore.}
While the exact solution can be easily obtained numerically using standard numerical
differential-equation solvers, in practice it is often
more convenient to have an approximate analytic solution that can be evaluated much faster than the numerical solution, which becomes
important when the scale $\mu$ is not fixed but dynamical. This is precisely the
case for the Sudakov evolution kernel, for which we will need to integrate $\alpha(\mu)$ over $\mu$.

In what follows, we review an iterative method to obtain an approximate analytic solution
for \eq{betageneral}.
At NLO, $\ord{\epsilon}$, the $\beta$-function RGE reads
%%%
\begin{align} \label{eq:betaNLO}
\frac{\df \alpha(\mu)}{\df\ln\mu}
&= -2\beta_0\frac{\alpha(\mu)^2}{4\pi} \biggl[1+\eps\, \frac{\alpha(\mu)}{4\pi}\, b_1 \biggr]
\nn \\
&= -2\beta_0\frac{\alpha(\mu)^2}{4\pi} \biggl[1+\eps\,\frac{\alpha(\mu_0)}{4\pi}\,
\frac{b_1}{X(\mu_0, \mu)}+\ord{\eps^2} \biggr]
\,.\end{align}
%%%
In the second line we substituted the LO solution in \eq{alphaLO} for the $\mu$ dependence
of $\alpha(\mu)$ in the $\ord{\eps}$ term. This induces an $\ord{\eps^2}$ error, since the difference between the LO and NLO $\mu$ dependence will itself be of $\ord{\eps}$.
Since the $\mu$ dependence in the term in square brackets is now explicit, it can be easily
integrated, yielding the NLO solution
%%%
\begin{align} \label{eq:alphaNLO}
\frac{1}{\alpha(\mu)}
&= \frac{X}{\alpha(\mu_0)} + \eps\, \frac{b_1}{4\pi}\, \ln X
\,, \nn \\
\Rightarrow\qquad
\alpha(\mu)
&= \alpha(\mu_0)\biggl[X + \eps\, \frac{\alpha(\mu_0)}{4\pi}\, b_1 \ln X\biggr]^{-1}
\,.\end{align}
%%%
We refer to this (and its higher-order analogues) as the ``iterative'' solution for $\alpha(\mu)$.
We can also expand the inverse in \eq{alphaNLO} in $\eps$ to obtain
%%%
\begin{align} \label{eq:alphaNLOinv}
\alpha(\mu) &= \frac{\alpha(\mu_0)}{X}\biggl[1 - \eps\,\frac{\alpha(\mu_0)}{4\pi}\, b_1\, \frac{\ln X}{X}+\ord{\eps^2} \biggr]
\,.\end{align}
%%%
We will refer to \eq{alphaNLOinv} (and its higher-order analogues) as the ``expanded'' solution.

One might wonder whether the iterative or expanded solution provides a better approximation
to the exact solution, and to this end it is instructive to see to what extent they satisfy
the original $\beta$-function RGE.
For the iterative solution in \eq{alphaNLO}, it is trivial to verify that upon differentiation with respect to
$\ln\mu$ it reproduces the RGE as given in the second line of \eq{betaNLO}.
On the other hand, taking the $\ln\mu$ derivative of the expanded NLO solution in \eq{alphaNLOinv} yields
%%%
\begin{align} \label{eq:alphaconsistencyinv}
\frac{\df\alpha(\mu)}{\df\ln\mu}
&=-2\beta_0\,\frac{\alpha(\mu_0)^2}{4\pi}\, \frac{1}{X^2}
\biggl[1 + \eps\, \frac{\alpha(\mu_0)}{4\pi}\, b_1\, \frac{1-2\ln X}{X} \biggr]
\nn \\
&=- 2\beta_0\, \frac{\alpha(\mu_0)^2}{4\pi}\, \frac{1}{X^2}
\biggl\{\biggl[1-2\eps\,\frac{\alpha(\mu_0)}{4\pi}\, b_1\, \frac{\ln X}{X} \biggl]
+\eps\, \frac{\alpha(\mu_0)}{4\pi}\,\frac{b_1}{X} \biggr\}
\,.\end{align}
%%%
Comparing this with \eqs{betaNLO}{alphaNLOinv}, we see that the term in square brackets
corresponds to expanding the overall $\alpha(\mu)^2$ in the $\beta$ function to $\ord{\eps}$.
This clearly amounts to a further approximation, so we can expect the expanded solution
in general to provide a worse approximation, which is indeed what we will find numerically below.

The iterative solution at NNLO, $\ord{\epsilon^2}$, follows analogously.
Starting from the exact NNLO RGE, we substitute the (expanded) NLO and LO solutions,
\eqs{alphaNLOinv}{alphaLO}, in the $\ord{\eps}$ and $\ord{\eps^2}$ terms,
keeping all terms up to $\ord{\eps^2}$ as well as the overall $\alpha(\mu)^2$,
%%%
\begin{align} \label{eq:betaNNLO}
\frac{\df\alpha(\mu)}{\df\ln\mu}
&=-2\beta_0\frac{\alpha(\mu)^2}{4\pi} \biggl[1+\eps\,\frac{\alpha(\mu)}{4\pi}\, b_1
+ \eps^2 \frac{\alpha(\mu)^2}{(4\pi)^2}\, b_2 \biggr]
\nn \\
&=-2\beta_0\frac{\alpha(\mu)^2}{4\pi}
\biggl[1+\eps\,\frac{\alpha(\mu_0)}{4\pi}\, \frac{b_1}{X}
+ \eps^2\, \frac{\alpha(\mu_0)^2}{(4\pi)^2} \frac{b_2 - b_1^2\, \ln X }{X^2}
+ \ord{\eps^3}
\biggr]
\,.\end{align}
%%%
The $\mu$-dependence in the square brackets on the second line, encoded in $X \equiv X(\mu_0, \mu)$, is again simple enough to be integrated analytically.%
\footnote{This would not be the case had we used the unexpanded NLO solution
from \eq{alphaNLO} in the $\ord{\eps}$ term.}
This yields the NNLO iterative solution
%%%
\begin{align} \label{eq:alphaNNLO}
\alpha(\mu)
&= \alpha(\mu_0)\biggl\{ X + \eps\, \frac{\alpha(\mu_0)}{4\pi}\, b_1 \ln X
+ \eps^2\,\frac{\alpha(\mu_0)^2}{(4\pi)^2} \biggl(b_2\, \frac{X-1}{X} + b_1^2\, \frac{1 -X +\ln X}{X} \biggr)
\biggr\}^{-1}
\,.\end{align}
%%%
Expanding the inverse in \eq{alphaNNLO} in $\eps$, we obtain the expanded NNLO solution,
%%%
\begin{align}  \label{eq:alphaNNLOinv}
\alpha(\mu)
&= \frac{\alpha(\mu_0)}{X} \biggl\{1- \eps\,\frac{\alpha(\mu_0)}{4\pi}\, b_1\, \frac{\ln X}{X}
\nn \\ & \quad
+ \eps^2\, \frac{\alpha(\mu_0)^2}{(4\pi)^2}\, \frac{1}{X^2}
\Bigl[b_2(1-X) + b_1^2\, (\ln^2 X - \ln X -1 + X)\Bigr]+\ord{\eps^3}
\biggr\}
\,.\end{align}
%%%

It is straightforward to extend the iterative solution to higher orders. 
To obtain the N$^k$LO solution, one simply inserts the N$^{k-1}$LO solution for $\alpha(\mu)$ into the N$^k$LO $\beta$ function and expands it to $\ord{\eps^k}$, while keeping the overall $\alpha(\mu)^2$ exact.
The N$^3$LO solution is given in \app{RGE}.

%-------------------------------------------------------------------------------
\begin{figure*}
\centering
\includegraphics[width=\WidthTwoSubfigs]{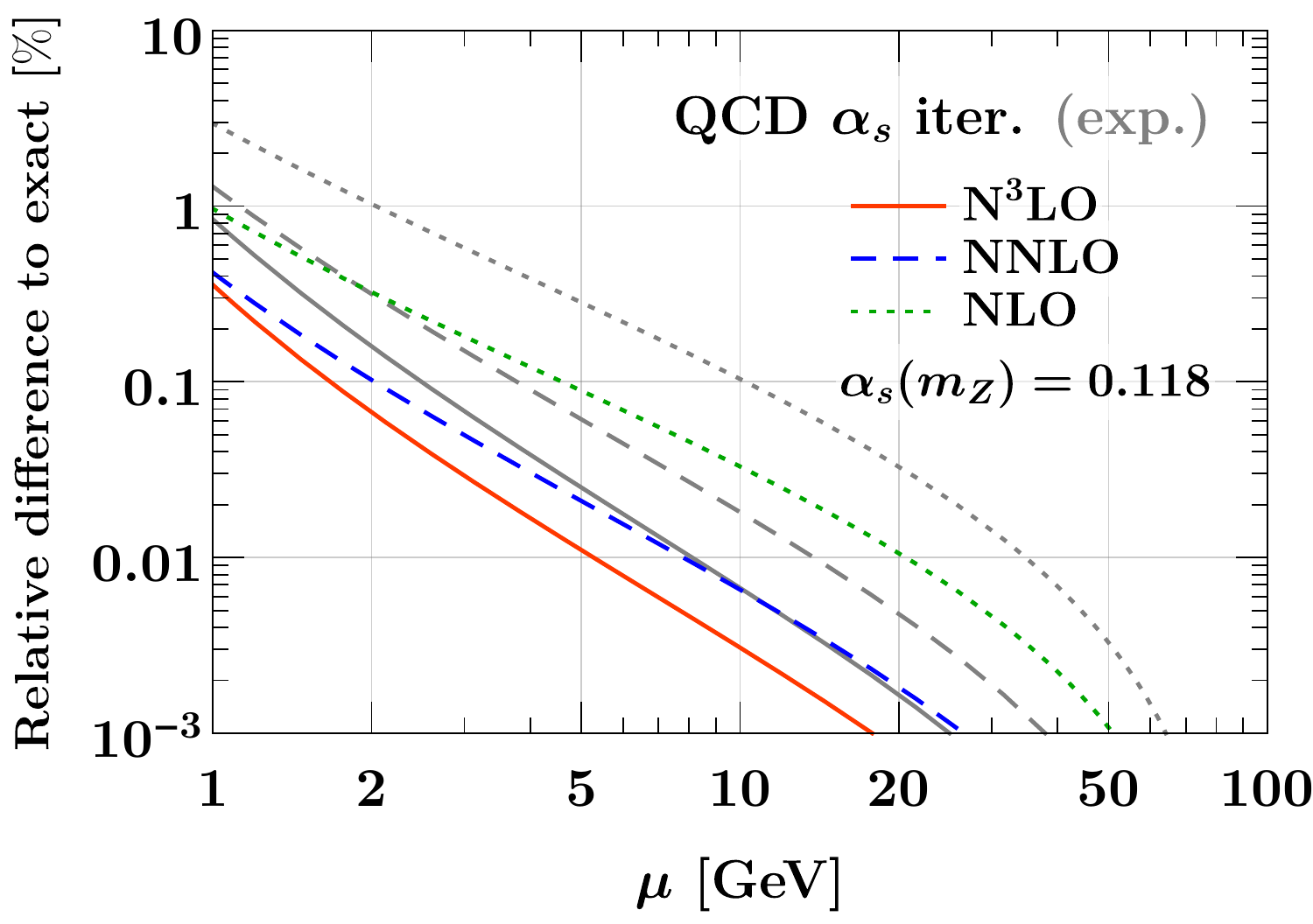}%
\includegraphics[width=\WidthTwoSubfigs]{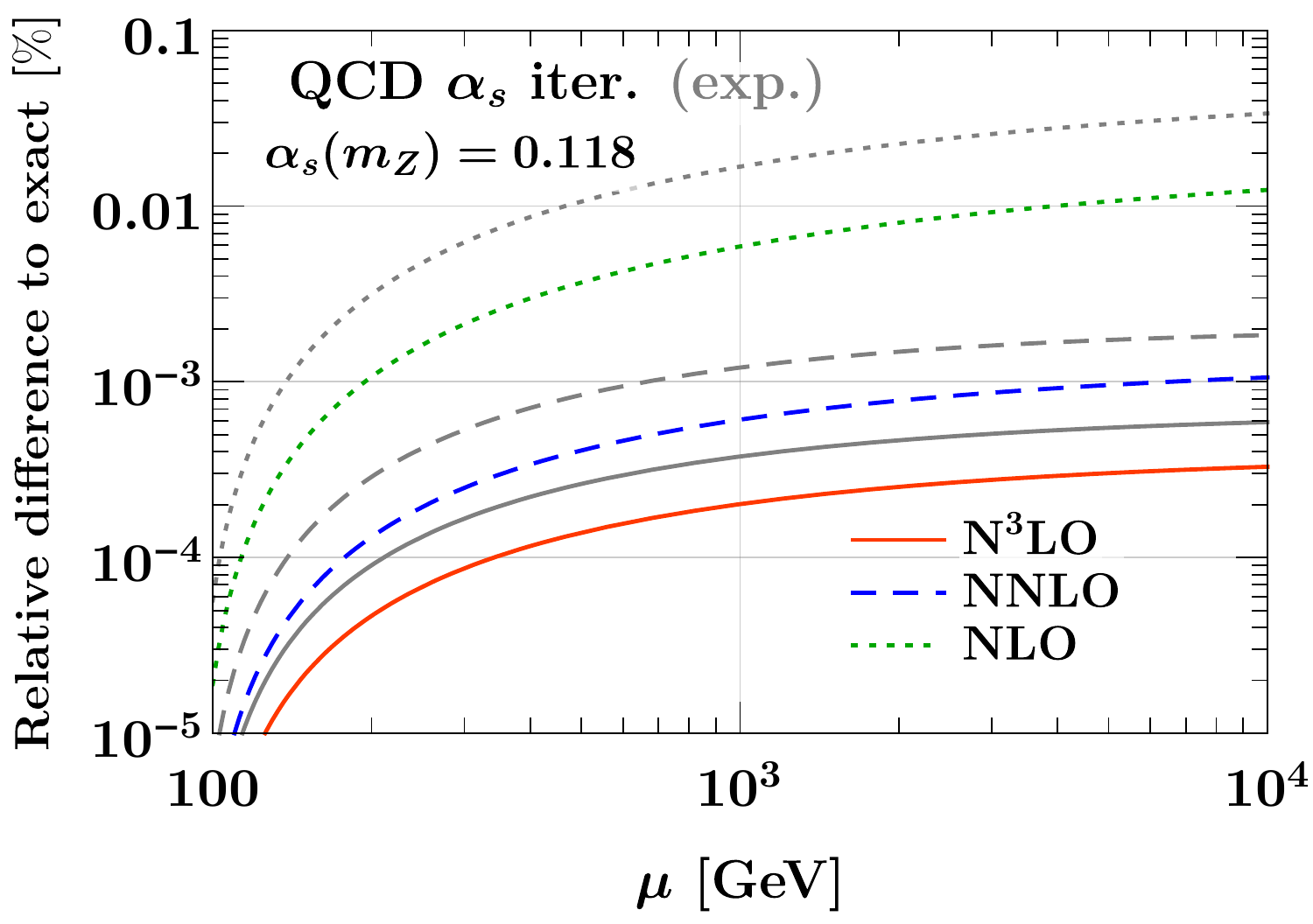}%
\caption{Relative deviation of the iterative (colored) and expanded solutions
(gray) from the exact solution for the running of $\alpha_{s}(\mu)$
at NLO (dotted), NNLO (dashed), and N$^3$LO (solid).}
\label{fig:beta1D}
\end{figure*}
%-------------------------------------------------------------------------------

To illustrate the numerical precision of the approximate analytic solutions, we
take the QCD coupling constant $\alpha_s(\mu)$ as an example, using
$\alpha_s(m_Z) = 0.118$ as our boundary condition. As we are primarily
interested in the numerical precision of the solution, we always use $n_f = 5$
massless flavors and do not consider any flavor thresholds. The difference of
the iterative and expanded solutions to the exact%
\footnote{We always perform the numerical solution with sufficiently high
numerical precision such that the numerical error is completely negligible for
our purposes.}
numerical solution, which we refer to as the approximation error, is shown in
\fig{beta1D} for different running orders.

The approximation error decreases as the order increases, as expected.
The approximation error is largest for running from $m_Z$ down to lower scales,
since here the running increases the coupling. The iterative solution still provides an
excellent approximation, with the error at NNLO and beyond reaching at most 0.1\% when running down to $\mu = 2\GeV$, and at most 0.3\% at $\mu = 1\GeV$. For running above $m_Z$, the
approximation error is much smaller due to asymptotic freedom.
Beyond the highest scale shown, $\mu = 10^4\GeV$, the error stops growing and at some point starts decreasing again.
We also observe that the approximation error for the expanded solution (gray lines) is
always 2-3 times larger than for the iterative solution.

%===============================================================================
\subsection{Two coupled gauge theories}
\label{subsec:beta2D}
%===============================================================================

We now consider the case of two gauge theories. Their $\beta$ functions become coupled as soon
as there are matter fields that are charged under both gauge interactions, since loops of
 matter particles can exchange the gauge bosons from both theories. The example
we will eventually consider is the mixed QCD$\otimes$QED running. For now, we keep the discussion
general and consider the following set of coupled $\beta$-function RGEs for two gauge couplings
$\alpha_a(\mu)$ and $\alpha_b(\mu)$,
%%%
\begin{align} \label{eq:2Dbetageneral}
\frac{\df \alpha_a(\mu)}{\df\ln\mu}
&\equiv \beta^{a}[\alpha_a(\mu), \alpha_b(\mu)]
= -2\alpha_a(\mu) \sum_{n,k=0}^{\infty} \eps^{n}_a\eps_b^k\,
\beta^{a}_{nk}\, \Bigl[\frac{\alpha_a(\mu)}{4\pi}\Bigr]^{n+1} \Bigl[\frac{\alpha_b(\mu)}{4\pi}\Bigr]^k
\,, \nn \\
\frac{\df \alpha_b(\mu)}{\df\ln\mu}
&\equiv \beta^{b}[\alpha_b(\mu),\alpha_a(\mu)]
= -2 \alpha_b(\mu)\sum_{n,k=0}^{\infty} \eps^{n}_b\eps_a^k\,
\beta^{b}_{nk}\, \Bigl[\frac{\alpha_b(\mu)}{4\pi}\Bigr]^{n+1}\Bigl[\frac{\alpha_a(\mu)}{4\pi}\Bigr]^k
\,.\end{align}
%%%
We have again introduced formal expansion parameters $\eps_{a,b}\equiv 1$ to easily
keep track of the evolution order. For future convenience, we also define the rescaled
coefficients
%%%
\begin{equation}
b^a_{nk} = \frac{\beta^a_{nk}}{\beta^a_{00}}
\,, \qquad
b^b_{nk} = \frac{\beta^b_{nk}}{\beta^b_{00}}
\,, \qquad\text{and}\qquad
b_0^a = \frac{\beta_{00}^a}{\beta^b_{00}}
\,, \qquad
b_0^b = \frac{\beta_{00}^b}{\beta^a_{00}}
\,.\end{equation}
%%%
Note that by definition $\beta^x_{n0} \equiv \beta_n$ and $b^x_{n0} \equiv b_n$ are
the coefficients of the individual gauge theories in the absence of the second.

As before, the order in $\eps_{a,b}$ to which
\eq{2Dbetageneral} is expanded is what defines the running
order of the couplings. Generically, we consider $\eps_{a,b}$ on equal footing and define
the N$^n$LO evolution by including in \eq{2Dbetageneral} all terms of
$\ord{\eps_a^k \eps_b^{n-k}}$ with $0\leq k\leq n$. Note that this also makes it easy to have
well-defined mixed orders, e.g., when there is a hierarchy between the two couplings,
as is the case for QCD and QED. For this purpose, one can simply specify the explicit 
combinations of powers of $\eps_a$ and $\eps_b$ that are included in \eq{2Dbetageneral}.

As in the single gauge scenario, an exact solution to the coupled system of differential
equations can be obtained straightforwardly at any given order by solving it numerically.
Our goal is to derive an approximate analytic solution for the coupled $\beta$-function RGE system
by extending the iterative method in \subsec{beta1D}. The key property of \eq{2Dbetageneral}
that allows us to do so is that, at leading order, $\ord{\eps_a^0, \eps_b^0}$, the RGE system
decouples, yielding exact LO solutions:
%%%
\begin{alignat}{2} \label{eq:2DalphaLO}
\alpha_a(\mu) &= \frac{\alpha_a(\mu_0)}{X_a}
\,, \qquad &
X_a &\equiv X_a(\mu_0,\mu) = 1 + \frac{\alpha_a(\mu_0)}{2\pi}\, \beta^{a}_{00}\, \ln \frac{\mu}{\mu_0}
\,, \nn \\
\alpha_b(\mu) &= \frac{\alpha_b(\mu_0)}{X_b}
\,,\qquad &
X_b &\equiv X_b(\mu_0,\mu) = 1 + \frac{\alpha_b(\mu_0)}{2\pi}\, \beta^{b}_{00}\, \ln \frac{\mu}{\mu_0}
\,.\end{alignat}
%%%
To obtain an approximate NLO solution, it then suffices to substitute the above LO solutions into
the $\ord{\eps_a}$ and $\ord{\eps_b}$ terms of the NLO RGE system, which induces
$\ord{\eps_a^2, \eps_a\eps_b, \eps_b^2}$ errors,
%%%
\begin{align}
\frac{\df\alpha_{a}(\mu)}{\df\ln\mu}
&= -2\beta_{00}^a \frac{\alpha_a(\mu)^2}{4\pi}
\biggl[1 + \eps_a\,\frac{\alpha_a(\mu)}{4\pi}\, b^a_{10}
+ \eps_b\, \frac{\alpha_b(\mu)}{4\pi}\, b^a_{01} \biggr]
\\ \nn
&= -2\beta_{00}^a\frac{\alpha_a(\mu)^2}{4\pi} \biggl[
1 + \eps_a\,\frac{\alpha_a(\mu_0)}{4\pi}\, \frac{b^a_{10}}{X_a(\mu_0, \mu)}
  + \eps_b\,\frac{\alpha_b(\mu_0)}{4\pi}\, \frac{b^a_{01}}{X_b(\mu_0, \mu)}
  + \ord{\eps_a^2, \eps_a\eps_b, \eps_b^2} \biggr]
\,.\end{align}
%%%
The terms in square brackets can now be explicitly integrated over $\ln\mu$ to
obtain the iterative NLO solution
%%%
\begin{align} 
\label{eq:2DalphaNLO}
\alpha_a(\mu) = \alpha_a(\mu_0)\biggl[
X_a + \frac{\alpha_a(\mu_0)}{4\pi} \Bigl(\eps_a\, b_{10}^a\, \ln X_a
 + \eps_b\, b^a_0\, b_{01}^a \, \ln X_b \Bigr)
\biggr]^{-1}
\,.\end{align}
%%%
The solution for $\alpha_b(\mu)$ is given by replacing
$a\leftrightarrow b$ everywhere.

Note that in \eq{2DalphaNLO} the utility of the $\eps_{a,b}$ counting parameters
becomes evident. Naively, one might have expected that the $\ord{\eps_a}$ and
$\ord{\eps_b}$ terms will be proportional to $\alpha_a(\mu_0)$ and
$\alpha_b(\mu_0)$ respectively, which however is not the case, as both are
proportional to $\alpha_a(\mu_0)$. Instead, the $\eps_{a,b}$ actually keep track
of the fact that the $\ln X_{a,b}$ factors respectively resum a series of
$\alpha_{a,b}^n \ln^n(\mu/\mu_0)$ terms.

The iterative NNLO solution is obtained by substituting the NLO solutions into the $\beta$-function RGE system and expanding it to $\ord{\epsilon_a^2,\epsilon_a\epsilon_b,\epsilon_b^2}$, while keeping the overall $\alpha_{a,b}(\mu)^2$ exact. We find
%%%
\begin{align} \label{eq:2DalphaaNNLO}
\frac{\alpha_a(\mu_0)}{\alpha_a(\mu)}
&=X_a
  + \eps_a\, \frac{\alpha_a(\mu_0)}{4\pi}\, b_{10}^a \ln X_{a}
  + \eps_a^2\, \frac{\alpha_a(\mu_0)^2}{(4\pi)^2}
 \biggl( b_{20}^{a}\, \frac{X_a -1}{X_a} + (b_{10}^a )^2 \frac{1-X_a+\ln X_a}{X_a} \biggr)
\nn \\ & \quad
 + \eps_b\, \frac{\alpha_a(\mu_0)}{4\pi}\, b^a_{0}\, b_{01}^a \ln X_b
 + \eps_b^2\, \frac{\alpha_a(\mu_0)\alpha_b(\mu_0)}{(4\pi)^2} b^a_0
 \biggl(b_{02}^a\, \frac{X_b - 1}{X_b} + b_{01}^a b_{10}^b\, \frac{1 - X_b + \ln X_b}{X_b} \biggr)
\nn \\ & \quad
  + \eps_a\eps_b\, \frac{\alpha_a(\mu_0)}{b_{0}^a\, \alpha_a(\mu_0) - \alpha_b(\mu_0)} \biggl[
\frac{\alpha_a^2(\mu_0)}{(4\pi)^2} (b^{a}_{0})^2 b_{10}^a b_{01}^a
\biggl(\frac{X_b}{X_a}\ln X_b - \frac{1-X_b}{1 - X_a}\ln X_a \biggr)
\nn \\ & \qquad
- \frac{\alpha_b^2(\mu_0)}{(4\pi)^2}\, b_{01}^a b_{01}^b
\biggl(\frac{X_a}{X_b}\ln X_a
-\frac{1 - X_a}{1 - X_b}\ln X_b \biggr)
+ \frac{\alpha_a(\mu_0)\alpha_b(\mu_0)}{(4\pi)^2} b^a_{0}\, b_{11}^a \ln\frac{X_a}{X_b} \biggr]
\,.\end{align}
%%%
As before, the solution for $\alpha_b(\mu)$ is obtained by replacing $a\leftrightarrow b$.
The terms in the first line correspond to the NNLO solution of $\alpha_a$ in the absence
of $\alpha_b$, while the remaining ones are the mixing contributions involving at least
one power of $\eps_b$. The corresponding expanded solution is obtained by inverting
\eq{2DalphaaNNLO} and expanding it in $\epsilon_i$. Note that when doing so, it becomes
essential to expand in terms of $\epsilon_i$ and not $\alpha_i(\mu_0)$.
The necessity of using $\epsilon_i$ as expansion parameter is also evident in the
mixed $\ord{\epsilon_a\epsilon_b}$ term in square brackets, which involves a
nontrivial rational function of both couplings.
The above iterative NNLO solution will be a key ingredient in the seminumerical
evaluation of the Sudakov evolution factor in \sec{2D}.

%-------------------------------------------------------------------------------
\begin{figure*}
\centering
\includegraphics[width=\WidthTwoSubfigs]{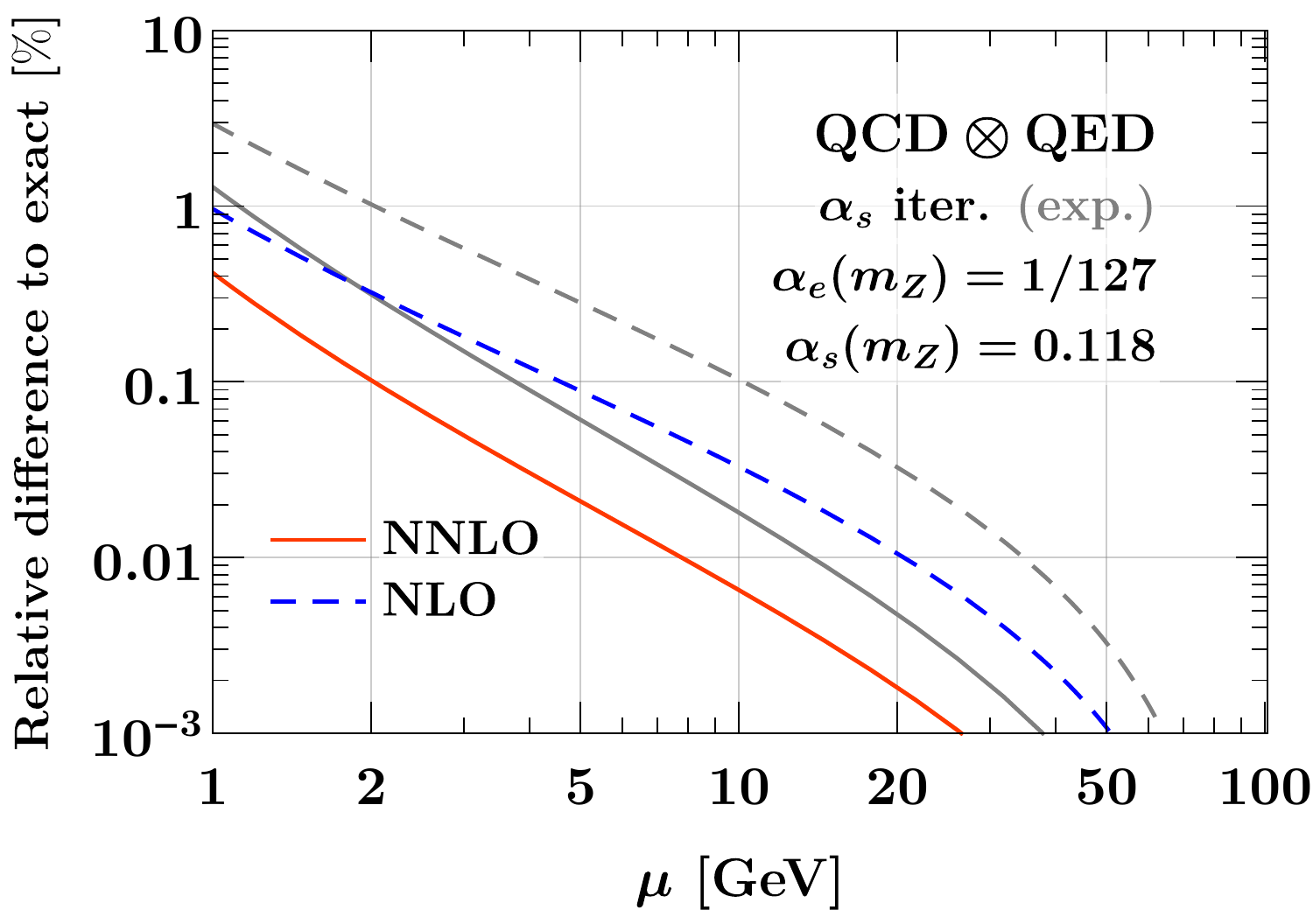}%
\hfil%
\includegraphics[width=\WidthTwoSubfigs]{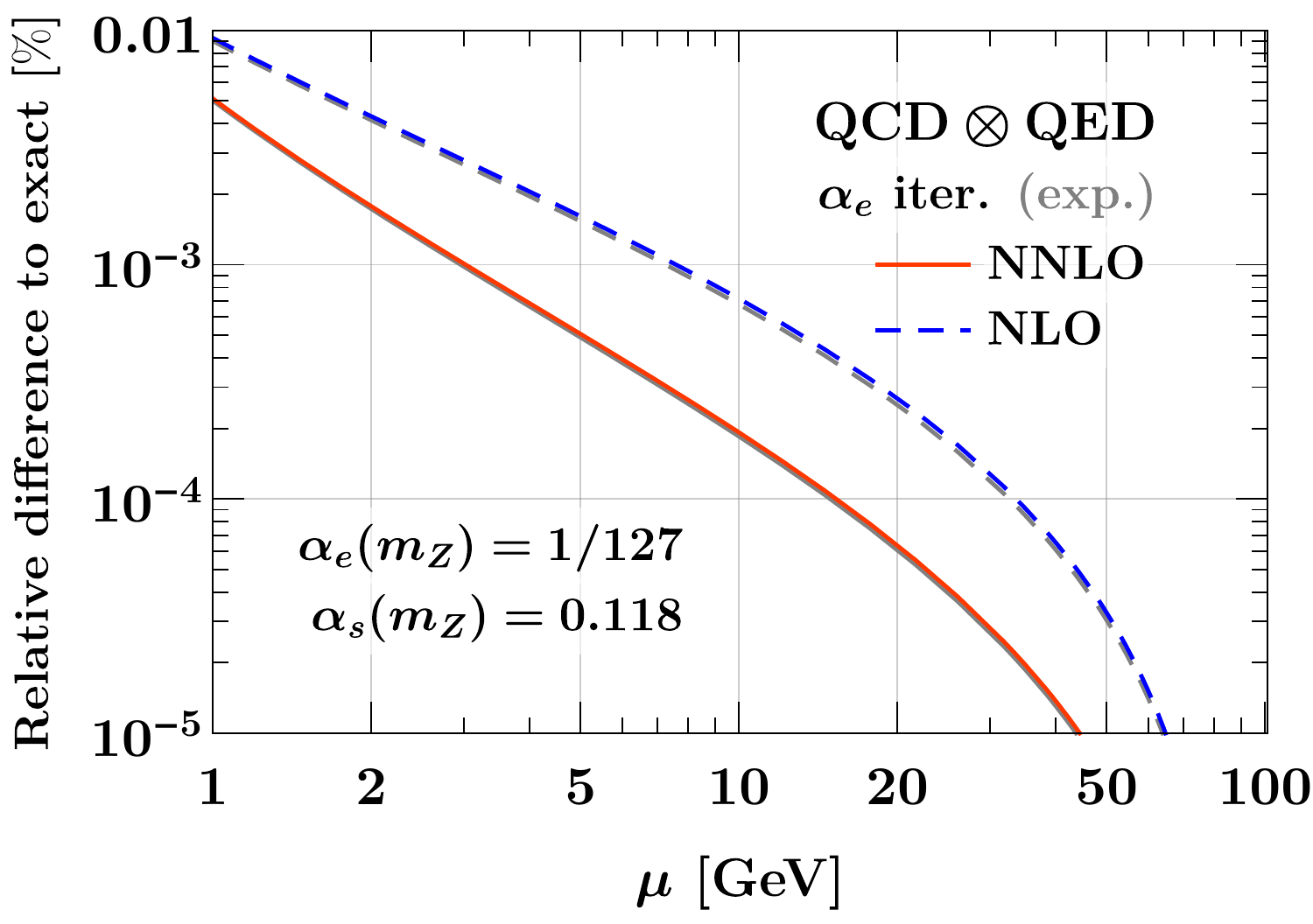}
\caption{Relative deviation from the exact solution for the running of the QCD (left)
and QED (right) coupling constants in QCD$\otimes$QED for the iterative (colored)
and expanded (gray) solutions.}
\label{fig:QCED_mixed}
\end{figure*}
%-------------------------------------------------------------------------------

%-------------------------------------------------------------------------------
\begin{figure*}
\centering
\includegraphics[width=\WidthTwoSubfigs]{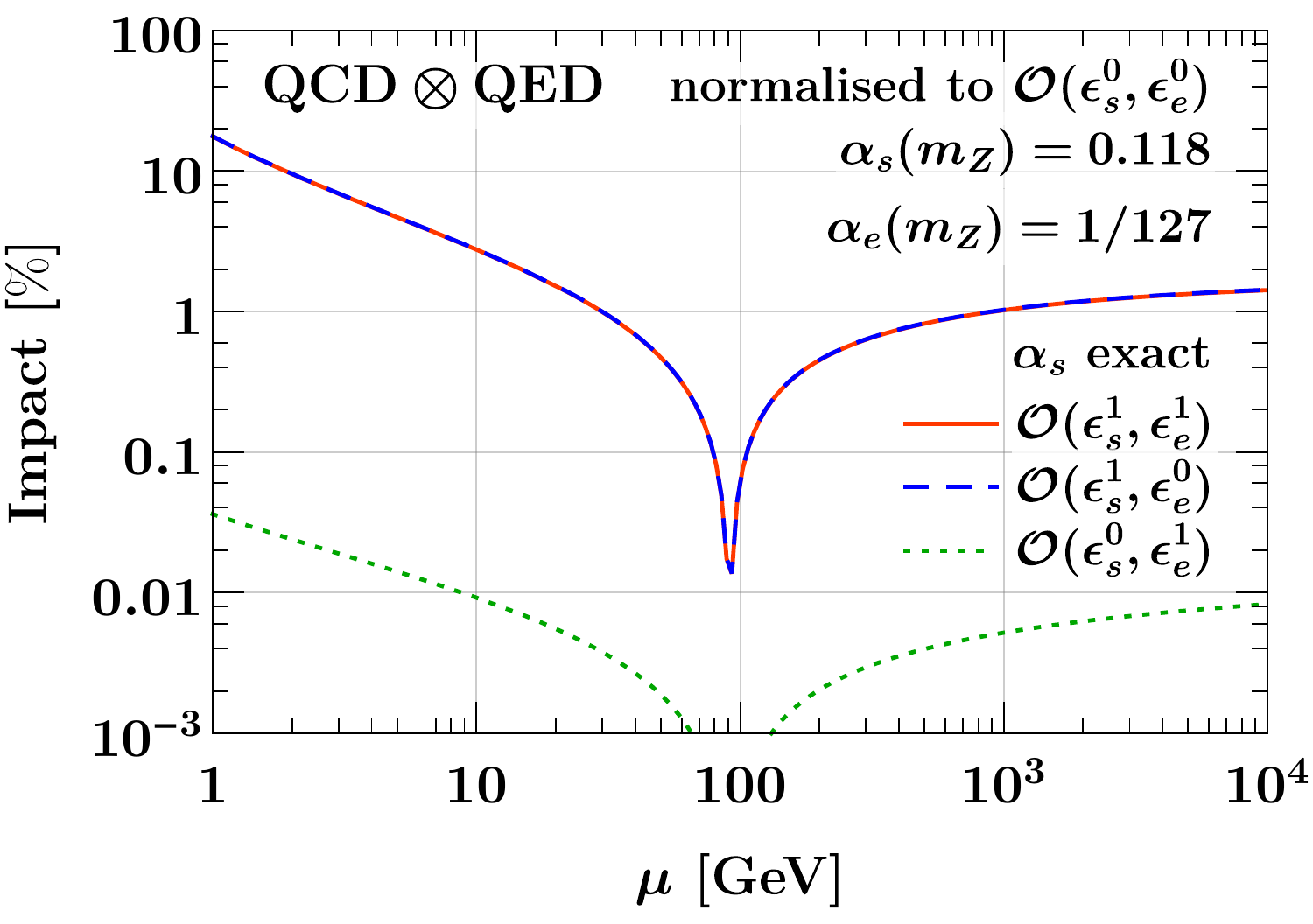}%
\hfill%
\includegraphics[width=\WidthTwoSubfigs]{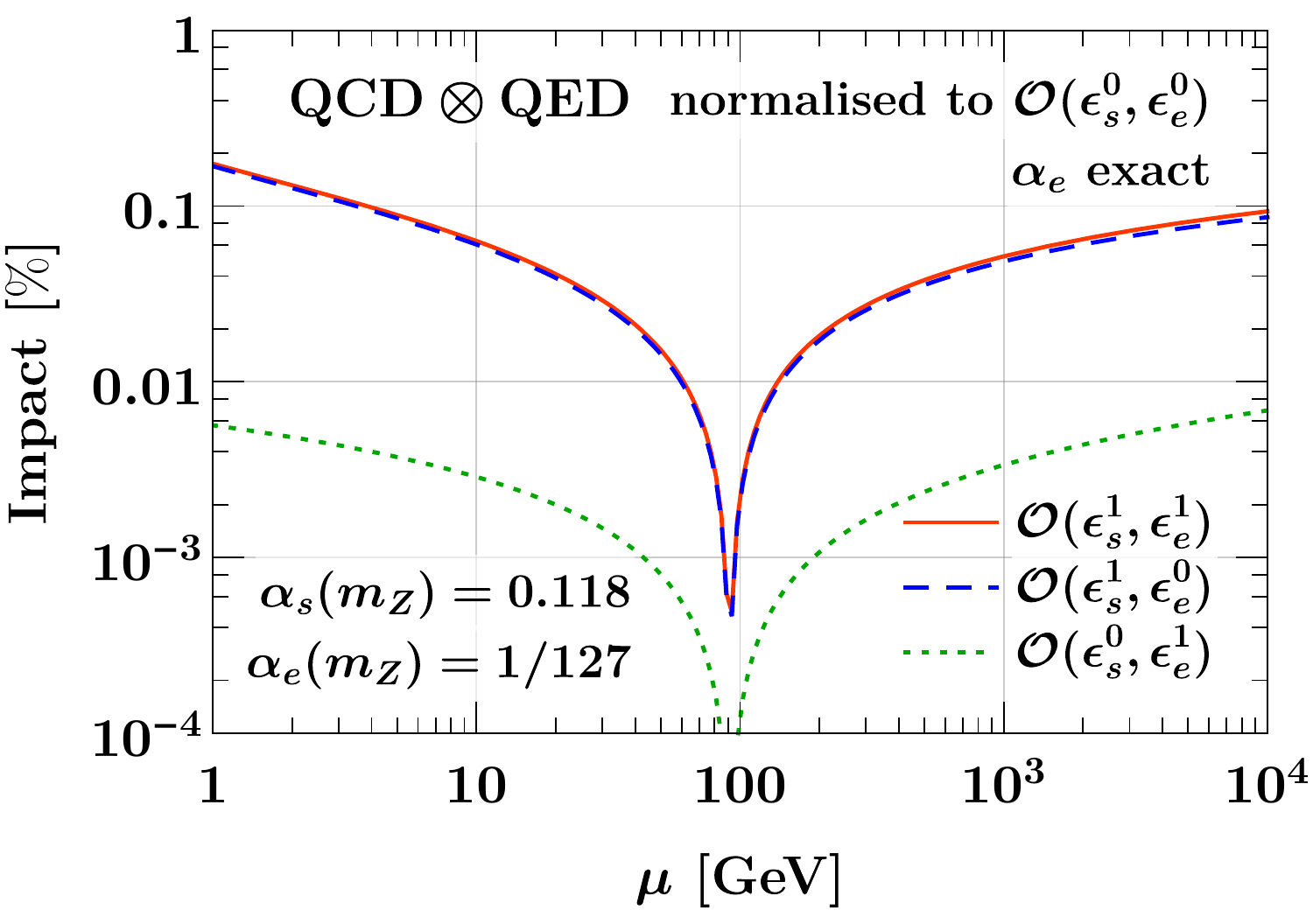}%
\caption{Impact of different higher-order terms on the running of the
QCD (left) and QED (right) couplings. Shown are the relative differences to
the LO running $\sim\ord{\eps_s^0,\eps_e^0}$.
}
\label{fig:QCEDimpact_mixed}
\end{figure*}
%-------------------------------------------------------------------------------

%-------------------------------------------------------------------------------
\begin{figure*}
\centering
\includegraphics[width=\WidthTwoSubfigs]{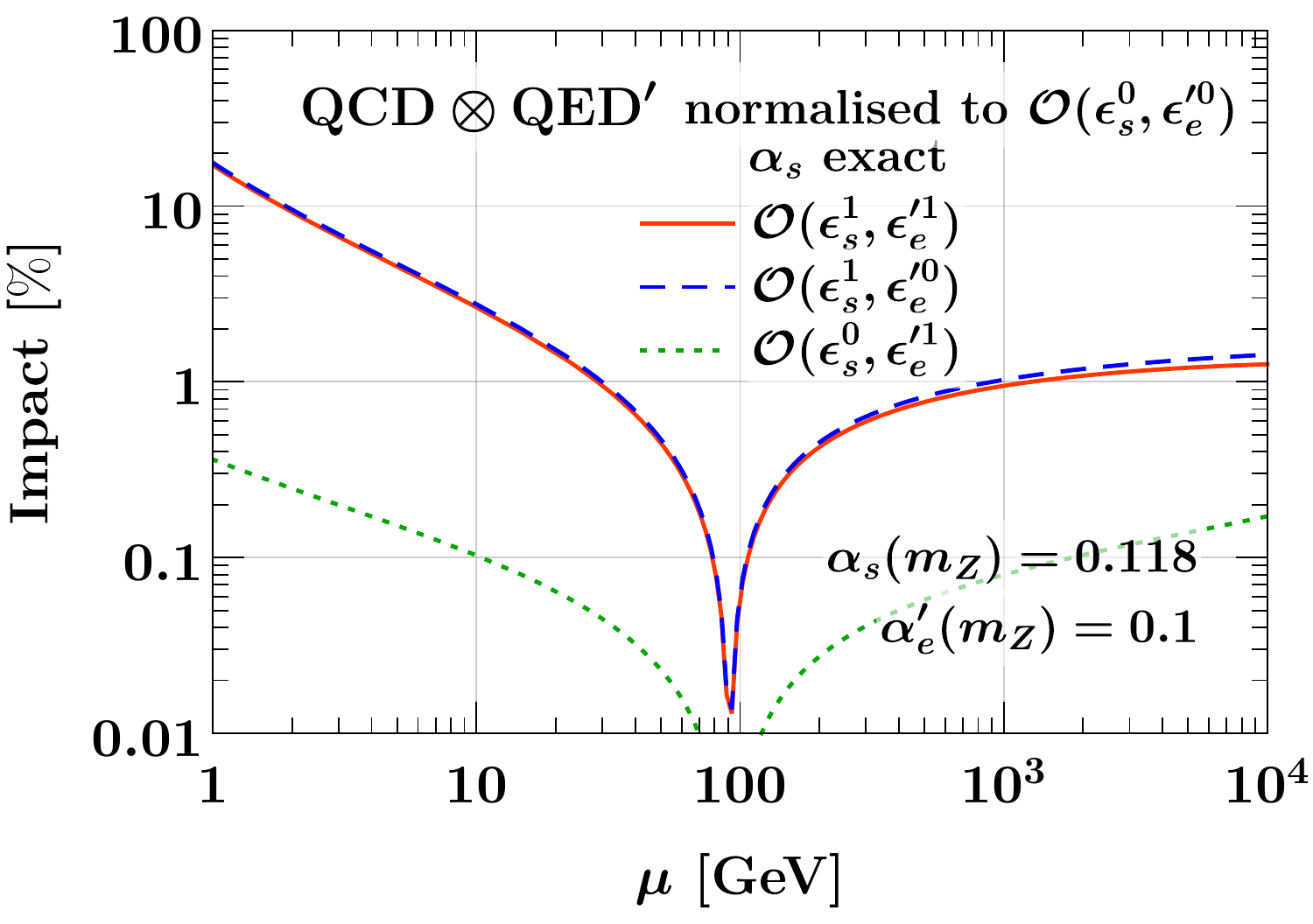}%
\hfill%
\includegraphics[width=\WidthTwoSubfigs]{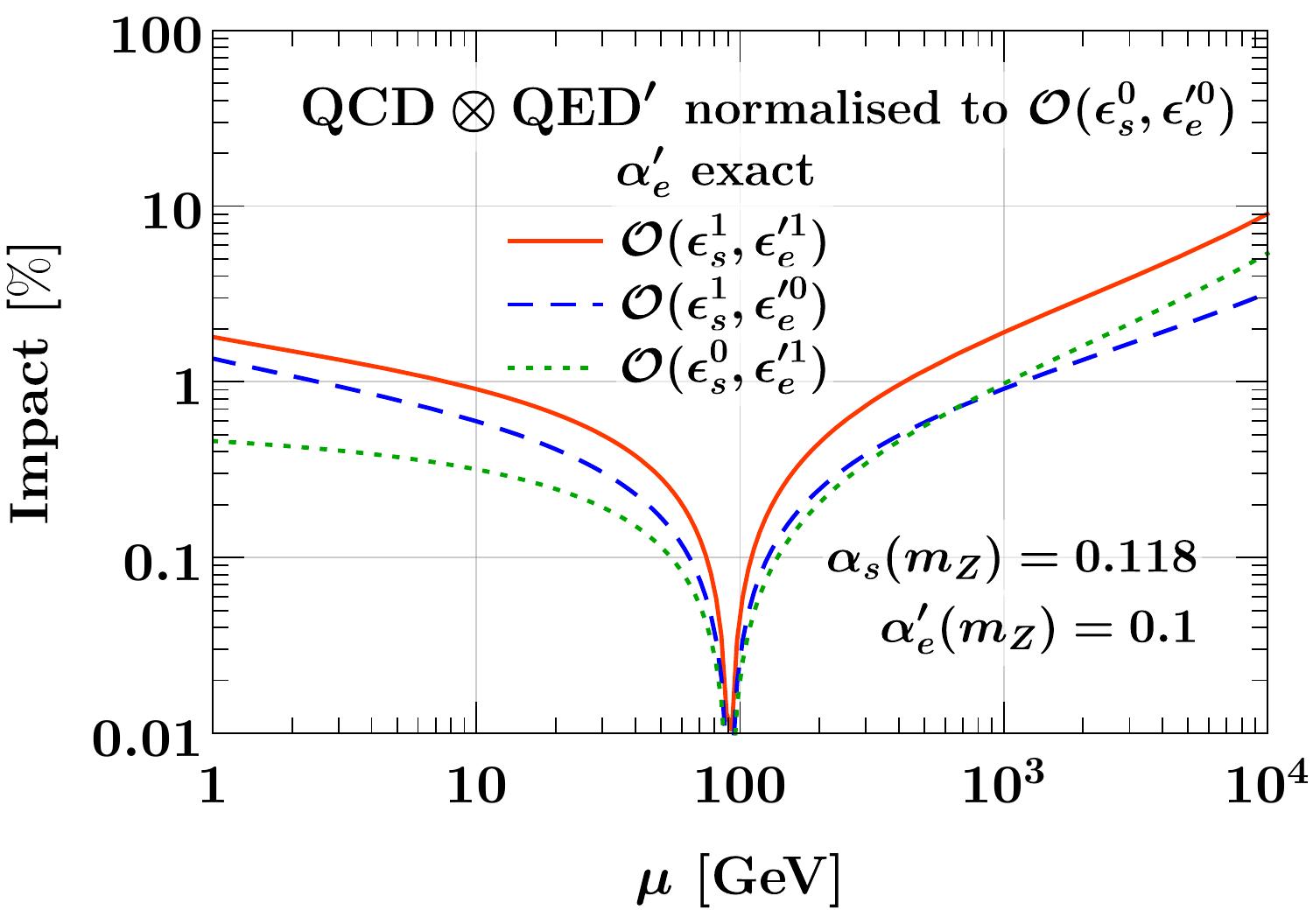}%
\caption{Same as \fig{QCEDimpact_mixed} but for QCD$\otimes$QED$'$, where
QED$'$ is a toy theory with a modified boundary condition of $\alpha_{e}^{\prime}(m_{Z}) = 0.1$.}
\label{fig:QCEDimpact_mixed_toy}
\end{figure*}
%-------------------------------------------------------------------------------

To illustrate the approximation error for the two analytic solutions, we
consider the case of QCD$\otimes$QED. The relevant coefficients for the coupled
$\beta$-function RGE to NNLO are given in \apps{anomdimQCD}{anomdimQED}.
Curiously, we were not able to find explicit expressions in the literature
for the mixed three-loop QED coefficient $\beta_{11}^{e}$.
We therefore performed an explicit extraction of all mixed three-loop coefficients
from the general results for a generic product group given in \refcite{Mihaila:2014caa},
as discussed in more detail in \app{betaextract}.

For the numerical results, we use $\alpha_s(m_Z)=0.118$ and $\alpha_e(m_Z)=1/127$
as boundary conditions, $n_f=5$ for the number of active quark flavors, and $n_{\ell}=3$ for the number
of active charged leptons. As before we do not consider any flavor thresholds.

The approximation error of the iterative (expanded) solution relative to the exact
numerical solution is shown in \fig{QCED_mixed} by the colored (gray) lines at
NLO (dashed) and NNLO (solid). For the strong coupling constant (left panel),
the approximation error for the iterative solution does not exceed  1\%, and it
is again 2-3 times larger for the expanded solution. For the QED coupling
constant (right panel), the approximation error is much smaller owing to the fact
that $\alpha_e$ is much smaller, and therefore both the iterative and expanded
solutions yield equally good approximations.

It is interesting to see the individual effects of various terms in the coupled
$\beta$ function. In \figs{QCEDimpact_mixed}{QCEDimpact_mixed_toy} we show them
for the case of QCD$\otimes$QED as well as for a toy QCD$\otimes$QED$^{\prime}$ for
which we set the boundary condition to $\alpha_{e}^{\prime}(m_Z)=1/10$.
In \fig{QCEDimpact_mixed}, as expected, the QED corrections $\sim\ord{\eps^0_s,\eps^1_e}$
have almost negligible effect on $\alpha_s$, with only the higher order QCD corrections
$\sim\ord{\eps_s^1,\eps_e^0}$ being relevant. This is also the case for the QED
coupling constant, whose evolution is almost entirely dictated by the QCD corrections.
But this is not the case for the toy QCD$\otimes$QED$^{\prime}$ scenario shown in
\fig{QCEDimpact_mixed_toy}. The fact that the QCD coupling constant is still almost
unaffected by the QED$^{\prime}$ corrections is somewhat accidental and due to the fact
that the $\beta$-function coefficients $\beta^s_{10}$ and $\beta^{s}_{01}$ differ
numerically by an order of magnitude. On the other hand, the QED$^{\prime}$ coupling
constant now has comparable higher-order QCD and QED$^{\prime}$ corrections since
here both $\beta$-function coefficients are of similar numerical size.

%%%%%%%%%%%%%%%%%%%%%%%%%%%%%%%%%%%%%%%%%%%%%%%%%%%%%%%%%%%%%%%%%%%%%%%%%%%%%%%%
\section{Sudakov evolution kernels with a single gauge interaction}
\label{sec:1D}
%%%%%%%%%%%%%%%%%%%%%%%%%%%%%%%%%%%%%%%%%%%%%%%%%%%%%%%%%%%%%%%%%%%%%%%%%%%%%%%%

In this section, we examine different strategies for evaluating the Sudakov
evolution kernel for the case of a single gauge interaction, with particular
emphasis on their numerical accuracy and reliability.
This will then serve as a guide when considering the two-dimensional case in \sec{2D}.

In \subsec{overview}, we start with a general discussion and give an overview
of the different methods we will study for evaluating the Sudakov evolution factor,
which are then elaborated on in \subsec{nummethod} to \subsec{reexpand}.
We then provide a detailed numerical comparison in \subsec{1Dkernelpheno}.

%===============================================================================
\subsection{General overview}
\label{subsec:overview}
%===============================================================================

One way to systematically resum Sudakov logarithms
is based on factorizing the perturbative series. The relevant factorization
can be derived diagrammatically or using effective field theories (EFTs).
All ingredients of the factorized cross section obey
(renormalization group) evolution equations of the form
%%%
\begin{align} \label{eq:RGE}
\frac{\df F(\mu)}{\df\ln\mu} = \gamma_F(\mu) \otimes F(\mu)
\,,\end{align}
%%%
where $\gamma_F(\mu)$ is the anomalous dimension, and the function $F(\mu)$
is any one of the factorized ingredients. The Sudakov resummation is then performed
by solving the resulting coupled system of RGE equations.

In general, $\gamma_F(\mu)$ and $F(\mu)$ both depend on an additional external kinematic quantity
(which appears as part of the argument of the Sudakov logarithms in the cross section).
The $\otimes$ denotes the fact that $\gamma_F$ and $F$ are not necessarily multiplied but could
be convolved in said kinematic variable.
It is worth nothing that the convolution structure can play a significant role
in the solution of \eq{RGE} (for a recent detailed discussion see e.g.~\refcite{Ebert:2016gcn}).
However, it does not play any role for the purpose of our discussion, so we consider
only the simplest multiplicative case. In case of a convolution, one can always transform
to a suitable conjugate space (e.g. Fourier or Laplace space), where the convolution
turns into a simple product. The Sudakov evolution factor in that conjugate space
then has the same general form we discuss here and all our conclusions apply equally.

The all-order expansion of the anomalous dimension is given by
%%%
\begin{align} \label{eq:gammaF}
\gamma_F(\mu) = \frac{1}{\epsilon}\,\Gamma_\cusp[\alpha(\mu)]\ln \frac{Q}{\mu} + \gamma[\alpha(\mu)]
\,,\end{align}
%%%
where $Q$ denotes the above-mentioned kinematic quantity,
$\Gamma_\cusp$ is (proportional to) the cusp anomalous dimension, and $\gamma$
is the noncusp anomalous dimension. They obey the perturbative expansions
%%%
\begin{equation} \label{eq:cuspexp}
\Gamma_\cusp(\alpha) = \sum_{n=0}^{\infty} \eps^{n+1}\,\Gamma_{n}\,\Bigl(\frac{\alpha}{4\pi}\Bigr)^{n+1}
\,,\qquad
\gamma(\alpha) = \sum_{n=0}^{\infty}\eps^{n+1}\, \gamma_{n}\, \Bigl(\frac{\alpha}{4\pi}\Bigr)^{n+1}
\,.\end{equation}
%%%
We have again introduced the formal expansion parameter $\eps \equiv 1$, which we will use to define
the resummation order. Since the $\mu$ dependence of $\gamma_F(\mu)$
primarily enters via the coupling constant's $\mu$ dependence, the $\beta$ RGE
for $\alpha(\mu)$ in \eq{betageneral} is an integral part of the full RGE system to be solved.
In particular, the $\eps$ parameter in \eqs{gammaF}{cuspexp} is the same
that appears in \eq{betageneral}.

As for the case of the coupling constant before, the truncation of
\eq{gammaF} together with \eq{betageneral} to a certain order in $\eps$
fundamentally \emph{defines} the resummation order. That is, keeping terms up to
$\ord{\eps^k}$ defines the Sudakov evolution at N$^k$LL order.
The explicit $1/\eps$ factor for the cusp term accounts
for the fact that it comes with an additional explicit logarithm relative to the noncusp term.
As a result, the noncusp term always enters at one lower order in perturbation theory than the
cusp term. And since the $\beta$ function in \eq{betageneral} starts at $\ord{\eps^0}$,
it enters at the same loop order as the cusp anomalous dimension.
So as usual, at N$^k$LL order, we require the $k+1$-loop cusp and beta function coefficients
and the $k$-loop noncusp coefficients.

Solving the RGE in \eq{RGE}, one finds
%%%
\begin{align} \label{eq:RGEsol}
F(\mu) = F(\mu_0) U(\mu_0,\mu)
\,,\end{align}
%%%
where $U(\mu_0,\mu)$ is the Sudakov evolution factor given by
%%%
\begin{align} \label{eq:evolutionkernel}
U(\mu_0,\mu)
= \exp\biggl\{\int_{\mu_0}^{\mu}\frac{\df\mu'}{\mu'} \gamma_F(\mu') \biggr\}
= \exp\biggl\{\int_{\mu_0}^{\mu}\frac{\df\mu'}{\mu'}\, \frac{1}{\epsilon}\,\Gamma_\cusp[\alpha(\mu')]\ln \frac{Q}{\mu'} +\gamma[\alpha(\mu')]
\biggr\}
\,.\end{align}
%%%
It resums the Sudakov logarithms appearing in the perturbative series of $F(\mu)/F(\mu_0)$.
One can easily check that counting powers of $\eps$ in the
anomalous dimension is equivalent to counting powers of logarithms in the Sudakov exponent,
simply because the entire structure of $U(\mu_0, \mu)$ is fully encoded by the anomalous
dimension. In the final resummed cross section, where $F(\mu)$ is combined with other
ingredients, $U(\mu_0,\mu)$ eventually appears with both scales $\mu$ and $\mu_0$ corresponding
to two different kinematic quantities such that $U(\mu_0, \mu)$ resums (part of) the
Sudakov logarithms of the ratio of these quantities.

We stress that while we obtained \eq{evolutionkernel} starting from the RGE in \eq{RGE},
a Sudakov evolution factor of the same structure (necessarily) appears in all the various
approaches for performing Sudakov resummation that exist in the literature. This includes
EFT-based and non-EFT-based approaches, and both analytic as well as numerical Monte-Carlo techniques
such as parton showers.

On the other hand, different implementations tend to follow different strategies
for evaluating the integral in the Sudakov exponent.
In the following sections we investigate several methods for doing so, paying
close attention to where additional assumptions and/or approximations are made.
As we will see, additional approximations that may appear mathematically justified
can still conspire to yield results for \eq{evolutionkernel} that exhibit
nontrivial numerical differences.
We investigate the following methods in the sections that follow:
%%%
\begin{itemize}
 \item {\bf{Numerical:}} In this method, both the $\beta$-function RGE \eq{betageneral} and the evolution kernel
 \eq{evolutionkernel} are evaluated fully numerically (with sufficiently high numerical precision that
 numerical integration errors are negligible).
 This provides the exact solution of the complete Sudakov RGE system at a given
 resummation order as defined above, and we will use it as the benchmark to compare the other methods against.
 As this method can be computationally expensive, it is often not very suitable
 for practical purposes.
 
 \item {\bf{Seminumerical:}} One option to speed up the fully numerically method is to employ
 an approximate analytic solution to the $\beta$-function RGE, but to perform the kernel integration numerically.
 In other words, we numerically integrate \eq{evolutionkernel} but with the iterative
 analytic solution for $\alpha(\mu)$ used in the perturbative expansion of the anomalous dimensions.
 This is described along with the fully numerical method in \subsec{nummethod}.
 
 \item {\bf{Unexpanded analytic:}} Approximate but fully analytic, closed-form expressions for
 \eq{evolutionkernel} can be obtained. One way to achieve this is to exploit the
 $\beta$ function to turn the $\mu$ integration into an integration over $\alpha$,
 which can be performed analytically after some expansion in $\eps$. This is
 combined with the analytic iterative solution for $\alpha_{s}(\mu)$ as input.
 We give details of the derivation and the resulting forms explicitly in \subsec{unexpanded}.

 \item {\bf{Expanded analytic:}}  Another way to obtain an approximate closed-form result
 for \eq{evolutionkernel} is to insert the analytic solution for $\alpha(\mu)$
 in \eq{alphaNNLOinv} in the perturbative expansions of $\Gamma(\alpha)$ and
 $\gamma(\alpha)$ and fully expanding the integrand in $\epsilon$. The
 integration can then be performed directly in terms of $\mu$.
 This method is described in more detail in \subsec{expanded}.

 \item {\bf{Reexpanded analytic:}} Finally, to connect to some of the available
 literature, we elaborate another analytic approach where, upon achieving the expanded
 analytic evolution kernels, one further expands $\alpha(\mu_0)$
 in terms of $\alpha(\mu_R)$ at a different reference scale $\mu_R$, assuming that
 $\mu_0 \sim \mu_R$, i.e., that there is no hierarchy between them.
 This method is discussed in more detail in \subsec{reexpand}.
\end{itemize}
%%%

Given these five methods of integration, we will probe their reliability in two
different ways in \subsec{1Dkernelpheno}: \emph{closure tests} and \emph{approximation errors}.
The latter simply tests the absolute difference between any one method and the
numerically exact method described above.
The closure test provides a test of the mathematical consistency of the evolution
factor. That is, it should satisfy the renormalization group property
%%%
\begin{equation} \label{eq:RGconsistency}
U(\mu_0, \mu_1)\,U(\mu_1, \mu_2) = U(\mu_0, \mu_2)
\,,\end{equation}
%%%
which is obvious from its definition in \eq{evolutionkernel}.
A simple way to test that \eq{RGconsistency} is satisfied is to consider the special case of
$\mu_2 = \mu_0$, which yields
%%%
\begin{align} \label{eq:closure2}
U(\mu_0,\mu)\, U(\mu,\mu_0)=1
\,,\end{align}
%%%
and simply expresses the fact that the RG evolution should close on itself.
Since the intermediate scale here is completely arbitrary, this property should
be satisfied exactly at any given order. However, deviations from unity
can arise due to simplifying assumptions or approximations made in evaluating
the integral in the exponent.

%===============================================================================
\subsection{Numerical and seminumerical methods}
\label{subsec:nummethod}
%===============================================================================

The most accurate method of integration is to perform the integration fully numerically.
The error introduced by the numerical integration routine can be made arbitrarily small
at the expense of computing time.
We always use a sufficiently high integration precision that the numerical integration error
is completely negligible.

Our strategy in the numerical and seminumerical approaches is summarized schematically as
%%%
\begin{align} \label{eq:Unum}
U(\mu_0,\mu)
= \exp\biggl\{\underbrace{\int_{\mu_0}^{\mu}\frac{\df \mu^{\prime}}{\mu^{\prime}}}_{\text{numerical}}
\sum_{n=0}^\infty \biggl[\eps^n \Gamma_n\!\!\! \overbrace{\Bigl(\frac{\alpha(\mu^{\prime})}{4\pi}\Bigr)^{n+1}}^{\textcolor{red}{\text{numerical}}\, / \, \textcolor{blue}{\text{\eq{alphaNNLO}}}} \!\!\!\ln\frac{Q}{\mu^{\prime}}
+ \eps^{n+1}\gamma_n\!\!\! \overbrace{\Bigl(\frac{\alpha(\mu^{\prime})}{4\pi}\Bigr)^{n+1}}^{\textcolor{red}{\text{numerical}}\, / \, \textcolor{blue}{\text{\eq{alphaNNLO}}}}
\biggr] \biggr\}
\,,\end{align}
%%%
where we have written the generic evolution kernel $U(\mu_{0},\mu)$ explicitly in terms of the perturbative series
of the anomalous dimensions from \eq{cuspexp}.
In both the numerical and seminumerical methods we truncate the series in parentheses at the desired
order in $\epsilon$, and evaluate the overall $\mu$-integration numerically.

For the numerical method, we use the exact numerical result for the solution of the running coupling
$\alpha(\mu')$, as indicated in red in \eq{Unum}.
For the seminumerical method we instead insert the iterative solution for the running of the coupling,
\eq{alphaNNLO}, as indicated in blue in \eq{Unum}. In both cases, the running of the coupling
is truncated at the appropriate order in $\eps$.

The seminumerical method turns out to provide a very good approximation of the
integral, since as we have seen in \sec{beta}, the iterative solution for the
running coupling provides a very accurate solution. At the same time, it is much
faster than the fully numerical method because it avoids calling the
computationally costly numerical solution for the running coupling in each
integrand call.

%===============================================================================
\subsection{Unexpanded analytic method}
\label{subsec:unexpanded}
%===============================================================================

We start from \eq{gammaF} and split the logarithm at $\mu_0$,
%%%
\begin{equation} \label{eq:logsplit}
\gamma_F(\mu) =
\frac{1}{\epsilon}\,\Gamma_\cusp[\alpha(\mu)]\biggl(\ln \frac{\mu_0}{\mu} + \ln\frac{Q}{\mu_0} \biggr) +\gamma[\alpha(\mu)]
\,.\end{equation}
%%%
To integrate it over $\mu$ we use the standard method of exploiting the $\beta$ function for $\alpha$,
%%%
\begin{align} \label{eq:mutoalpha}
\df\ln\mu = \frac{\df \alpha(\mu)}{\beta[\alpha(\mu)]}
\,,\end{align}
%%%
to change the integration variable from $\mu$ to $\alpha$.
To replace the explicit logarithm
of $\mu$ in \eq{logsplit}, we integrate \eq{mutoalpha} once to obtain
%%%
\begin{equation}
\ln\frac{\mu_0}{\mu} = \int_{\alpha(\mu)}^{\alpha(\mu_0)}\frac{\df\alpha}{\beta(\alpha)}
\,.\end{equation}
%%%
The evolution kernel in \eq{evolutionkernel} then takes the form
%%%
\begin{align} \label{eq:evolutionkerneldecomposed}
U(\mu_0,\mu)
= \exp\Bigl\{-K_{\Gamma}(\mu_0,\mu)+\eta_{\Gamma}(\mu_0,\mu)\ln\frac{Q}{\mu_0}+K_{\gamma}(\mu_0,\mu)\Bigr\}
\,,\end{align}
%%%
where the individual functions are defined as
%%%
\begin{align} \label{eq:KGamma}
K_{\Gamma}(\mu_0,\mu)
&= \frac{1}{\epsilon}\int_{\alpha(\mu_0)}^{\alpha(\mu)} \frac{\df \alpha}{\beta(\alpha)}\, \Gamma_\cusp(\alpha)\,
\int_{\alpha(\mu_0)}^{\alpha}\frac{\df\alpha'}{\beta(\alpha')}
\,,\\
\label{eq:etaGamma}
\eta_{\Gamma}(\mu_0,\mu)
&= \frac{1}{\epsilon}\int_{\alpha(\mu_0)}^{\alpha(\mu)}\frac{\df\alpha}{\beta(\alpha)}\, \Gamma_\cusp(\alpha)
\,,\\
\label{eq:Kgamma}
K_{\gamma}(\mu_0,\mu)
&= \int_{\alpha(\mu_0)}^{\alpha(\mu)} \frac{\df\alpha}{\beta(\alpha)}\, \gamma(\alpha)
\,.\end{align}
%%%

The integrals in \eqss{KGamma}{etaGamma}{Kgamma} yield an evolution kernel
which manifestly depends on $\mu$ and $\mu_0$ only via $\alpha(\mu)$ and $\alpha(\mu_0)$.
To illustrate this, for $K_{\Gamma}$ at LL order, $\sim\ord{\eps^0}$, we have
%%%
\begin{align} \label{eq:KGammaLLevkern}
K_{\Gamma}(\mu_0,\mu)
= \frac{\Gamma_0}{4\beta_0^2}\int_{a(\mu_0)}^{a(\mu)}\frac{\df a}{a}
\int_{a(\mu_0)}^{a}\frac{\df a'}{a^{\prime\,2}}
\,,\qquad
a(\mu) \equiv \frac{\alpha(\mu)}{4\pi}
\,.\end{align}
%%%
The integrations can be carried out easily to yield
%%%
\begin{align} \label{eq:KGammaLLsol}
K_{\Gamma}(\mu_0,\mu)
= - \frac{\Gamma_0}{4\beta_0^2}\,\frac{4\pi}{\alpha(\mu_0)}\Bigl(1-\frac{1}{r}-\ln r\Bigr)
\,,\qquad
r = \frac{\alpha(\mu)}{\alpha(\mu_0)}
\,.\end{align}
%%%

At NLL, $\ord{\eps}$, we have
%%%
\begin{align} \label{KGammaNLL}
K_{\Gamma}(\mu_0,\mu)
= \frac{\Gamma_0}{4\beta_0^2} \int_{a(\mu_0)}^{a(\mu)}\frac{\df a}{a}
\frac{1+ \eps\,\hat{\Gamma}_1 a}{1+\eps\,b_1 a}\int_{a(\mu_0)}^{a}\frac{\df a'}{a^{\prime\,2}}\,\frac{1}{1+\eps\,b_1 a'}
\,,\qquad
\hat{\Gamma}_n = \frac{\Gamma_n}{\Gamma_0}
\,.\end{align}
%%%
The approach followed in the unexpanded analytic method is to expand the denominators in $\eps$
keeping terms up to the $\ord{\eps}$, which induces an approximation error of $\ord{\eps^2}$, but
in turn allows one to easily perform the integration,
%%%
\begin{align} \label{eq:KGammaNLLexpansion}
K_{\Gamma}(\mu_0,\mu)
&= \frac{\Gamma_0}{4\beta_0^2}\biggl\{
\int_{a(\mu_0)}^{a(\mu)} \frac{\df a}{a} \Bigl[1+\eps(\hat{\Gamma}_1-b_1)a \Bigr]
\int_{a(\mu_0)}^{a}\frac{\df a'}{a^{\prime\,2}}
-\eps\, b_1 \int_{a(\mu_0)}^{a(\mu)}\frac{\df a}{a}\int_{a(\mu_0)}^{a}\frac{\df a'}{a'}
+ \ord{\eps^2} \biggr\}
\nn \\
&= -\frac{\Gamma_0}{4\beta_0^2} \biggl\{ \frac{4\pi}{\alpha(\mu_0)} \Bigl(1-\frac{1}{r} - \ln r \Bigr)
+ \eps\, \Bigl[(\hat{\Gamma}_1- b_1)(1-r+\ln r) + \frac{b_1}{2}\ln^2 r\Bigr] \biggr\}
\,.\end{align}
%%%

The other two integrals, $\eta_{\Gamma}$ and $K_{\gamma}$, are obtained in a
completely analogous fashion. At NLL, for $\eta_\Gamma$ we find
%%%
\begin{align} \label{eq:etaGammaNLL}
\eta_\Gamma(\mu_0,\mu)
& =-\frac{\Gamma_0}{2\beta_0 }\int_{a(\mu_0)}^{a(\mu)}\frac{\df a}{a} \bigl[1+\eps\,(\hat{\Gamma}_1-b_1) a + \ord{\eps^2} \bigr]
\nn \\
&= -\frac{\Gamma_0}{2\beta_0}\Bigl[\ln r + \eps\,\frac{\alpha(\mu_0)}{4\pi} (\hat{\Gamma}_1- b_1)(r-1)\Bigr]
\,.\end{align}
%%%
We stress, that in this method the anomalous dimensions in $\eta_\Gamma$ are kept
to the same loop order as in $K_\Gamma$, despite the fact that $\eta_\Gamma$
has an additional power of $\alpha(\mu_0)$ compared to $K_\Gamma$ at each order in $\eps$.
From the above derivation it should be clear that the separation of the cusp
term into $K_\Gamma$ and $\eta_\Gamma$ is arbitrary and merely a technical tool
to perform the integration. In other words, the fact that we used $\mu_0$ on the right-hand side
of \eq{logsplit} was merely for convenience and we could have used any other arbitrary scale.
Dropping the highest term in $\eta_\Gamma$ (as is sometimes done) amounts to multiplying
the $\ln(Q/\mu_0)$ in \eq{logsplit} by $\eps$, which introduces an artificial
dependence on $\mu_0$ into $\gamma_F$. In particular, doing so would lead to an
explicit violation of the RGE consistency in \eq{RGconsistency}.

From the expressions in \eqs{etaGamma}{Kgamma} it is clear that $K_\gamma$ can be
obtained from $\eta_\Gamma$ by replacing $\Gamma_n\to \gamma_n$ and multiplying by
an overall $\eps$. Hence, it does not contribute at LL, and at NLL we have
%%%
\begin{align} \label{eq:KgammaNLL}
K_{\gamma}(\mu_0,\mu) = -\frac{\gamma_0}{2\beta_0}\, \eps\, \ln r
\,.\end{align}
%%%

The calculation at higher orders proceeds in exactly the same fashion. The results
up to N$^3$LL are given in \app{RGE} in \eqss{KGammaN3LL}{etaGammaN3LL}{KgammaN3LL}.

To fully specify the method, we also have to specify how to obtain the integration
limits $\alpha(\mu)$ and $\alpha(\mu_0)$. To be consistent, we have to use
a solution for the $\beta$ RGE to the same order in $\eps$ to which we performed
the integration. To render the method fully analytic, we use the unexpanded
iterative solution to the corresponding order.

The Sudakov evolution factor expressed in terms of $K_\Gamma$, $\eta_\Gamma$,
$K_\gamma$ obtained as described above and combined with the iterative solution for $\alpha$
is precisely the form that is commonly used to perform Sudakov resummation in QCD
in much of the SCET literature; a few examples from a variety of applications
include e.g.~\refscite{Fleming:2007xt, Ligeti:2008ac, Abbate:2010xh, Berger:2010xi,
Stewart:2013faa, Almeida:2014uva, Hornig:2016ahz, Ebert:2017uel, Kang:2017cjk,
Chen:2018pzu, Procura:2018zpn, Bell:2018gce, Baumgart:2018yed, Lustermans:2019plv}.

%===============================================================================
\subsection{Expanded analytic method}
\label{subsec:expanded}
%===============================================================================

If, instead of changing integration variables from $\mu$ to $\alpha$ via \eq{mutoalpha},
one inserts the expanded solution \eq{alphaNNLOinv} for $\alpha(\mu)$ into the
integrand of the evolution kernel \eq{evolutionkernel} and expands in $\eps$,
the resulting integral exhibits an explicit $\mu$ dependence which can be integrated analytically.

Equivalently, the same result is obtained by starting from the unexpanded analytic
integrals of the previous subsection
[\eqss{KGammaN3LL}{etaGammaN3LL}{KgammaN3LL}], substituting the expanded solution
\eq{alphaNNLOinv} for $\alpha(\mu)$ in terms of $\alpha(\mu_0)$
and expanding everywhere in $\eps$.
We therefore refer to these kernels as ``expanded analytic'' as they involve a
further expansion in $\eps$ compared to the unexpanded ones of the previous subsection.

The expanded analytic evolution kernels to NNLL, $\ord{\eps^2}$, are given by
%%%
\begin{align} \label{eq:NNLLKGammaexp}
K_{\Gamma}(\mu_0,\mu)
&= -\frac{\Gamma_0}{(2\beta_0)^2} \biggl\{\frac{4\pi}{\alpha(\mu_0)}
\biggl[(1-X+\ln X) + \eps\, \frac{\alpha(\mu_0)}{4\pi}\, b_1 \frac{1-X}{X} \ln X
\nn \\ & \qquad
+\eps^2\, \frac{\alpha(\mu_0)^2}{(4\pi)^2} \frac{1}{X^2}\Bigl[ b_1^2 \Bigl((1-X)^2 + (1 - X)\ln X-\frac{1}{2}\ln^2\! X\Bigr)
- b_2 (1-X)^2\Bigr]\biggr]
\nn \\& \quad
+\eps\,(\hat{\Gamma}_1- b_1)\biggl[1-\frac{1}{X}-\ln X +\eps \frac{\alpha(\mu_0)}{4\pi}b_1 \frac{1-X}{X^2}\ln X \biggr]
\nn \\& \quad
+ \eps\,b_1 \biggl[\frac{1}{2}\ln^2\! X + \eps\, \frac{\alpha(\mu_0)}{4\pi}\,b_1 \frac{\ln^2\! X}{X}\biggr]
\nn \\ & \quad
+\eps^2\,\frac{\alpha(\mu_0)}{4\pi}\biggl[(b_1^2-b_2)\biggr(\frac{X^2-1}{2X^2}-\ln X\biggr)+(b_1\hat{\Gamma}_1-b_1^2)
 \frac{X-1-\ln X}{X}
\nn \\ & \qquad
-(\hat{\Gamma}_2-b_1\hat{\Gamma}_1)\frac{(X-1)^2}{2X^2}\biggr]
\biggr\}
\,,\end{align}
%%%
and
%%%
\begin{align} \label{eq:NNLLetaGammaexp}
\eta_{\Gamma}(\mu_0,\mu)
&= -\frac{\Gamma_0}{2\beta_0}\biggl\{
\ln\frac{1}{X} - \eps\,\frac{\alpha(\mu_0)}{4\pi} b_1\frac{\ln X}{X}
+ \eps^2\, \frac{\alpha(\mu_0)^2}{(4\pi)^2}\, \biggl[\frac{b_1^2}{2}\frac{\ln^2 X}{X^2}
-\frac{b_2}{X}\Bigl(1-\frac{1}{X}\Bigr)
\nn \\ & \qquad
-\frac{b_1^2}{X}\Bigl(\frac{\ln X}{X}+\frac{1}{X}-1\Bigr)\biggr]
\nn \\ & \quad
+\eps\,\frac{\alpha(\mu_0)}{4\pi}(\hat{\Gamma}_1-b_1) \biggr[\frac{1-X}{X}
-\eps\frac{\alpha(\mu_0)}{4\pi}b_1\frac{\ln X}{X^2}\biggr]
\nn \\ & \quad
+\eps^2\,\frac{\alpha(\mu_0)^2}{(4\pi)^2}(\hat{\Gamma}_2- b_1\hat{\Gamma}_1+ b_1^2- b_2)\frac{1-X^2}{2X^2}
\biggr\}
\,,\end{align}
%%%
where as before $b_i=\beta_i/\beta_0$, $\hat\Gamma_i = \Gamma_i/\Gamma_0$, and
$X\equiv X(\mu_0,\mu)=1+\frac{\alpha(\mu_0)}{2\pi}\beta_0\ln\frac{\mu}{\mu_0}$.
For illustration, we kept explicit the additional terms compared to the unexpanded
results that correspond to the expansion of
$r = \alpha(\mu)/\alpha(\mu_0)$ in terms of $\eps$.
The $K_{\gamma}$ kernel is again obtained from $\eta_\Gamma$ in \eq{NNLLetaGammaexp}
by replacing $\gamma_n\rightarrow\Gamma_n$ and multiplying by an overall $\eps$.
The N$^3$LL results are obtained analogously. Since they are rather lengthy and not very
illuminating we do not give them explicitly here.

As one can see, in this method the dependence on the two scales appears explicitly
in the argument of the logarithm in $X(\mu_0, \mu)$. One can still choose how
to obtain the value for $\alpha(\mu_0)$, for which we use by default
the expanded solution as it is closer in spirit to this method. In our
numerical results we will however also show the effect of using the iterative
solution for $\alpha(\mu_0)$.

%===============================================================================
\subsection{Reexpanded analytic method}
\label{subsec:reexpand}
%===============================================================================

This method is related to the expanded analytic method of the previous
subsection but uses a different treatment for the remaining dependence on $\alpha$.
That is, starting from the expanded results of the previous subsection,
all powers of $\alpha(\mu_0)$ are reexpanded in terms of $\alpha(\mu_R)$
evaluated at a reference scale $\mu_R$,
which is typically chosen equal or proportional to the (hard) kinematic variable $Q$.%
\footnote{In our notation, taking Drell-Yan as an example, the kinematic
variable $Q$ would be equivalent to the $Z$-boson or dilepton invariant mass.
It should not be confused with what is sometimes called the resummation scale
and also denoted as $Q \equiv Q_{\rm res}$, and which is the same as our
$\mu_0$, i.e.\ $\mu_0 \equiv Q_{\rm res}$.}

To illustrate this for the simplest case, we start from the LL expanded analytic result, which
written out explicitly is given by
%%%
\begin{align} \label{eq:reexpandedevolkern}
K_\Gamma = -\frac{\Gamma_0}{(2\beta_0)^2}\frac{4\pi}{\alpha(\mu_0)} \biggl[ \alpha(\mu_0) \frac{\beta_0}{2\pi} \ln \frac{\mu_0}{\mu} + \ln \Bigl(1-\alpha(\mu_0)\frac{\beta_0}{2\pi} \ln\frac{\mu_0}{\mu} \Bigr) \biggr]
\,.\end{align}
%%%
One now reexpands $\alpha(\mu_0)$ in terms of $\alpha(\mu_R)$ using its fixed-order expansion
%%%
\begin{align} \label{eq:fixedcoupling}
\alpha(\mu_0)
= \alpha(\mu_R) \Bigl[ 1 - \epsilon\,\alpha(\mu_R)\frac{\beta_0}{2\pi} \ln\frac{\mu_0}{\mu_R} + \ord{\epsilon^2} \Bigr]
\,.\end{align}
%%%
In doing so one assumes that $\mu_0 \sim \mu_R$ or, more precisely,
one explicitly chooses that the logarithms of $\mu_0/\mu_R$ are not resummed via the evolution
of $\alpha$ but are instead treated at fixed order. Formally, this is implemented
by multiplying any $\ln(\mu_0/\mu_R)$ in the relation between $\alpha(\mu_0)$ in terms
of $\alpha(\mu_R)$ by $\eps$ as in \eq{fixedcoupling} and expanding in $\eps$.
Substituting \eq{fixedcoupling} into \eq{reexpandedevolkern} and reexpanding to $\ord{\epsilon^0}$,
we obtain the ``reexpanded analytic'' result at LL,
%%%
\begin{align} \label{eq:LLreexpandedKernel}
K_\Gamma = -\frac{\Gamma_0}{2\beta_0}\, \frac{\lambda + \ln(1-\lambda)}{\lambda}\,L
\,,\qquad
\lambda \equiv \alpha(\mu_R)\, \frac{\beta_0}{2\pi}\, L
\,, \qquad L \equiv \ln \frac{\mu_0}{\mu}.
\end{align}
%%%

At this order, the result only involves the resummed logarithms $\ln(\mu_0/\mu)$.
The results at higher orders also contain explicit fixed-order logarithms $\ln(\mu_0/\mu_R)$
induced by the fixed-order expansion of $\alpha(\mu_0)$ in terms of $\alpha(\mu_R)$.
Since the higher-order results are very lengthy we do not give them here.
Their calculation is discussed for example in appendix C of \refcite{Catani:2003zt}.
The LL result for $K_\Gamma$ in \eq{LLreexpandedKernel} is equivalent to the
$L g^{(1)}(\alpha L)$ term in the notation there, and we also explicitly verified
that we reproduce the NLL $g^{(2)}(\alpha L)$ term. The N$^3$LL expressions are
taken from \refcite{Bizon:2017rah} translated to our conventions.

Another important difference in this method is the treatment of the $\eta_\Gamma$ term,
%%%
\begin{equation}
\eta_\Gamma(\mu_0, \mu)\, \ln \frac{Q}{\mu_0}
\,,\end{equation}
%%%
in the Sudakov exponent. Since $\mu_R$ and $Q$ are either identified or considered
of similar size, the explicit $\ln(Q/\mu_0)$ here is treated analogously to
the $\ln(\mu_0/\mu_R)$ in \eq{fixedcoupling} and multiplied by $\eps$, such that
the $\eta_\Gamma$ term is treated like the noncusp term and
included at one order lower than $K_\Gamma$. This is typically achieved by absorbing
it into the noncusp term as
%%%
\begin{align} \label{eq:gammabar}
\overline{\gamma}(\alpha) = \gamma(\alpha) + \Gamma_\cusp(\alpha) \ln\frac{Q}{\mu_0}
\,.\end{align}
%%%

For the specific choice $\mu_0 = \mu_R = Q$, this method reduces to the expanded
analytic method, since all logarithms that are treated differently vanish exactly.
However, as soon as these scales are chosen not to coincide the different
treatment of the $\mu_0$ dependence can have a sizeable numerical
effect on the evolution kernel, as we will see below.
It is also important to note that since the first and second arguments of
$U(\mu_0, \mu)$ are explicitly treated differently, and $\mu_0$ is assumed to be of
order $Q$, the group property in \eq{RGconsistency} is lost, and similarly
the closure condition in \eq{closure2} becomes meaningless. In other words, with
the above modifications the evolution factor explicitly targets a specific situation
and cannot (and is not meant to) be used to evolve between two arbitrary scales.

This reexpanded analytic expression for the Sudakov evolution factor
is also commonly used, in particular in the formalism of \refscite{Catani:2000vq,
Catani:2003zt, Bozzi:2005wk, Bozzi:2010xn} and (presumably most) implementations
following it, and also in the formalism of \refscite{Banfi:2012jm, Banfi:2014sua, Bizon:2017rah}.

%===============================================================================
\subsection{Numerical analysis of the evolution kernel}
\label{subsec:1Dkernelpheno}
%===============================================================================

Having elaborated various methods for evaluating the Sudakov evolution factor,
we now study their numerical behaviour. We consider the relative deviation from
the exact closure condition in \eq{closure2} as well as the approximation
error as given by the relative difference of each method to the exact numerical
method.

In general, the Sudakov evolution factor depends on the process and observable
under consideration, so we have to specify a concrete example.
Here, we consider the hard evolution for two colored partons in QCD, namely
the $q\bar q$ vector current (corresponding to Drell-Yan production or $e^+e^- \to$ dijets)
and the $gg$ scalar current (corresponding to $gg\to$ Higgs production).
The relevant hard evolution kernel is given by
%%%
\begin{align} \label{eq:hardevolution}
U^i(\mu_0,\mu)
&= \exp\biggl\{\int_{\mu_0}^{\mu}\frac{\df\mu'}{\mu'}\, 4\Gamma^i_\cusp[\alpha_s(\mu')]\ln \frac{Q}{\mu'}
\biggr\}
\times  \exp\biggl\{\int_{\mu_0}^{\mu}\frac{\df\mu'}{\mu'} \gamma_H^i[\alpha_s(\mu')]
\biggr\}
\nn \\
&\equiv U^{i}_{\Gamma}(\mu_0,\mu) \times U^{i}_{\gamma}(\mu_0,\mu)
\,,\end{align}
%%%
where $i=q,g$ denotes the quark and gluon cases,
and we have separately defined the cusp $U^{i}_{\Gamma}$ and noncusp $U^{i}_{\gamma}$
evolution factors. All relevant anomalous dimension coefficients are given in
\app{anomdimQCD}.

In this case, the kinematic quantity $Q$ appearing in \eq{hardevolution} is the
invariant mass $Q = \sqrt{q^2}$ of the momentum $q$ flowing
through the current, which we take to be $Q = 100\GeV$ (i.e.\ typical of Drell-Yan
or Higgs production). We consider the case of evolving from $\mu_0 \sim Q$
to an arbitrary scale $\mu$ over two decades above and below. As before, we
use $n_f = 5$ and ignore any flavor thresholds.

Of course, the takeaway message of our analysis would be the same if we were to
use anomalous dimension coefficients associated with soft or collinear quantities.
The advantage of considering the hard evolution is that it is multiplicative and independent
of the low-energy observable, so it provides a simple and generic use case, while
the only process dependence of the evolution is via the color channel.
Furthermore, the hard evolution factor $U^i(\mu_0, \mu)$ in this scenario has
a direct correspondence in various resummation formalisms.
In the formalisms of \refscite{Catani:2000vq, Catani:2003zt, Bozzi:2005wk, Bozzi:2010xn}
and \refscite{Banfi:2012jm, Banfi:2014sua, Bizon:2017rah} it corresponds to the
Sudakov form factor or the Sudakov radiator with $\mu_0 \equiv Q_{\rm res}$
being the resummation scale. In the context of SCET, it constitutes the evolution
factor for the $q\bar q$ and $gg$ hard functions with $\mu_0\equiv\mu_H$ being
the renormalization scale of the hard function. In all cases, $\mu$ would then be associated
with an appropriate low-energy quantity, e.g.\ $\mu \sim b_0/b_T \sim p_T$ in the case
of $p_T$ resummation.

%~~~~~~~~~~~~~~~~~~~~~~~~~~~~~~~~~~~~~~~~~~~~~~~~~~~~~~~~~~~~~~~~~~~~~~~~~~~~~~~
\subsubsection{Closure tests}
%~~~~~~~~~~~~~~~~~~~~~~~~~~~~~~~~~~~~~~~~~~~~~~~~~~~~~~~~~~~~~~~~~~~~~~~~~~~~~~~

%-------------------------------------------------------------------------------
\begin{figure*}
\centering
\includegraphics[width=\WidthTwoSubfigs]{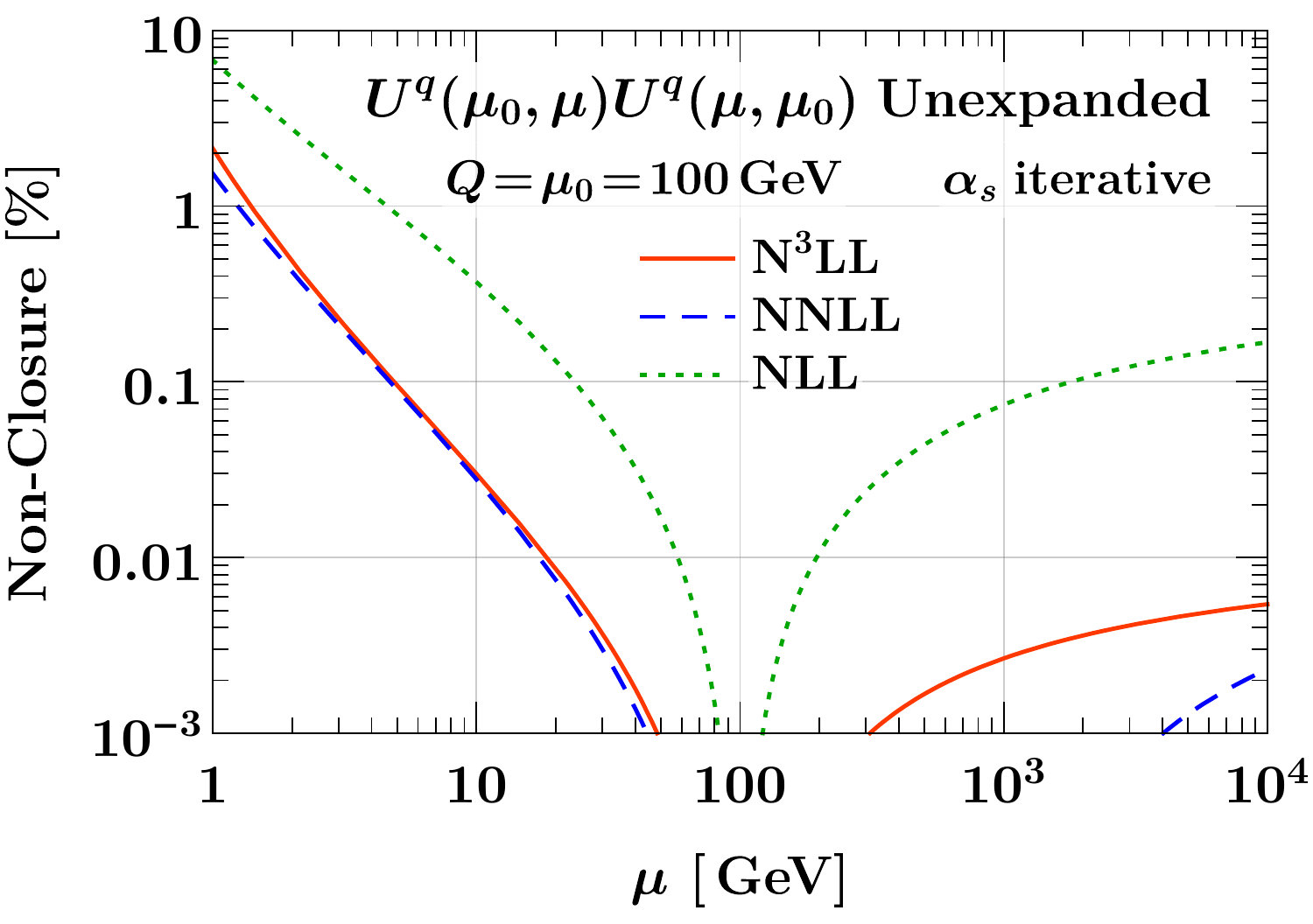}%
\hfill%
\includegraphics[width=\WidthTwoSubfigs]{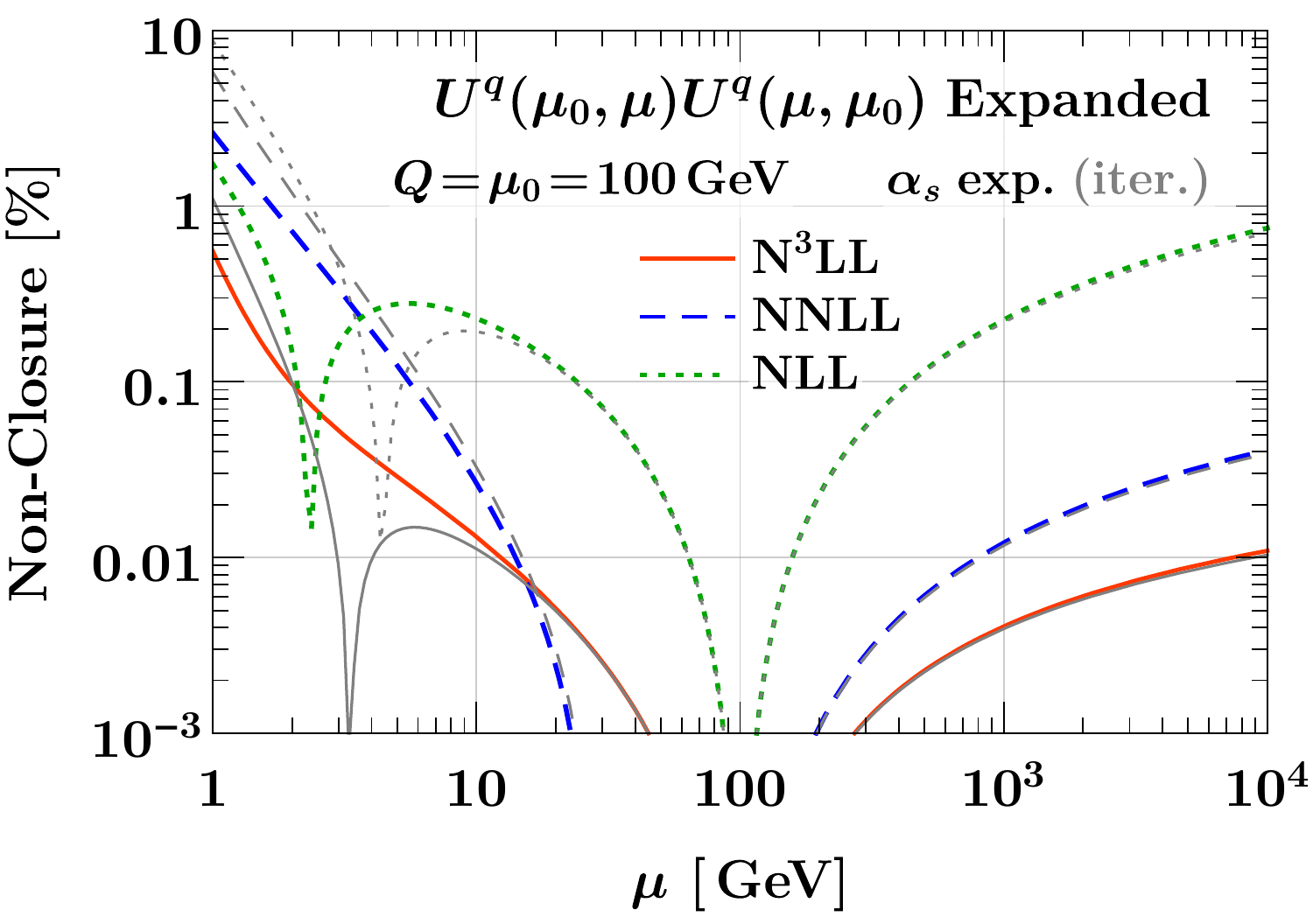}%
\caption{Deviation from the closure condition $U^q(\mu_0, \mu) U^q(\mu, \mu_0) = 1$
for the quark evolution kernels from NLL to N$^{3}$LL for
the unexpanded method (left) and expanded method (right).
For the expanded method, the gray lines show the result of using the iterative
instead of the expanded $\alpha_s(\mu_0)$.}
\label{fig:closurequark}
\end{figure*}
%-------------------------------------------------------------------------------

%-------------------------------------------------------------------------------
\begin{figure*}
\centering
\includegraphics[width=\WidthTwoSubfigs]{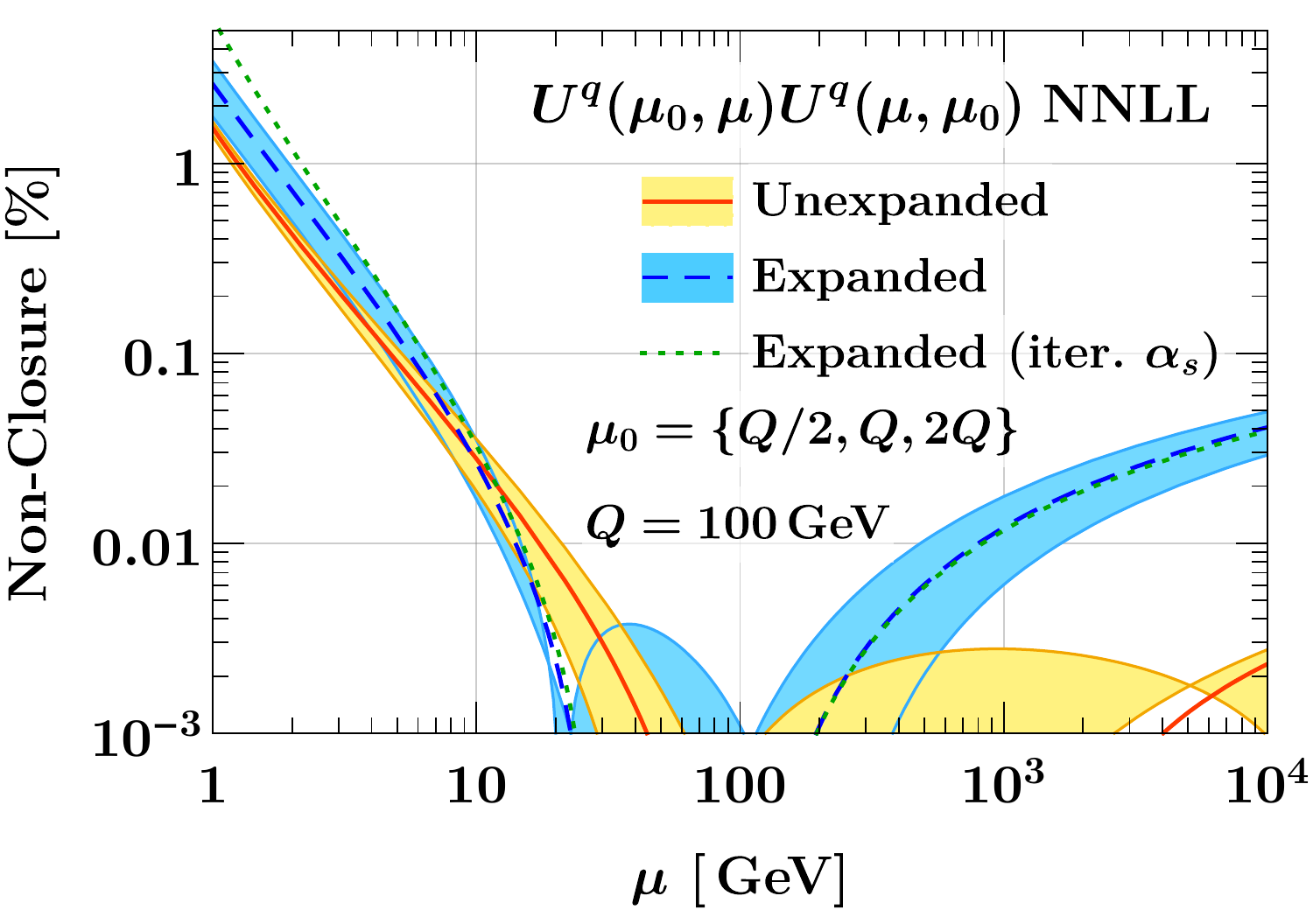}%
\hfill%
\includegraphics[width=\WidthTwoSubfigs]{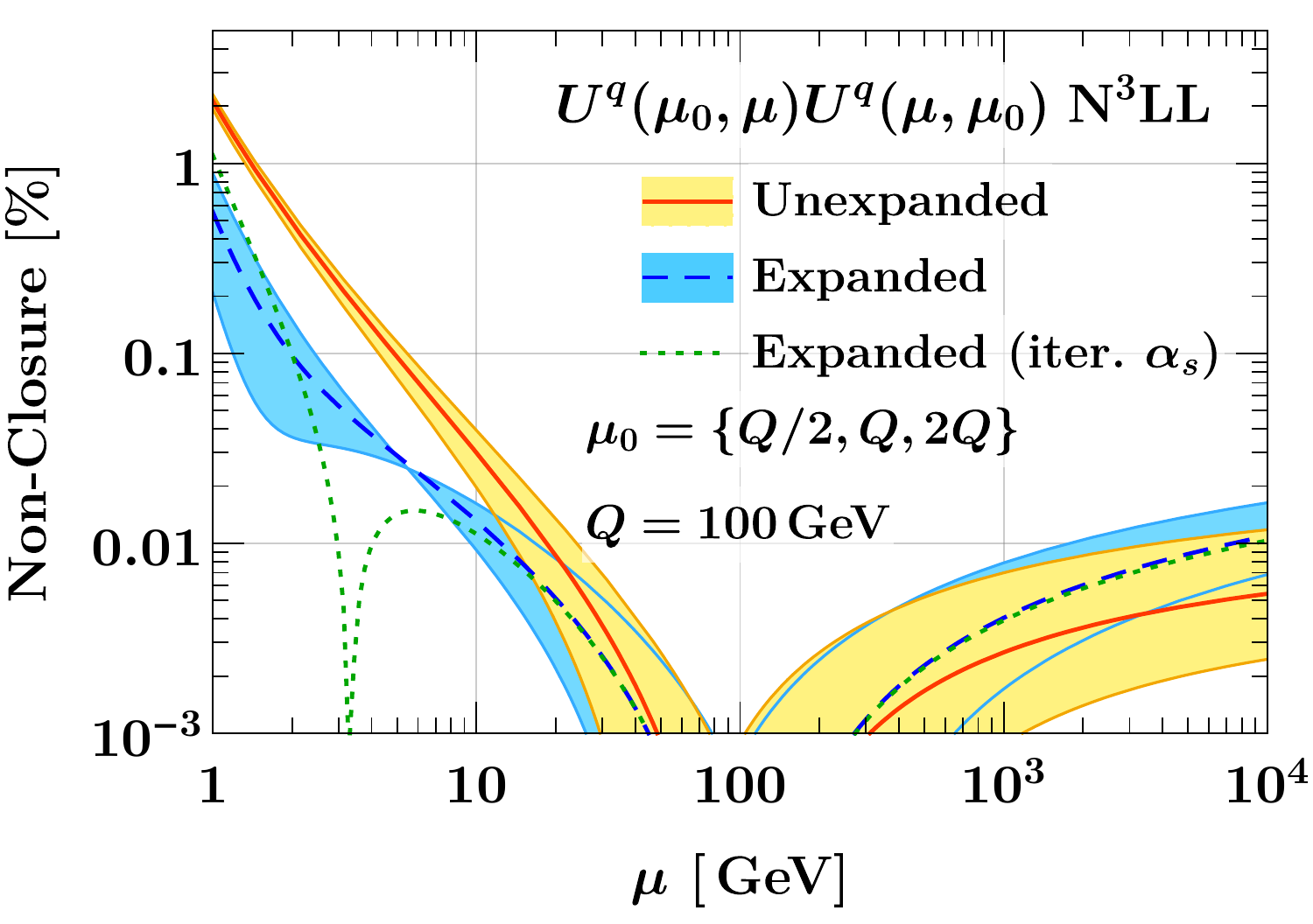}%
\caption{Deviation from the closure condition $U^q(\mu_0, \mu) U^q(\mu, \mu_0) = 1$
at NNLL (left) and N$^3$LL (right) for the unexpanded method (solid) and expanded
method (dashed). The central lines are for $\mu_0 = Q$, while the bands show the variation
when changing $\mu_0$ to $Q/2$ and $2Q$. The green dotted line shows the result of using
the iterative $\alpha_s(\mu_0)$ in the expanded kernels (for $\mu_0 = Q$ only).}
\label{fig:closurequarkVARY}
\end{figure*}
%-------------------------------------------------------------------------------

By construction, the seminumerical and numerical methods satisfy the closure
condition exactly, since they treat the integral and integration limits exactly.
As already mentioned in \subsec{reexpand}, the closure condition cannot be
applied to the reexpanded method. We therefore only consider the unexpanded and
expanded methods here. The reason these methods do not satisfy the closure
condition exactly is due to the expansions in the integrand after the variable
transformation from $\mu$ to $\alpha$, which means that the change of variables
in the integration limits and the integrand are not in exact correspondence.%
\footnote{Note that in \refcite{Bell:2018gce} modified unexpanded kernels were
constructed that restore exact closure.}

In \fig{closurequark}, we compare the non-closure for the full quark evolution
factor at different orders for the unexpanded (left panel) and expanded (right
panel) methods. At LL, the integrals are trivially exact and satisfy exact closure, so
we do not show them. At NLL, the non-closure can be quite sizeable, exceeding
$\gtrsim 5\%$ when running to low scales. For the expanded kernels it even
reaches $\sim 1\%$ when running to high scales. While the non-closure effect
is reduced at higher orders, it can still reach $1-2\%$ at the lowest scales, and
for the unexpanded kernels it does not reduce from NNLL to N$^3$LL. For the
expanded kernels, the non-closure reduces to below $1\%$ at N$^3$LL, but this is
likely accidental, since the expanded kernels are very sensitive to numerical
cancellations when evolving to scales $\mu \lesssim 10\GeV$. This is evident
when comparing the effect of using the iterative instead of the expanded
solution for $\alpha_s(\mu_0)$: Even a small change in $\alpha_s(\mu_0)$
causes large changes in the observed level of non-closure.

In \fig{closurequarkVARY}, we compare the unexpanded and expanded methods to each
other at NNLL (left panel) and N$^3$LL (right panel). In addition we vary $\mu_0$
away from $Q$ to $Q/2$ and $2Q$, as one would do in practical applications to
estimate a resummation uncertainty. Here, this should however not be considered
as an uncertainty estimate on the non-closure. Rather, it illustrates the effect of
$\eta_\Gamma$, which contributes when $\mu_0 \neq Q$, and the level at which
the non-closure may influence such uncertainty estimates. The (non-)closure of
the unexpanded kernels tends to be less sensitive to the choice of $\mu_0$ than
the expanded ones.

Overall, we might say that the unexpanded kernels show a somewhat better closure
behaviour. However, considering that at N$^3$LL one is aiming for perturbative
precision in the several percent range, their non-closure at this order is
uncomfortably large.

For brevity we have only shown results for the quark evolution kernels here.
The gluon evolution kernels have the same qualitative behaviour. The only difference
is that the overall non-closure effect is about a factor of two larger for gluons
than for quarks, corresponding to their larger color factor.

%~~~~~~~~~~~~~~~~~~~~~~~~~~~~~~~~~~~~~~~~~~~~~~~~~~~~~~~~~~~~~~~~~~~~~~~~~~~~~~~
\subsubsection{Approximation errors}
%~~~~~~~~~~~~~~~~~~~~~~~~~~~~~~~~~~~~~~~~~~~~~~~~~~~~~~~~~~~~~~~~~~~~~~~~~~~~~~~

%-------------------------------------------------------------------------------
\begin{figure*}
\centering
\includegraphics[width=\WidthTwoSubfigs]{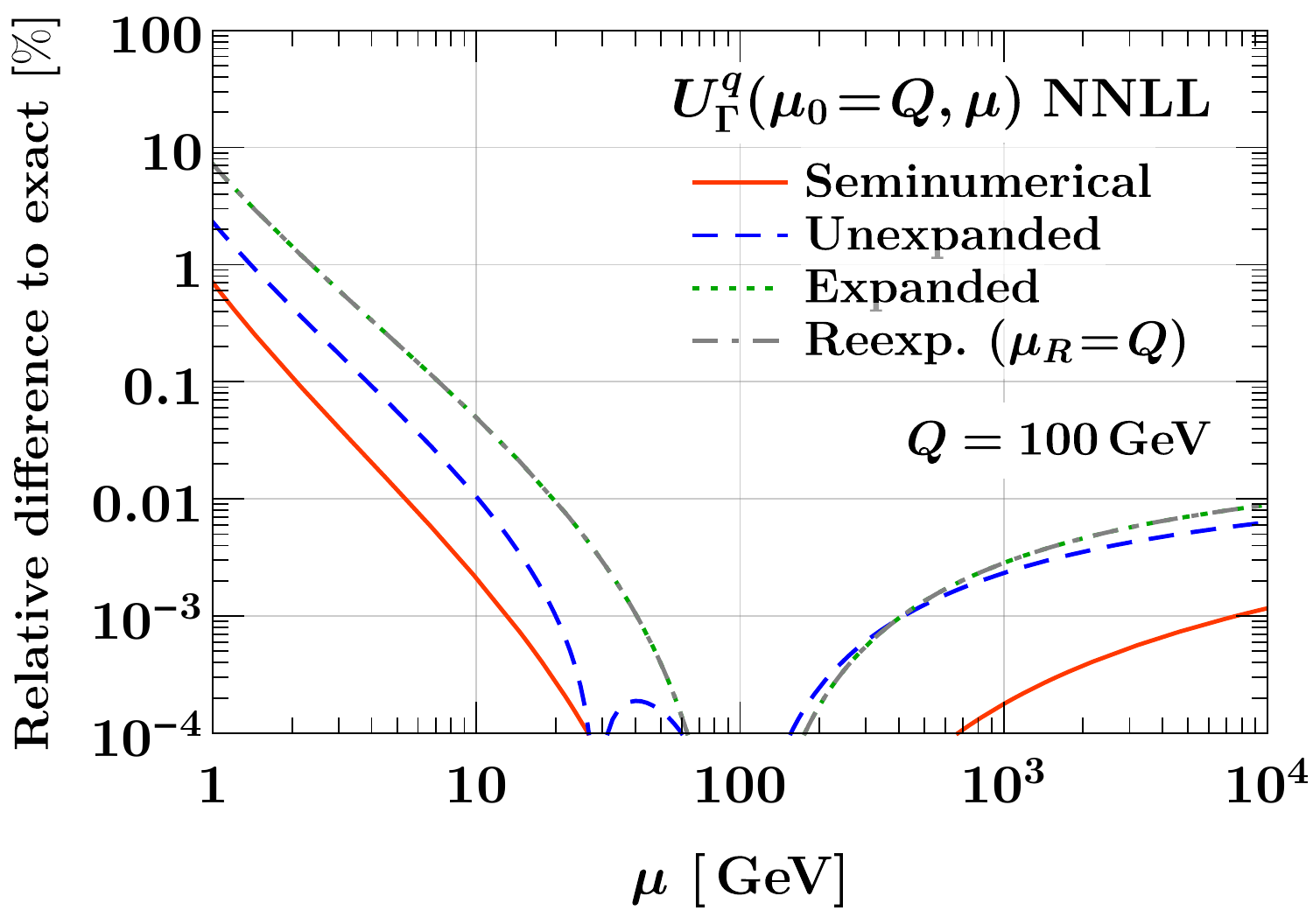}%
\hfill%
\includegraphics[width=\WidthTwoSubfigs]{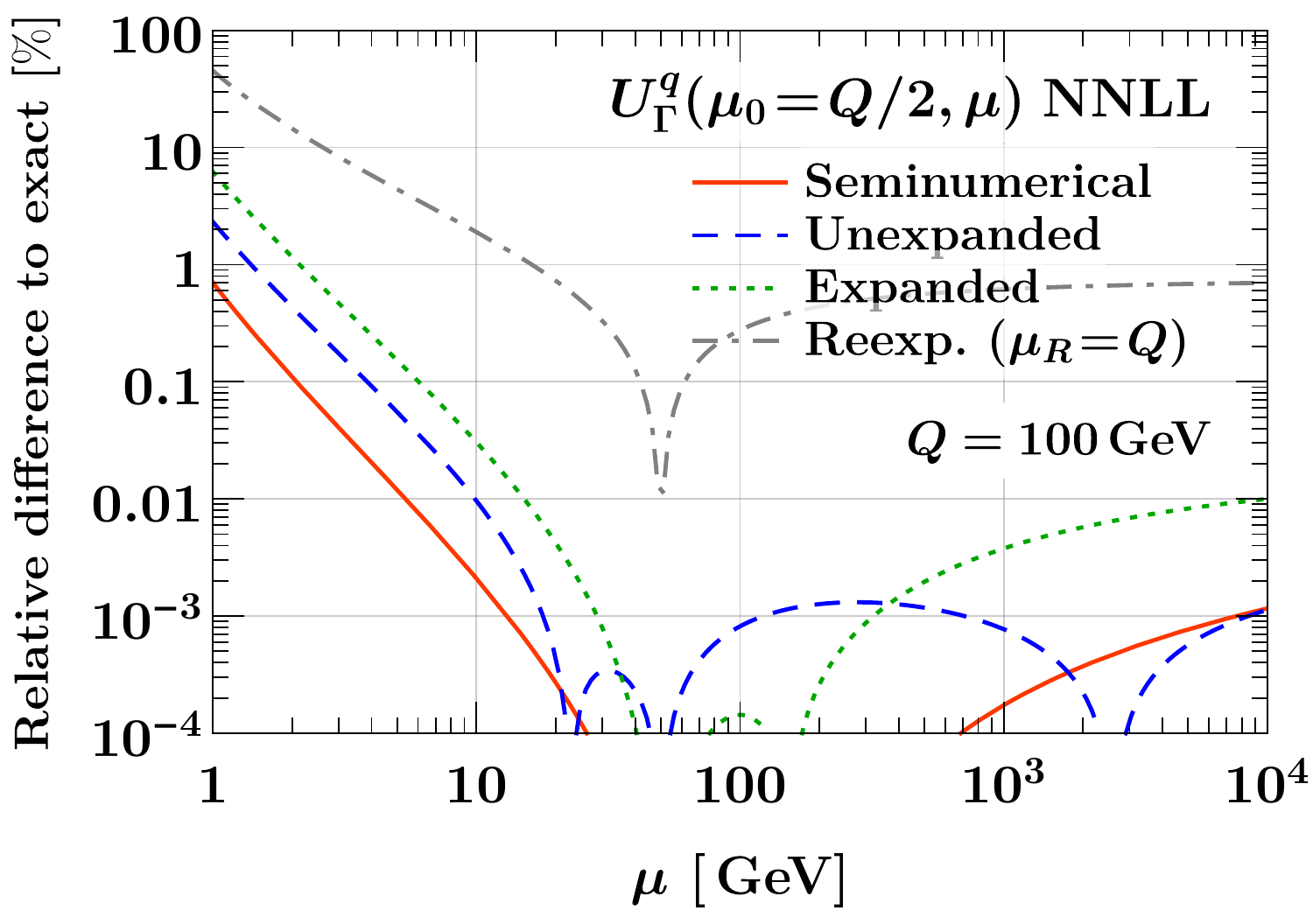}%
\hfill%
\includegraphics[width=\WidthTwoSubfigs]{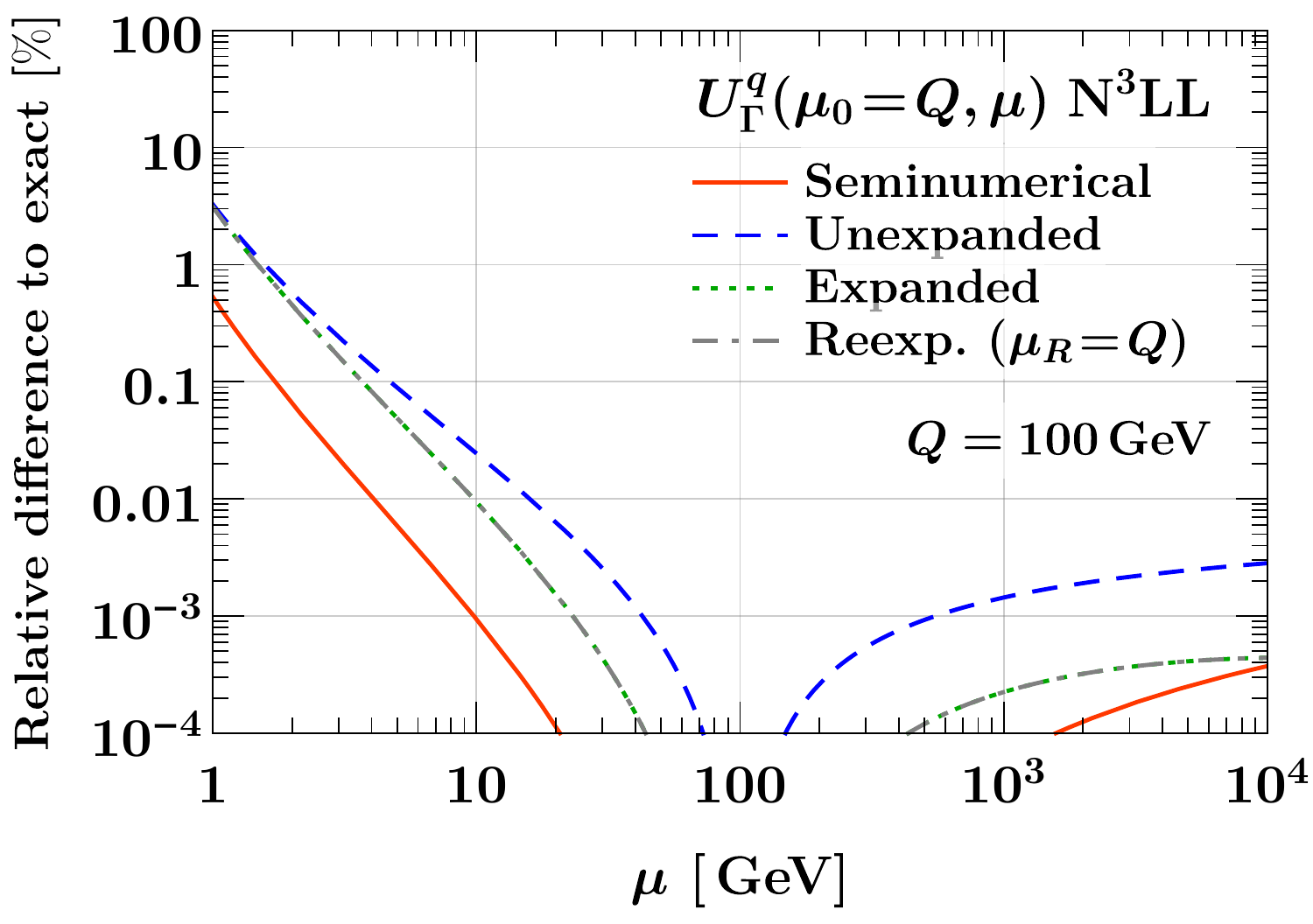}%
\hfill%
\includegraphics[width=\WidthTwoSubfigs]{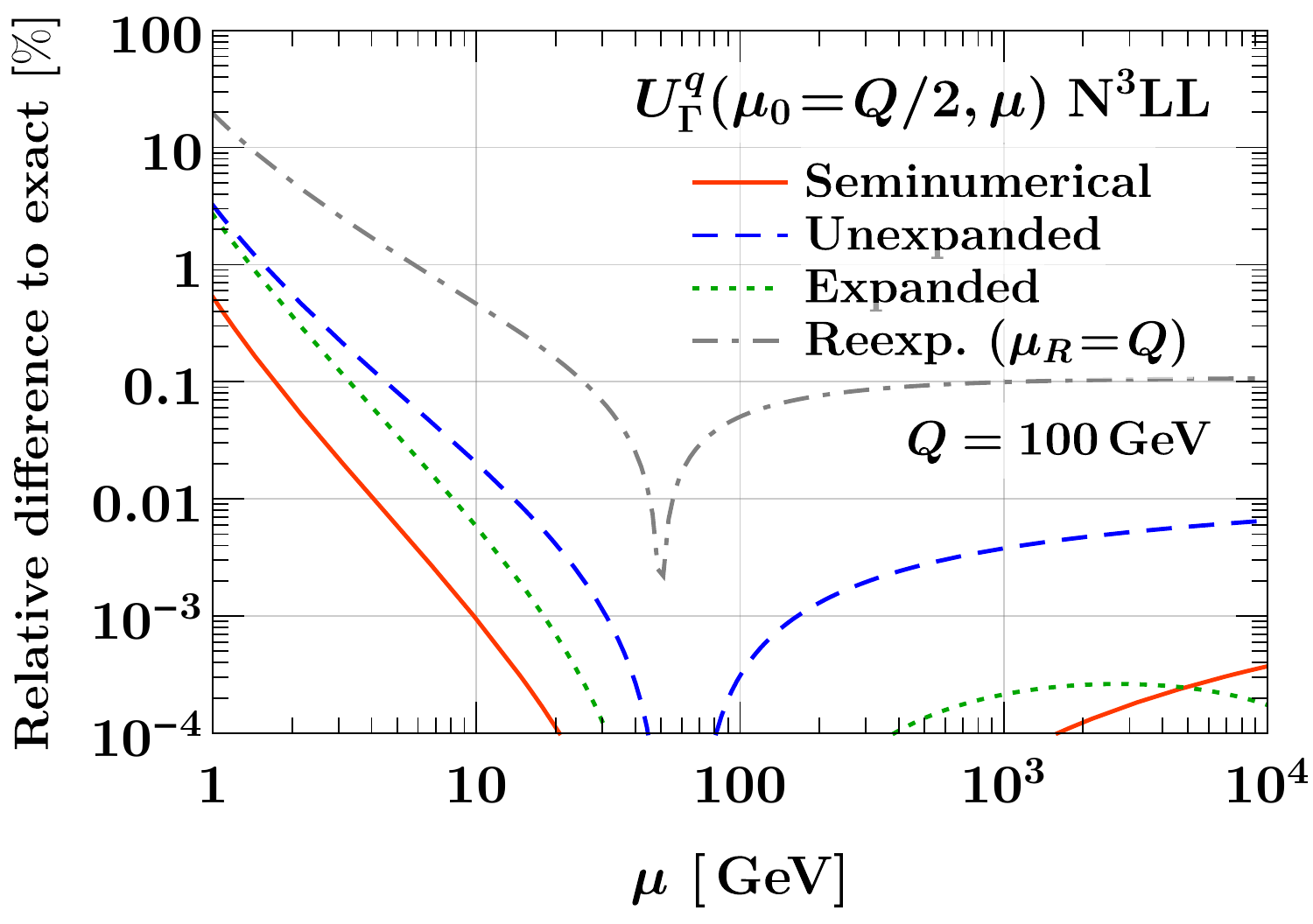}%
\caption{Deviation from the exact result for the quark cusp evolution kernel at NNLL (top) and N$^3$LL (bottom) for all four approximate integration methods. On the left we use $\mu_0 = Q$,
for which the expanded and reexpanded kernels are equivalent.
On the right we use $\mu_0 = Q/2$, with $\mu_R = Q$ for the reexpanded method.}
\label{fig:cusperr}
\end{figure*}
%-------------------------------------------------------------------------------

%-------------------------------------------------------------------------------
\begin{figure*}
\centering
\includegraphics[width=\WidthTwoSubfigs]{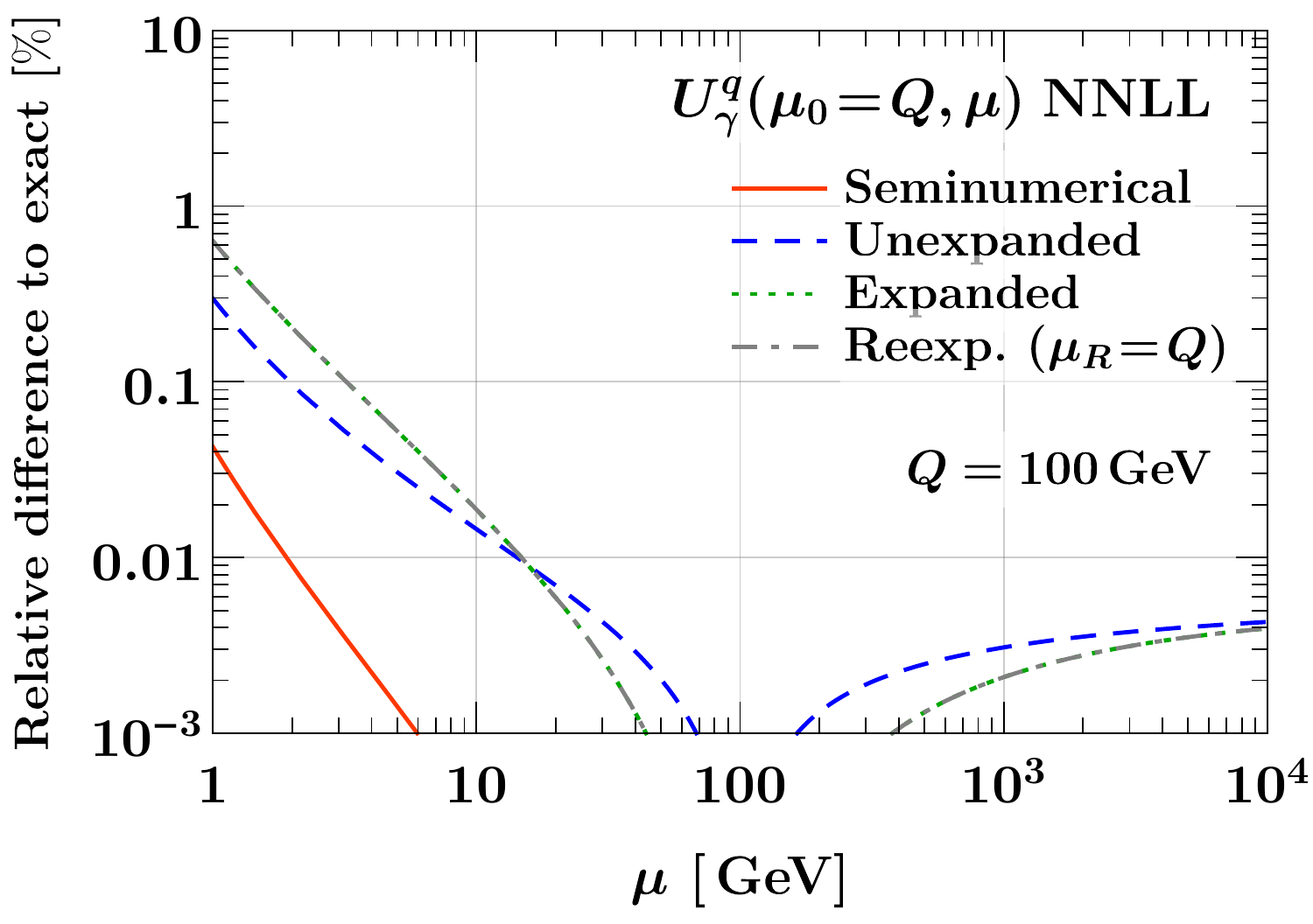}%
\hfill%
\includegraphics[width=\WidthTwoSubfigs]{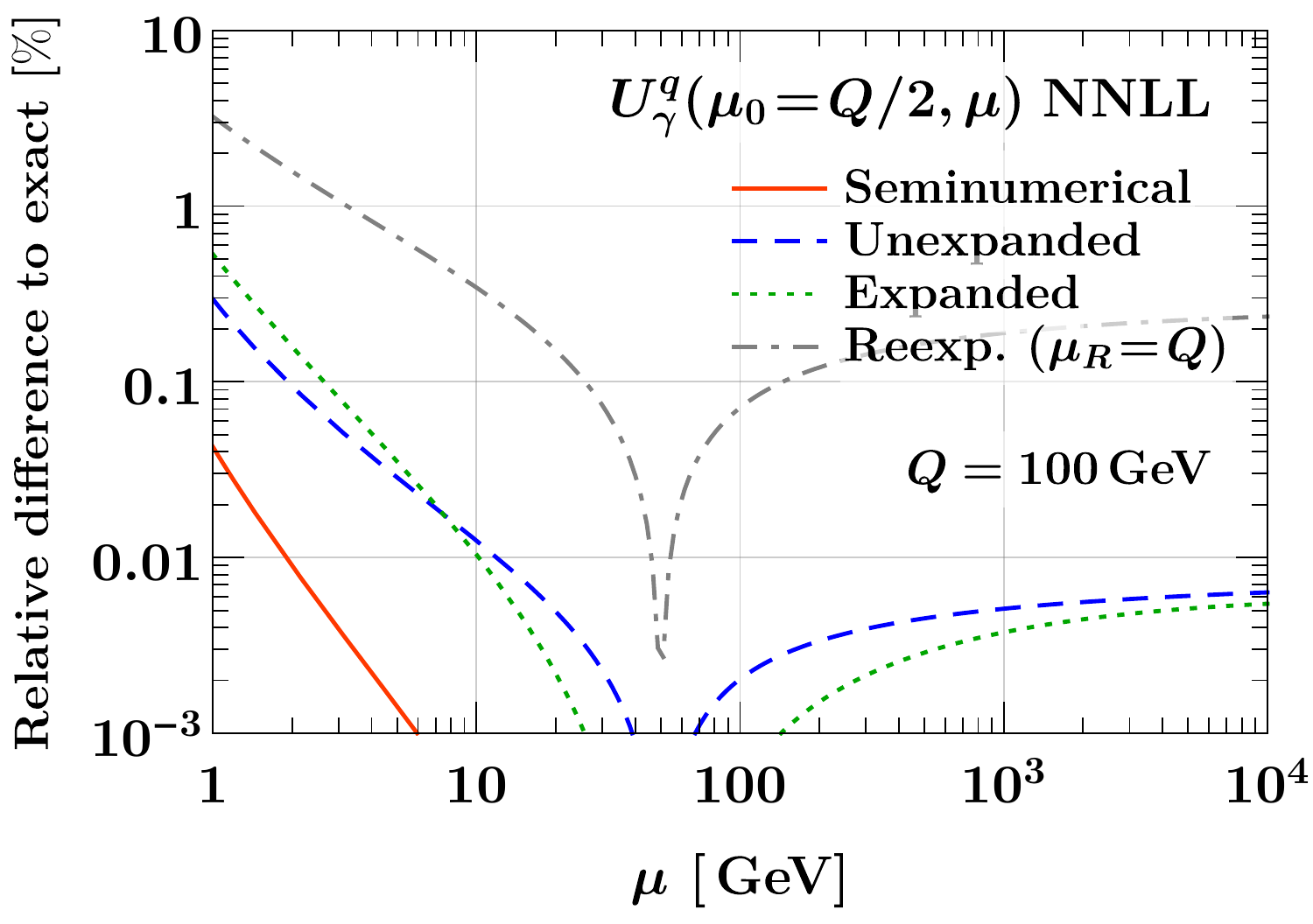}%
\hfill
\includegraphics[width=\WidthTwoSubfigs]{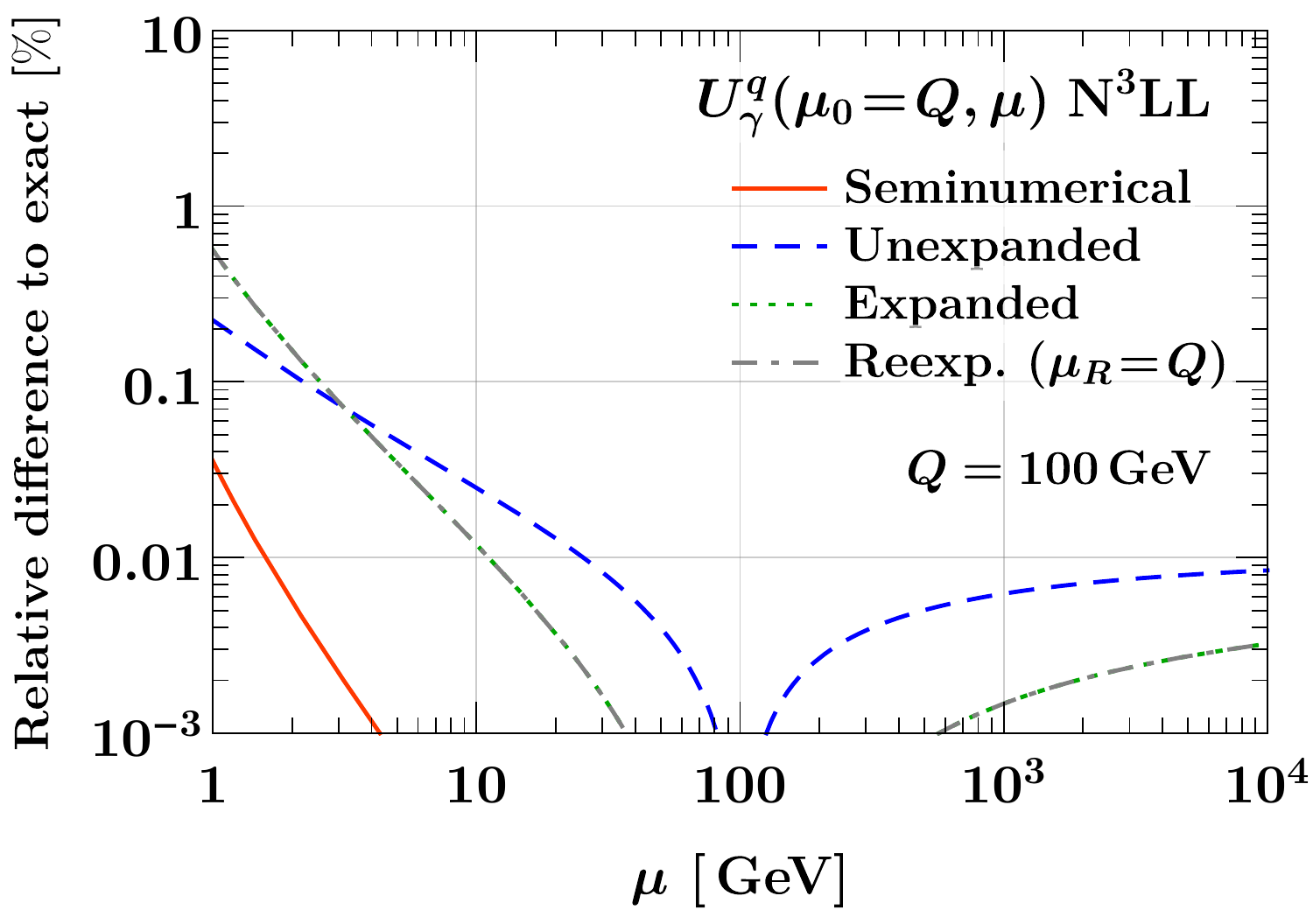}%
\hfill%
\includegraphics[width=\WidthTwoSubfigs]{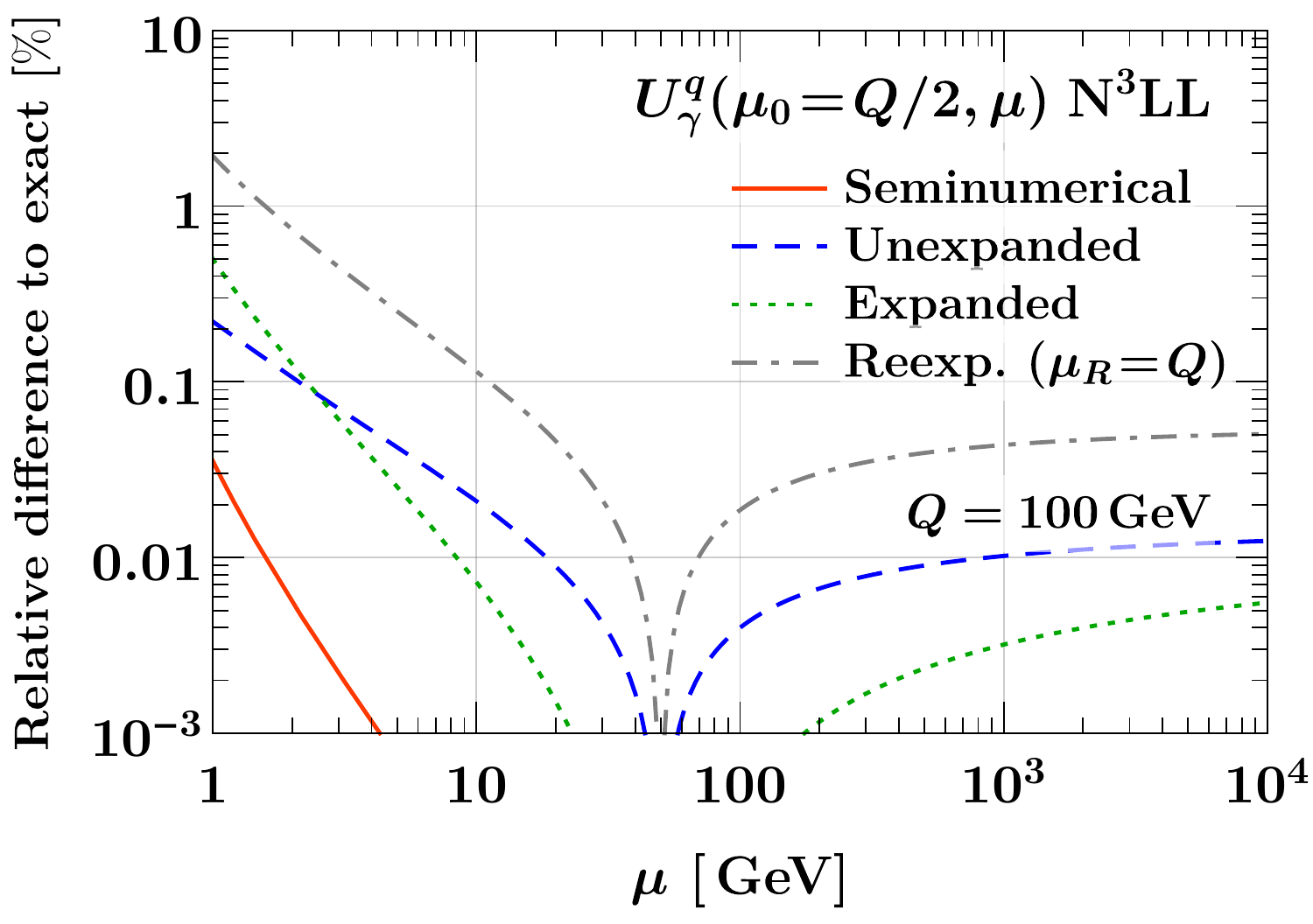}%
\caption{Same as \fig{cusperr}, but for the quark noncusp evolution kernels.}
\label{fig:noncusperr}
\end{figure*}
%-------------------------------------------------------------------------------

%-------------------------------------------------------------------------------
\begin{figure*}
\centering
\includegraphics[width=\WidthTwoSubfigs]{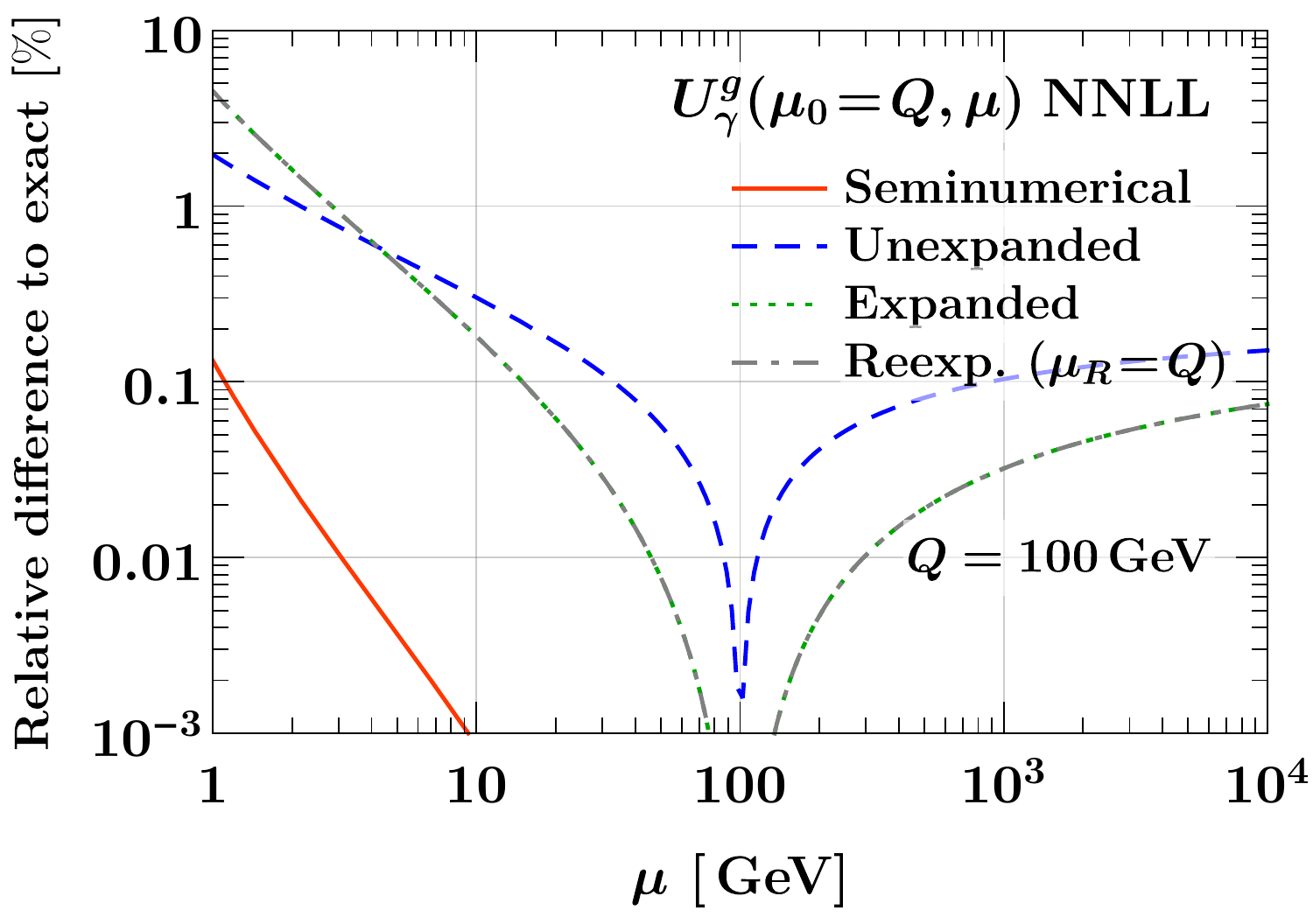}%
\hfill%
\includegraphics[width=\WidthTwoSubfigs]{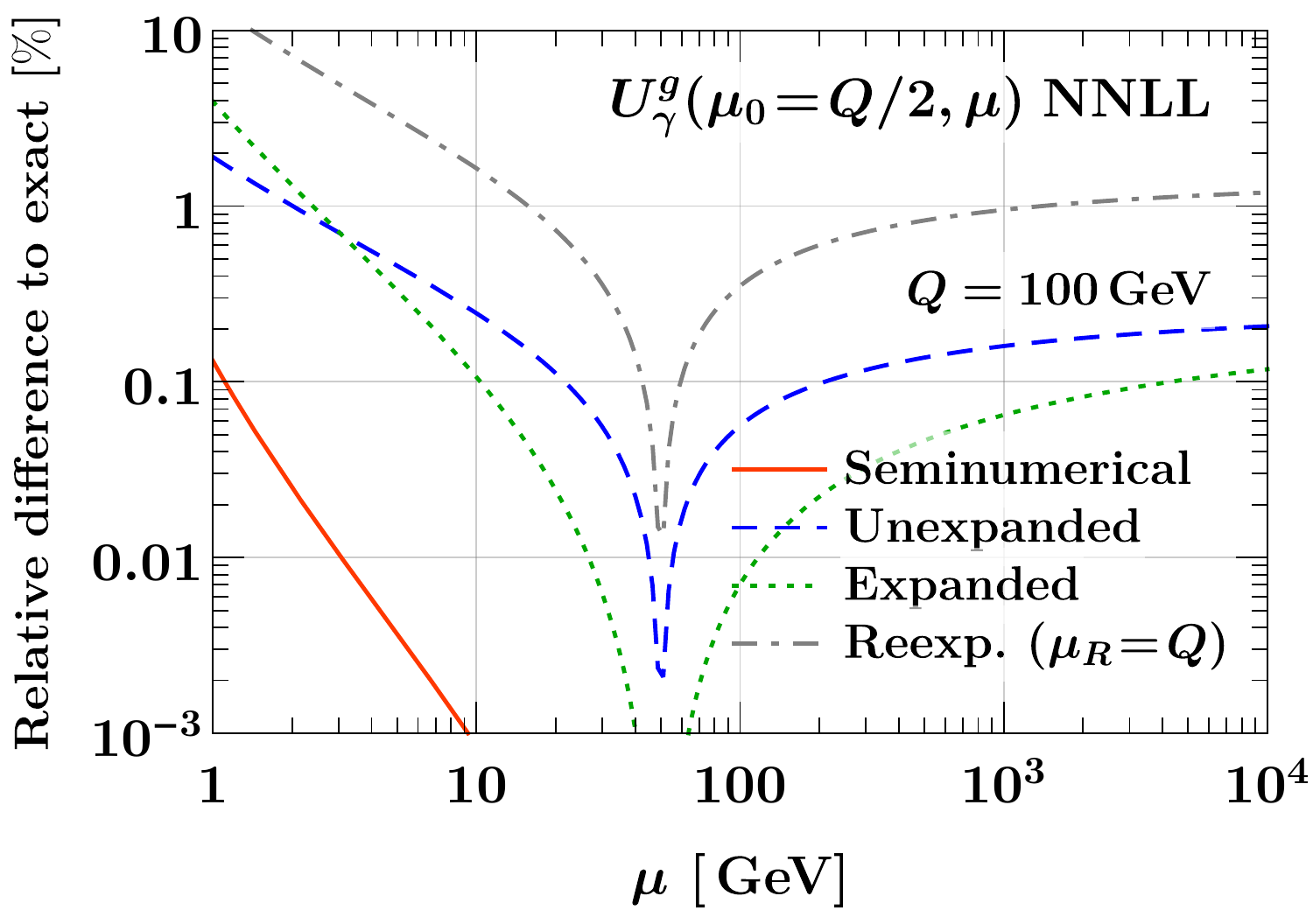}%
\hfill
\includegraphics[width=\WidthTwoSubfigs]{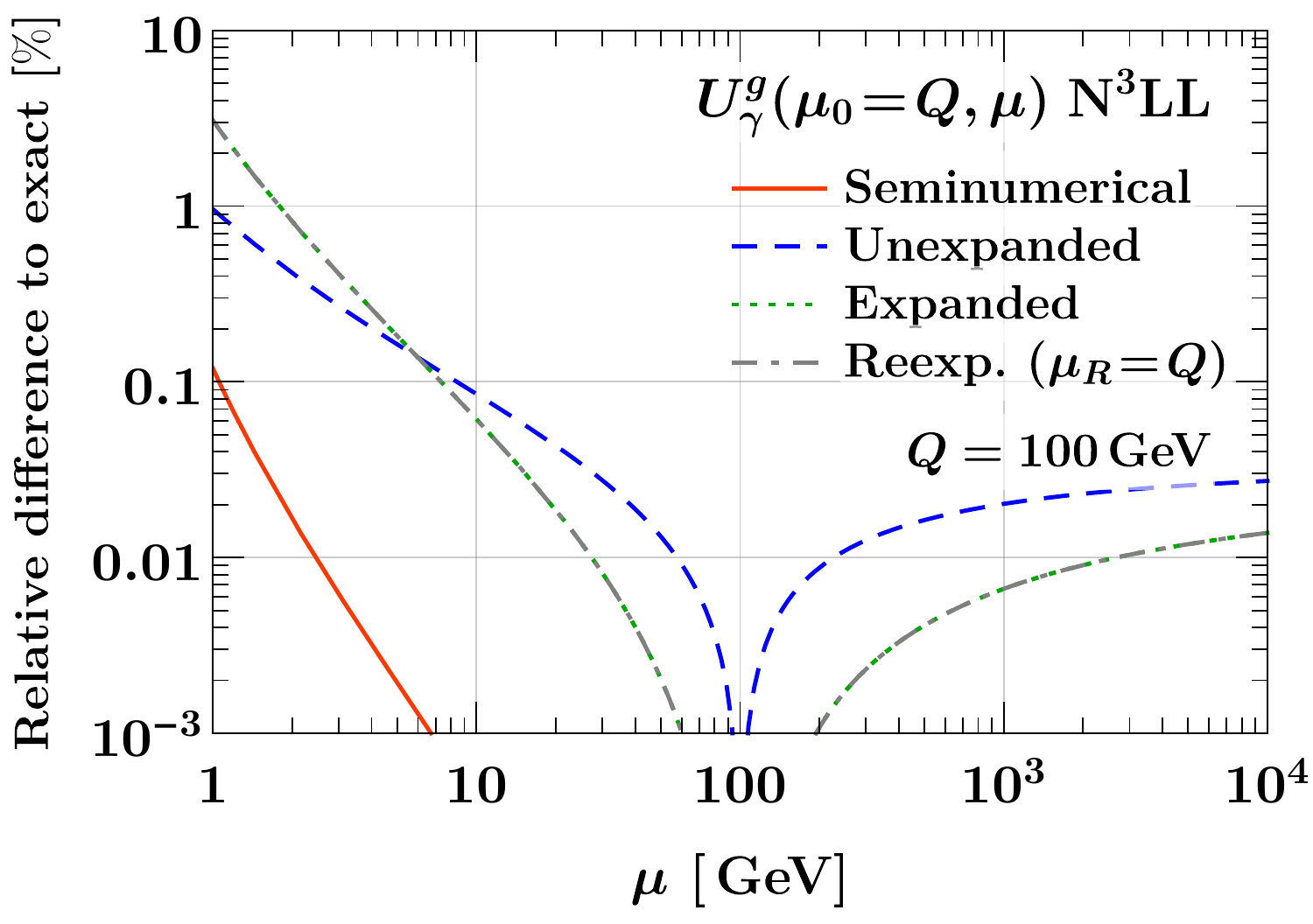}%
\hfill%
\includegraphics[width=\WidthTwoSubfigs]{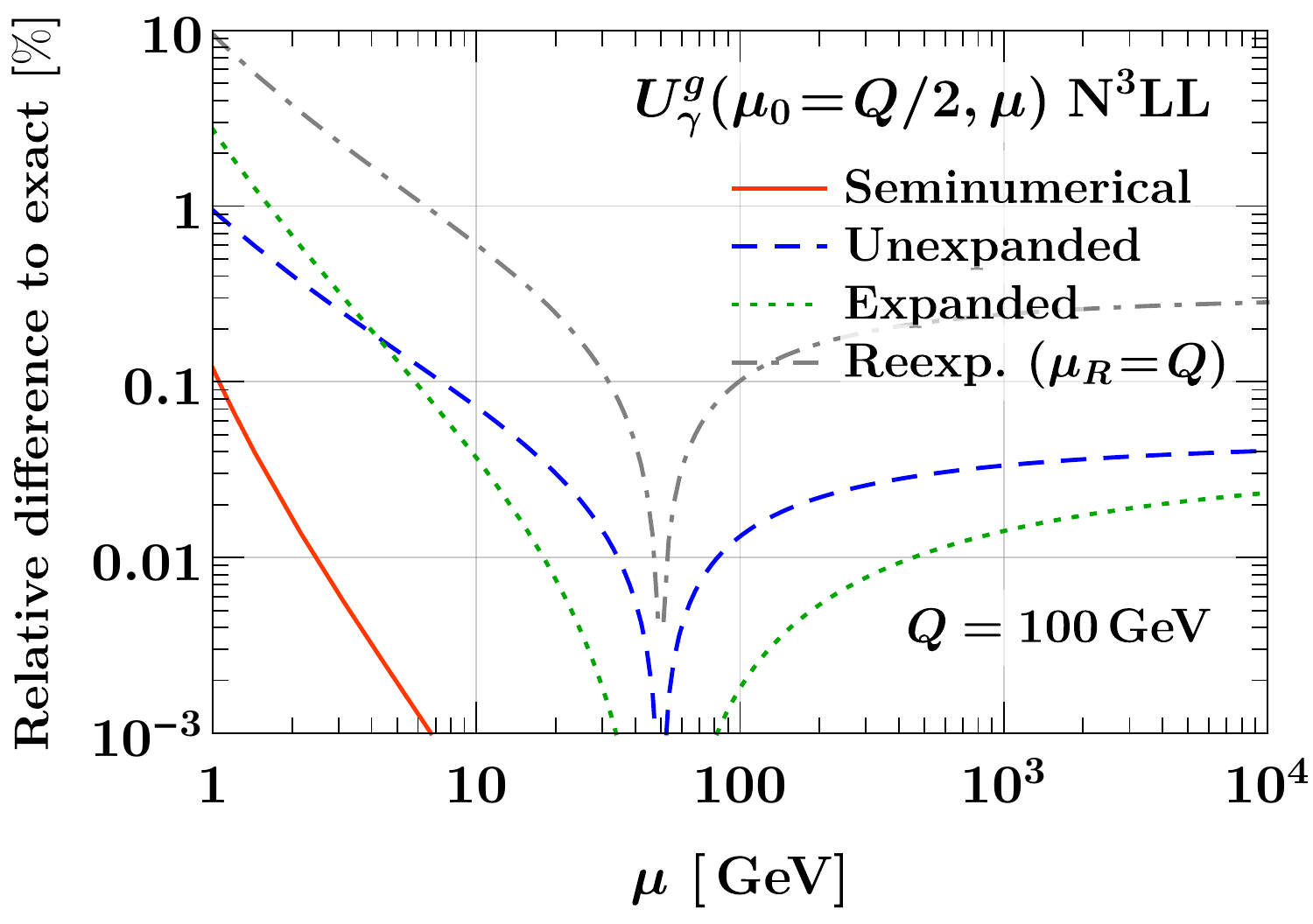}%
\caption{Same as \fig{cusperr}, but for the gluon noncusp evolution kernels.}
\label{fig:noncusperrgluon}
\end{figure*}
%-------------------------------------------------------------------------------

We now study the approximation errors of all four approximate methods relative
to the exact numerical method. The results for the quark cusp contribution, $U^q_\Gamma$,
are shown in \fig{cusperr}, and for the quark and gluon noncusp contributions, $U^{q,g}_\gamma$,
in \figs{noncusperr}{noncusperrgluon}. In all cases, we show the results at NNLL
(top rows) and N$^3$LL (bottom rows), and at $\mu_0 = Q$ (left panels) and $\mu_0 = Q/2$
(right panels).

At NNLL, we see a clear hierarchy, with the seminumerical method having the smallest
approximation errors, followed by the unexpanded, and then the expanded methods, which
is as expected given the increasing level of approximation involved in
each method. At N$^3$LL, the seminumerical method again performs best. The unexpanded
and expanded kernels have similar errors for the cusp term, while the unexpanded ones
fare better for the noncusp terms. The approximation error for the cusp term is always much
larger than for the noncusp term, which is not surprising due to the additional
$\ln(\mu)$ in its integrand. The errors for the gluon cusp contribution are about a
factor of two larger than for quarks.

We find, somewhat surprisingly, that at NNLL the approximation error for the expanded kernels
can exceed several percent when evolving to low scales, while at N$^3$LL it still exceeds
the percent level for both the unexpanded and expanded  kernels alike. Overall, the picture
is quantitatively quite similar to what we observed with the closure test.

We stress that the systematic differences we observe are solely due to the method of
integrating the RGE for identical perturbative inputs. Typically, we would want
such systematic effects to be much smaller than the perturbative precision we are
aiming at, as was the case for the running of the coupling. In other words, we clearly
want to avoid the method of integration to bias the result in any way. This is clearly
not the case here, since at NNLL and N$^3$LL one would typically aim at a
perturbative precision of order several to few percent.

For $\mu_0 = \mu_R = Q$, the reexpanded and expanded kernels are still equivalent.
In the right panels, we therefore also show the results when reducing the hard
scale of the evolution to $\mu_0 = Q/2$. This has essentially no effect on the
approximation error of the seminumerical, unexpanded, and expanded kernels. For the reexpanded
kernels, we keep $\mu_R = Q$, which is a commonly used choice in resummation applications.
The impact of treating the logarithms of $\mu_0/\mu_R$ and $Q/\mu_0$ at
fixed-order in the reexpanded kernels compared to the expanded ones now becomes visible
and turns out to be very large, which is quite unexpected.
It easily exceeds the percent level even at N$^3$LL,
and not just for the cusp but even the noncusp contributions. For the cusp term, it exceeds
$\gtrsim 10\%$ at the lowest scales. Note that roughly half of the observed difference
is due to the reexpansion around $\alpha(\mu_R)$ and half is due to the lower-order
treatment of the $\eta_\Gamma$ term.  This effect is not just due to the reduced
amount of evolution from $Q/2$ to $\mu$, as that is present for all methods
including the exact numerical result. Given the large numerical impact it has,
it might be worthwhile to reconsider the reasons for performing this additional
reexpansion around $\alpha(\mu_R)$. Certainly, from the point of view of the
evolution, the appropriate scale for $\alpha$ when starting the evolution from $\mu_0$ is $\mu_0$.

Of course, the differences due to the various approximations involved in all
methods, including the reexpanded method, might be considered as higher-order
effects. Nevertheless, given that they are not negligible or even exceed the
perturbative precision they have to be addressed and accounted for. One option
would be to include them as an additional systematic uncertainty in the
theoretical uncertainty estimate. On the other hand, there is no fundamental
theoretical reason for using any specific approximate solution. Hence, the best option
would be to avoid incurring this additional uncertainty by using the unique
exact solution of the defining RGE system resulting from the truncation of the
perturbative series of the anomalous dimensions. If that is computationally
prohibitive (or technically inconvenient), the seminumerical method offers a
good compromise, since its approximation error is always well below the percent
level, and so it is sufficiently accurate even for high-precision predictions.

%%%%%%%%%%%%%%%%%%%%%%%%%%%%%%%%%%%%%%%%%%%%%%%%%%%%%%%%%%%%%%%%%%%%%%%%%%%%%%%%
\section{Sudakov evolution kernels with two gauge interactions}
\label{sec:2D}
%%%%%%%%%%%%%%%%%%%%%%%%%%%%%%%%%%%%%%%%%%%%%%%%%%%%%%%%%%%%%%%%%%%%%%%%%%%%%%%%

Having investigated the integration of the one-dimensional kernels, we now consider
the extension to two gauge interactions, in which case also
mixed effects involving both gauge couplings $\alpha_{a,b}$ require resummation.
In \sec{struct2D}, we review the general analytic structure for this case and
the methods for evaluating them, based on what we learned from our exhaustive
analysis of the one-dimensional case. We then present numerical results for the
example of QCD$\otimes$QED in \sec{eval2D}.

%===============================================================================
\subsection{Structure of the two-dimensional evolution kernel}
\label{sec:struct2D}
%===============================================================================

We consider the Sudakov resummation for the direct product of two gauge groups
$G_{a} \otimes G_{b}$. The extension to more groups is then straightforward.
One key difference from the one-dimensional case is that the $\beta$ functions
that govern the evolution of the couplings $\alpha_a$ and $\alpha_b$ now become
a set of nonlinear coupled differential equations, as discussed in \subsec{beta2D}.

The generic Sudakov RGE structure remains the same as in \eq{RGE}, except that
the perturbative expansions for all quantities now involve a double series in
$\alpha_a$ and $\alpha_b$, including mixed terms corresponding to the emissions
of two distinct gauge bosons. Hence, the all-order structure of the anomalous
dimension is now given by%
\footnote{For the general EW case, the complete factorized structure of the
cross section can become more involved, because the  masses of the EW gauge
bosons introduce an additional scale. The generic Sudakov RGE however still has
the form of \eq{gammaF2D} with at most a single logarithm
$\ln(\mu)$~\cite{Chiu:2007dg}.}
%%%
\begin{align} \label{eq:gammaF2D}
\gamma_F(\mu)
= \frac{1}{\epsilon}\,\Gamma_\cusp[\alpha_a(\mu), \alpha_b(\mu)]\ln \frac{Q}{\mu}
+ \gamma[\alpha_a(\mu), \alpha_b(\mu)]
\,,\end{align}
%%%
with
%%%
\begin{align} \label{eq:2Dcusp}
\Gamma_\cusp(\alpha_{a},\alpha_{b})
&\equiv \mathop{\sum_{n,m}}_{n+m\ge1} \eps_a^n \eps_b^n\, \Gamma_{(n,m)}\Bigl(\frac{\alpha_a}{4\pi}\Bigr)^{n}\Bigl(\frac{\alpha_b}{4\pi}\Bigr)^{m}
\,,\\ \label{eq:2Dnoncusp}
\gamma(\alpha_{a},\alpha_{b})
&\equiv \mathop{\sum_{n,m}}_{n+m\ge1} \eps_a^n \eps_b^m\, \gamma_{(n,m)}\Bigl(\frac{\alpha_a}{4\pi}\Bigr)^{n}\Bigl(\frac{\alpha_b}{4\pi}\Bigr)^{m}
\,.\end{align}
%%%
The bookkeeping parameters $\eps_{a,b}\equiv 1$ are the same as in the $\beta$
functions in \eq{2Dbetageneral}, and $\eps \sim \eps_a \sim \eps_b$, i.e.\ we
make no assumption about the relative hierarchy of the two coupling constants.
It is also important to note the correspondence to the typical notation for a
single gauge theory $G_{a}$,%
\footnote{In contrast to the $\beta$ functions, $\alpha_a$ and $\alpha_b$
appear on equal footing in \eqs{2Dcusp}{2Dnoncusp}, so we have no choice
but to increment the meaning of $n$ in the subscript compared to the one-dimensional case.}
%%%
\begin{equation}
\Gamma_n \equiv \Gamma_{(n+1,0)}
\,,\qquad
\gamma_n \equiv \gamma_{(n+1,0)}
\end{equation}
%%%

The Sudakov evolution kernel is given by the two-dimensional analogue of \eq{evolutionkernel},
%%%
\begin{equation} \label{eq:2Dgenerickernel}
U(\mu_0,\mu) =
\exp\biggl\{
\int_{\mu_0}^{\mu}\frac{\df\mu'}{\mu'} \frac{1}{\epsilon}\,\Gamma_\cusp[\alpha_a(\mu'),\alpha_b(\mu')]\ln \frac{Q}{\mu'} +\gamma[\alpha_a(\mu'),\alpha_b(\mu')]
\biggr\}
\,.\end{equation}
%%%
It is clear from \eqs{2Dcusp}{2Dnoncusp} that the sums implicit in
\eq{2Dgenerickernel} include pure $G_a$ terms $\ord{\eps_a^n}$, pure $G_b$ terms
$\ord{\eps^m_b}$ terms, and mixed terms $\ord{\eps^n_a\eps^m_b}$.

%===============================================================================
\subsection{Evaluation of the two-dimensional evolution kernel}
\label{sec:eval2D}
%===============================================================================

Evaluating \eq{2Dgenerickernel} does not correspond to a simple extension of the
single gauge interaction scenario, since mixed terms $\sim\ord{\alpha_a\alpha_b}$
appear in conjunction with the coupled $\beta$ functions.

The fully numerical method is of course still applicable, although it is even
more computationally demanding now, since multiple coupled differential equations
for $\alpha_{a,b}$ must be solved. As before we use the fully numerical method
to provide the exact reference result to which other methods are compared.

The unexpanded analytic method is not easily extendable, since the coupled $\beta$
functions do not allow an analogous change of variables along the lines of
of $\df\ln\mu\rightarrow \df\alpha_a/\beta^a(\alpha_a,\alpha_b)$, because of the
dependence on the second coupling. Doing so would require one to express the
$\mu$ dependence of $\alpha_b$ in terms of $\alpha_a$, which in turn requires that one
treats $\alpha_b$ as in the expanded analytic method, at which point the advantage
of the unexpanded method is lost.
In principle, this could still be an option for cases where there is a clear
hierarchy between two couplings to justify treating them on unequal footing,
as would be the case for QCD$\otimes$QED.
However, we do not pursue this option further here for the reasons given below.

The expanded analytic method can still be applied to evaluate \eq{2Dgenerickernel}.
This is achieved by using the expanded solution of the coupled $\beta$ RGE for
$\alpha_a$ and $\alpha_b$ obtained from \eq{2DalphaaNNLO}, substituting it into
the perturbative expansions of the anomalous dimensions, and then explicitly
performing the integration in terms of $\ln\mu$. This was done in
\refcite{Cieri:2018sfk}.

In either case, the obtained analytic kernels will inevitably suffer
from the same-sized approximation errors already seen in the pure QCD case in
\subsec{1Dkernelpheno}. The generalization to the product group $G_{a} \otimes G_{b}$
does not alter the ultimate source, which are the additional approximation(s)
made in the integrand. By including EW or QED corrections, one is looking
for percent-level effects, and based on our findings and discussion in the
previous section, the analytic methods do not appear to provide sufficient
numerical accuracy.%
\footnote{One might consider using a more accurate (semi)numerical method
for the dominant QCD contributions, while including the smaller
mixed and pure EW corrections via an analytic approximation. However, we do not
see any gain in doing so compared to e.g.\ using the seminumerical method
everywhere.}

We therefore take the seminumerical method as our method of choice for
evaluating the two-dimensional evolution kernels. As already seen in
\subsec{1Dkernelpheno}, it features exact closure and very small
approximation errors (well below the percent-level at NNLL), at a reasonable
computational cost. The extension to the two-dimensional case \eq{2Dgenerickernel}
only requires two steps:
%%%
\begin{enumerate}
 \item We solve the coupled $\beta$ functions in \eq{2Dbetageneral} via the iterative
 method, up to the required order in $\eps$, which yields the closed-form analytic
 expressions for $\alpha_a(\mu)$ and $\alpha_b(\mu)$ in \eq{2DalphaaNNLO}.
 \item We then evaluate the evolution kernel $U(\mu,\mu_0)$ by employing a
 numerical integration routine in \eq{2Dgenerickernel}, using the analytic
 expressions for $\alpha_{a,b}(\mu)$ obtained in step 1 in the integrand.
\end{enumerate}
%%%
Schematically, this procedure is illustrated by
%%%
\begin{align} \label{eq:2Dgenericevaluate}
U(\mu_0, \mu)
 = \exp\biggl\{
\underbrace{\int_{\mu_0}^{\mu} \frac{\df \mu'}{\mu'}}_{\text{numerical}} \sum_{\substack{n,m\\n+m\geq 1}}
\biggl[
& \frac{\eps_a^n \eps_b^m}{\eps}  \Gamma_{(n,m)} \overbrace{\Bigl(\frac{\alpha_a(\mu')}{4\pi}\Bigr)^n \Bigl(\frac{\alpha_b(\mu')}{4\pi}\Bigr)^m}^{{\color{blue}{\text{\eq{2DalphaaNNLO}}}}}\ln\frac{Q}{\mu'}
\nn \\*
& +
\eps_a^n \eps_b^m \gamma_{(n,m)} \underbrace{\Bigl(\frac{\alpha_a(\mu')}{4\pi}\Bigr)^n \Bigl(\frac{\alpha_b(\mu')}{4\pi}\Bigr)^m}_{\color{blue}{\text{\eq{2DalphaaNNLO}}}} \biggr]
\biggr\}
\,.\end{align}

%-------------------------------------------------------------------------------
\begin{figure*}
\centering
\includegraphics[width=\WidthTwoSubfigs]{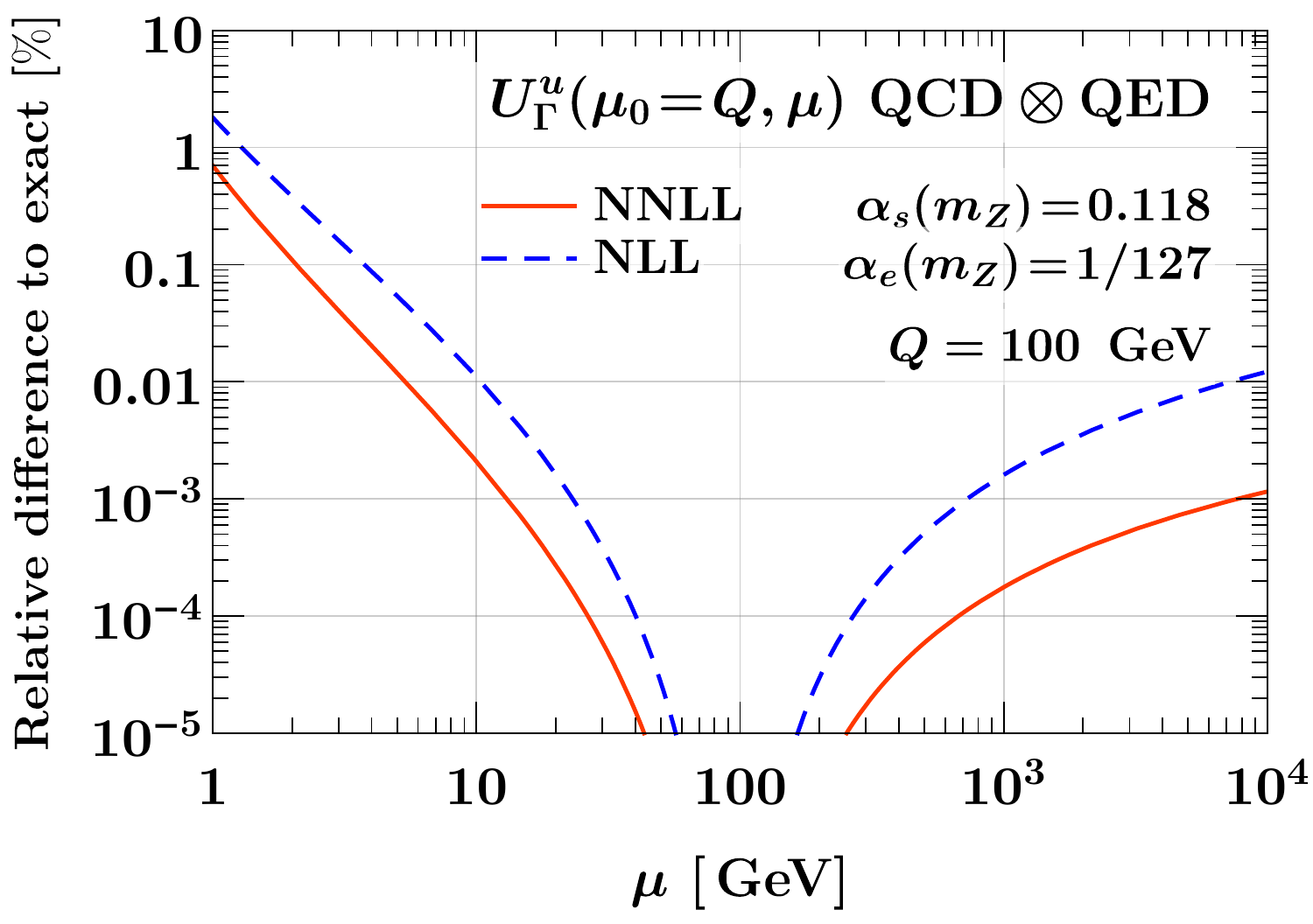}%
\hfill%
\includegraphics[width=\WidthTwoSubfigs]{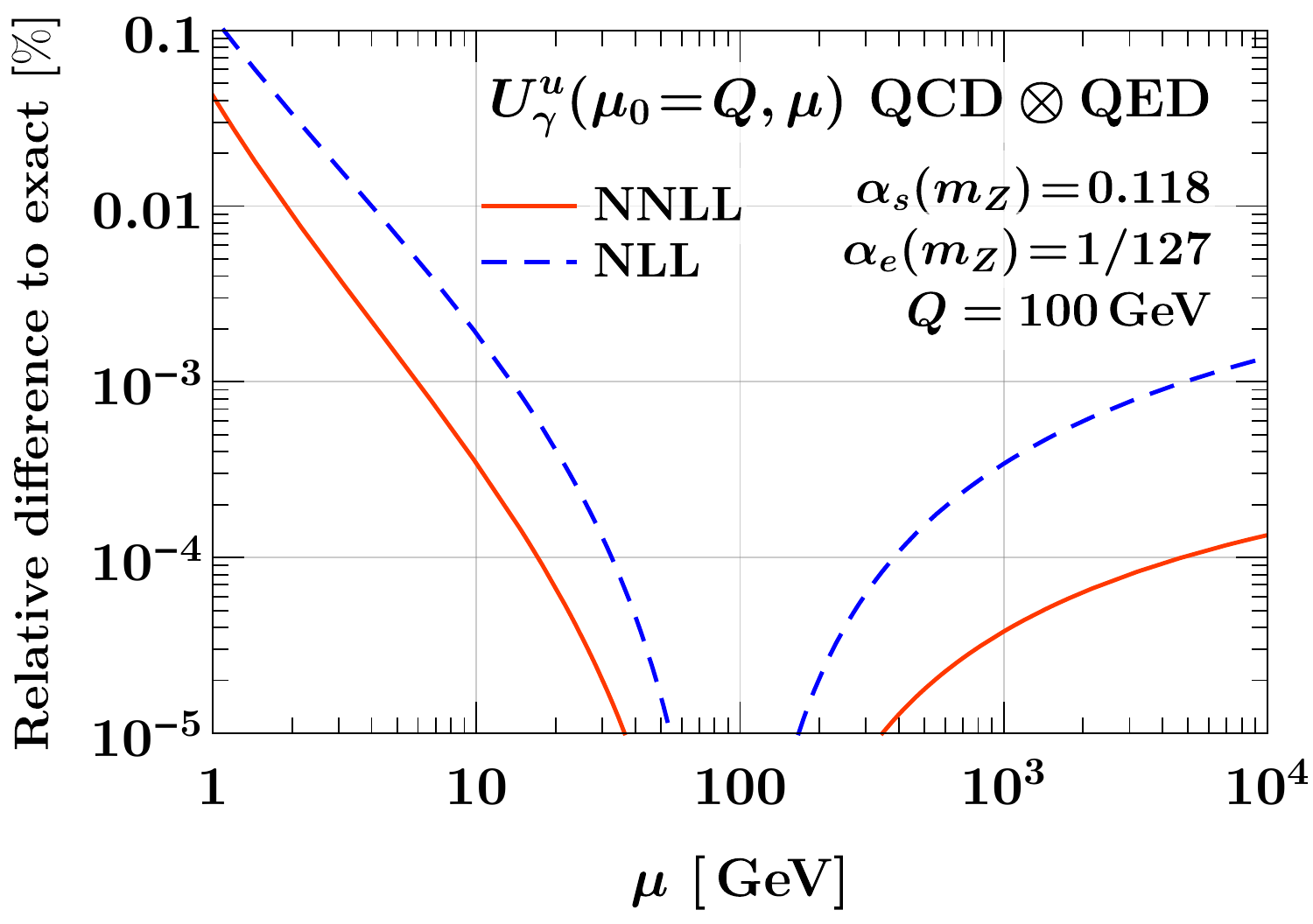}%
\caption{Deviation of the seminumerical kernels from the exact result for the
joint QCD$\otimes$QED $u$-quark evolution kernel at NLL (dashed) and
NNLL (solid). The cusp contribution is shown on the left and
the noncusp contribution on the right.}
\label{fig:mixederror}
\end{figure*}
%-------------------------------------------------------------------------------

In what follows, we apply this method to the mixed gauge QCD$\otimes$QED scenario.
We consider the $u\bar u$ hard function as a concrete example. The relevant coefficients
up to three loops can be found in \apps{anomdimQCD}{anomdimQED}, allowing us to obtain
the complete NNLL joint QCD$\otimes$QED Sudakov evolution.

In \fig{mixederror}, we show the approximation errors for the seminumerical method
at NLL and NNLL for the cusp (left panel) and noncusp (right panel) contributions.
As for the pure QCD case, the former has a larger approximation error, while overall
it remains well below 1\% everywhere (except at NLL for the cusp term at the very
lowest $\mu$ values).

\begin{figure*}
\centering
\includegraphics[width=\WidthTwoSubfigs]{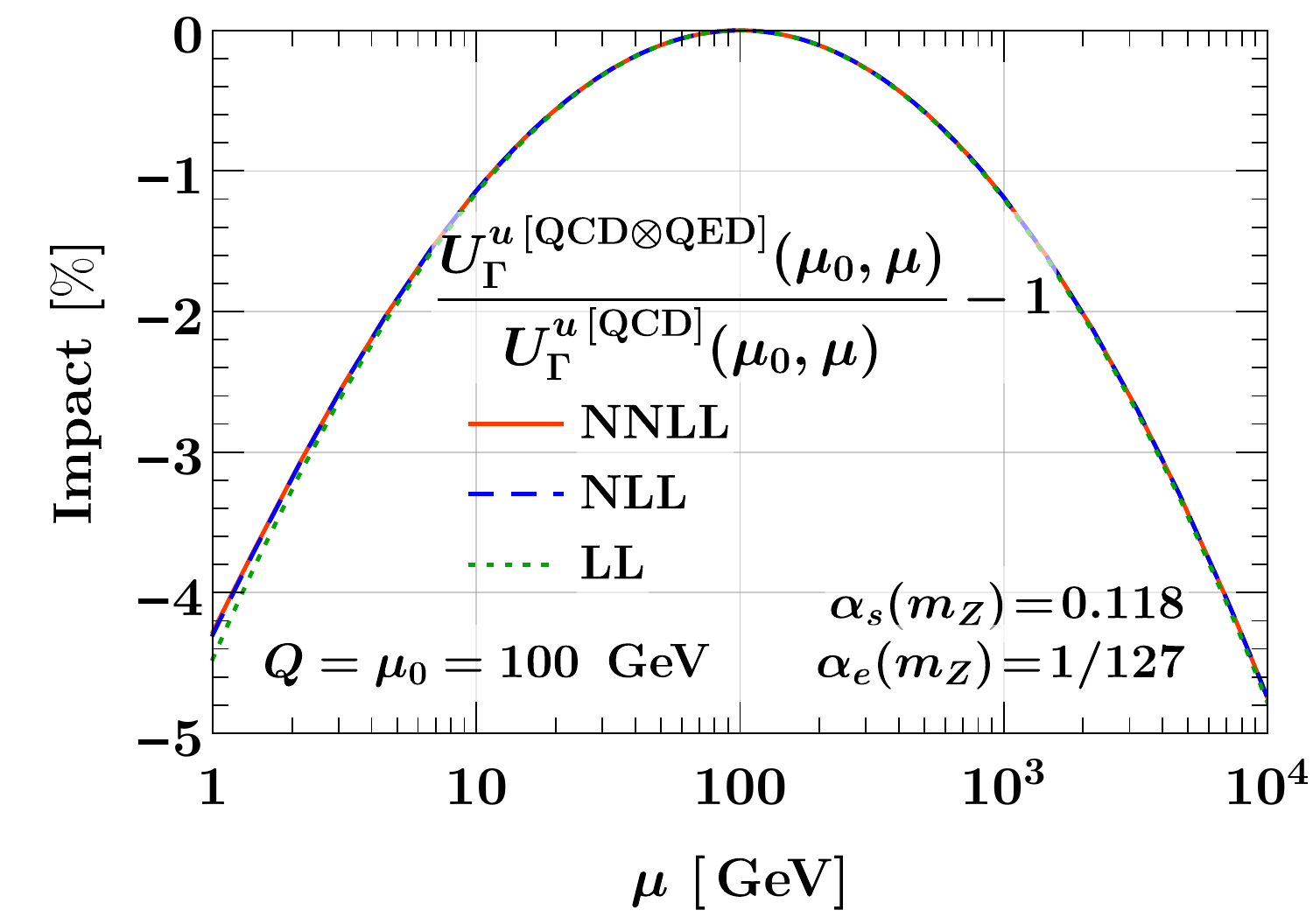}%
\hfill%
\includegraphics[width=\WidthTwoSubfigs]{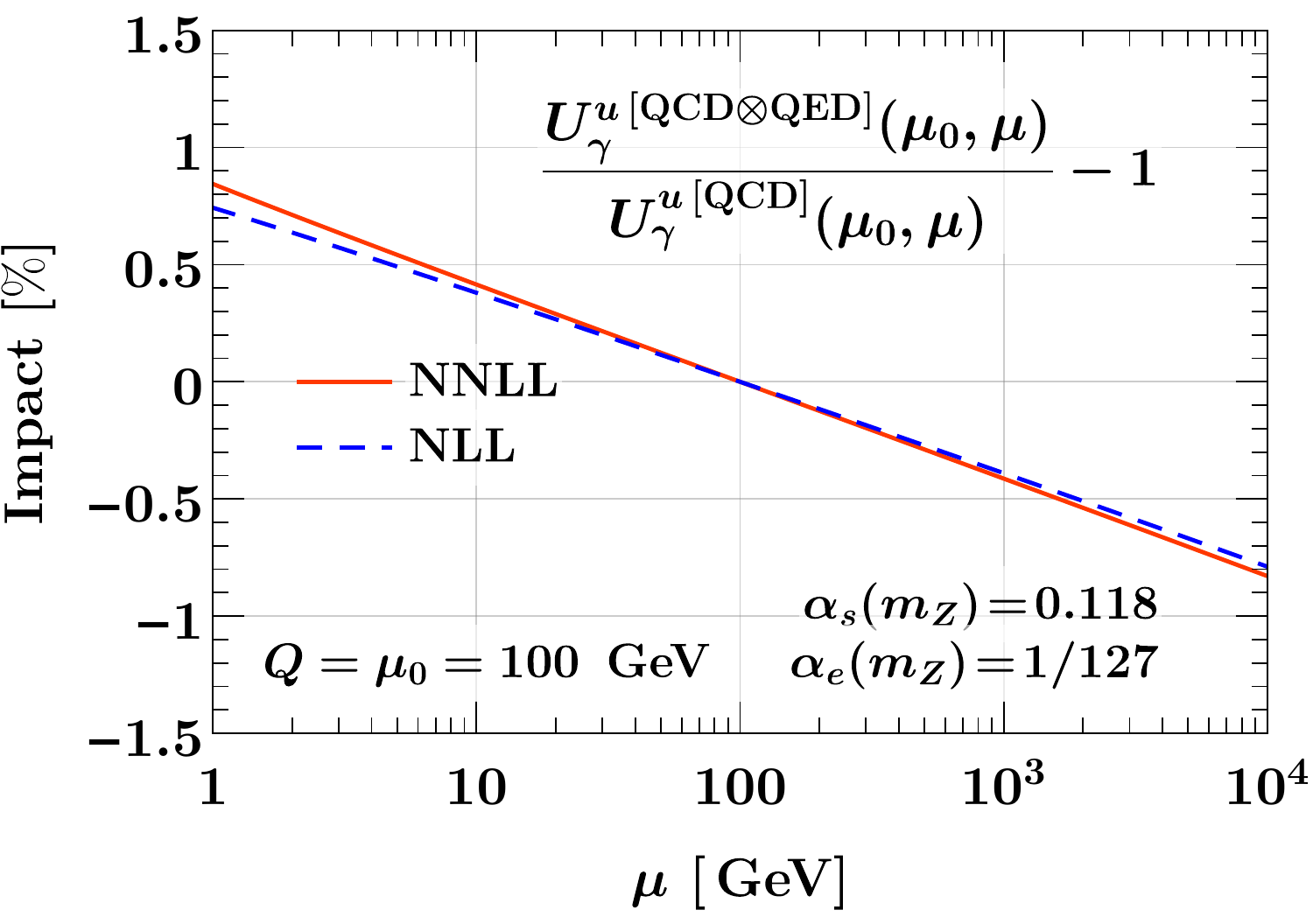}%
\caption{Impact of the QED corrections on the $u$-quark Sudakov evolution kernel
at LL, NLL, and NNLL for the cusp (left panel) and noncusp (right panel) contributions.
We show the relative difference of the full QCD$\otimes$QED evolution to the pure QCD case
at the corresponding order.}
\label{fig:impact2D}
\end{figure*}

\begin{figure*}
\centering
\includegraphics[width=\WidthTwoSubfigs]{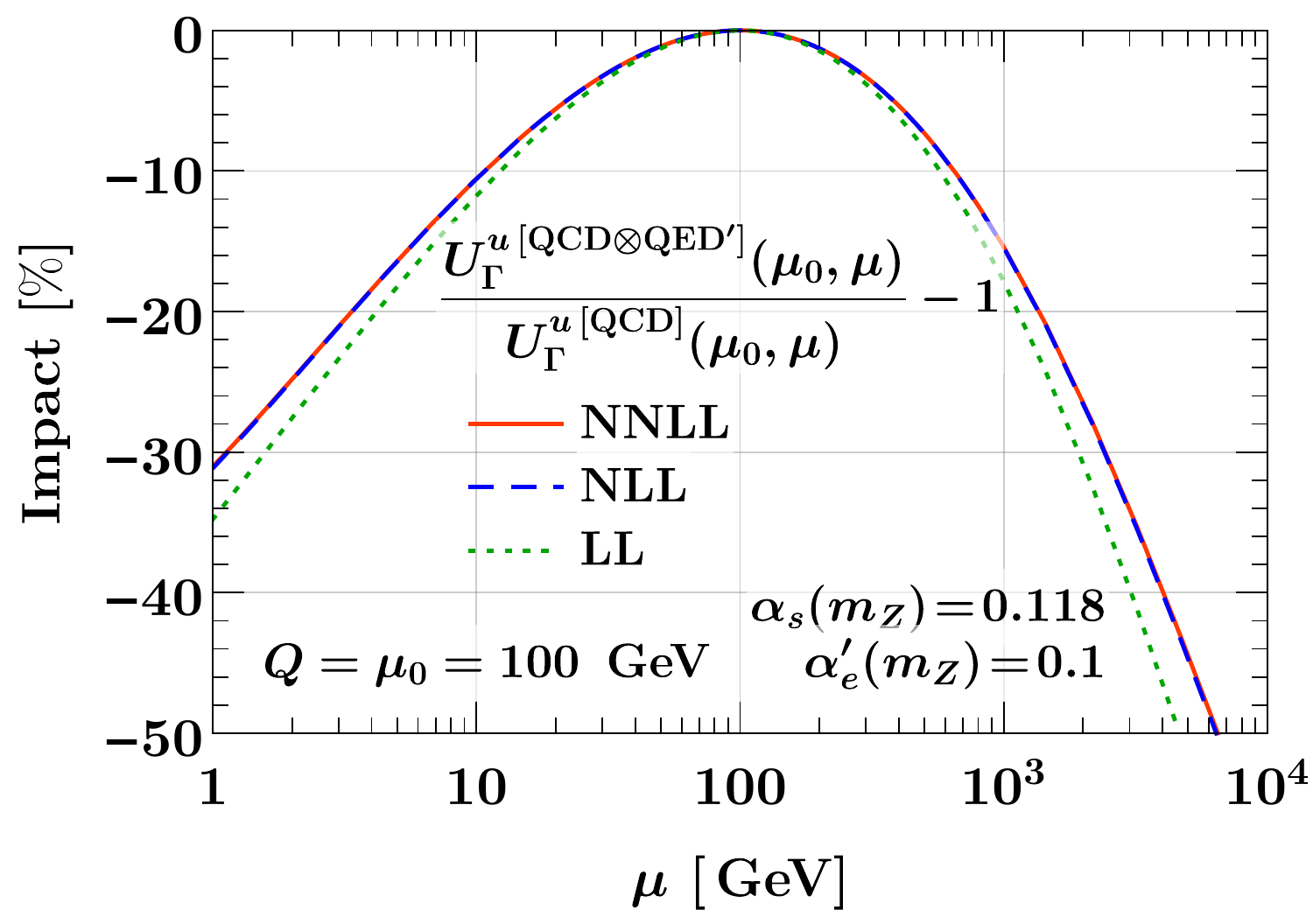}%
\hfill%
\includegraphics[width=\WidthTwoSubfigs]{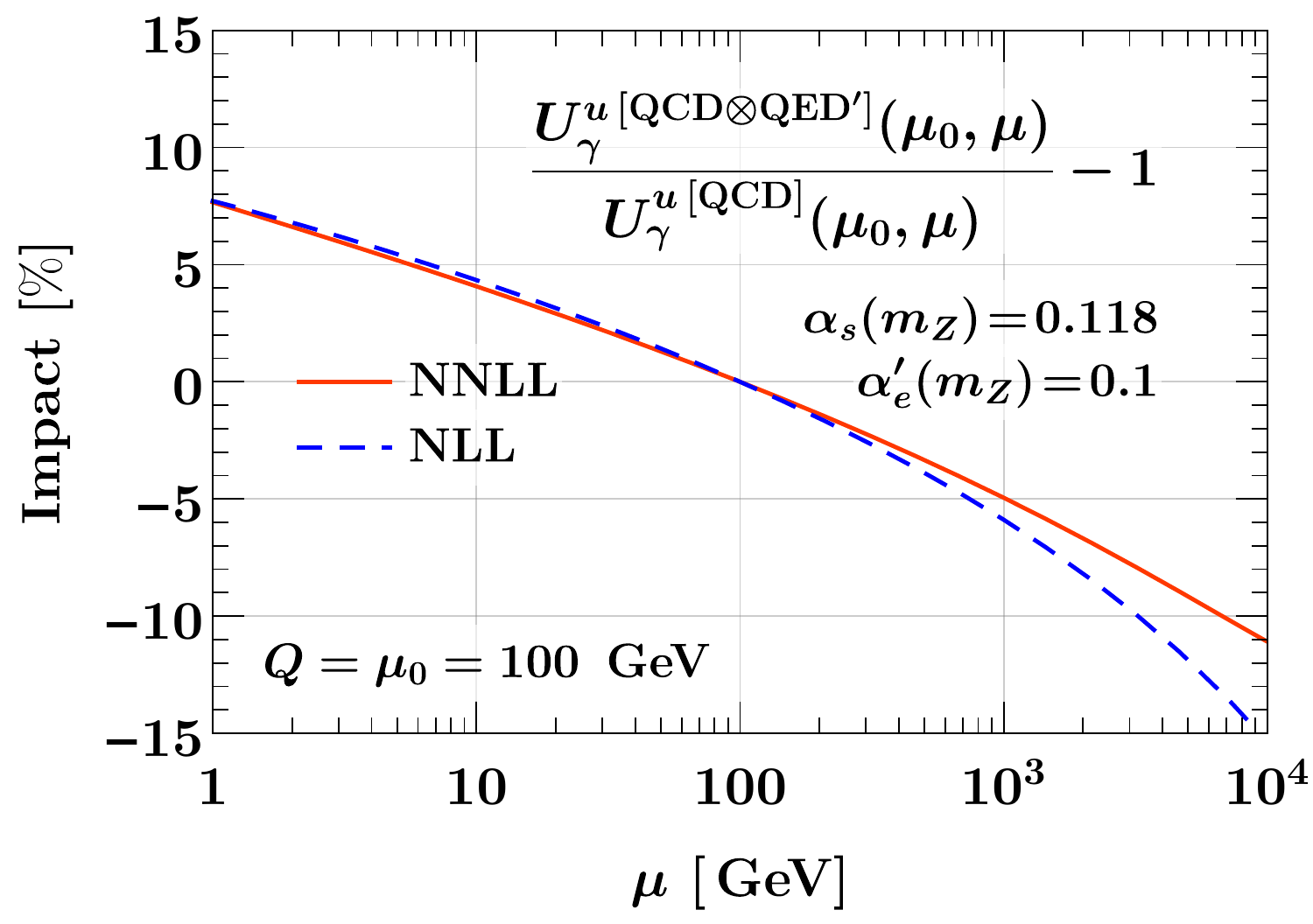}%
\caption{Same as \fig{impact2D}, but for a toy QED$^{\prime}$
with $\alpha_{e}^{\prime}(m_{Z}) = 0.1$.}
\label{fig:impact2DQEDp}
\end{figure*}

In \fig{impact2D}, we show the relative impact of the QED corrections by comparing the full
QCD$\otimes$QED to the pure QCD resummation kernels at each order.
For the cusp piece (left panel) the impact reaches $4-5\%$ over two decades of evolution,
while for the noncusp piece (right panel) it reaches close to $1\%$.
The overall effect for QED is expectedly small due to the smallness of the electromagnetic
coupling.
Considering a toy QCD$\otimes$QED$^{\prime}$ model with $\alpha_{e}^{\prime}(m_{Z}) = 0.1$
in \fig{impact2DQEDp},
the corrections become much larger, $30-50\%$ for the cusp piece and $5-15\%$ for the noncusp
piece.
Note that the impact is driven not only by pure QED and mixed corrections in the expansion of the anomalous dimensions, but also by the effects induced by the mixed $\beta$ functions in the coupling evolution.
Note also that the impact for the cusp term is practically identical at NLL and NNLL
even for QED$'$. This is somewhat accidental and due to the fact that the cusp anomalous dimension
does not yet receive any mixed contributions at three loops while its pure QCD and QED
three-loop coefficients happen to be very small.

%-------------------------------------------------------------------------------
%-------------------------------------------------------------------------------

%-------------------------------------------------------------------------------
%-------------------------------------------------------------------------------

%%%%%%%%%%%%%%%%%%%%%%%%%%%%%%%%%%%%%%%%%%%%%%%%%%%%%%%%%%%%%%%%%%%%%%%%%%%%%%%%
\section{Conclusions}
\label{sec:conclude}
%%%%%%%%%%%%%%%%%%%%%%%%%%%%%%%%%%%%%%%%%%%%%%%%%%%%%%%%%%%%%%%%%%%%%%%%%%%%%%%%

High-precision experimental measurements of (e.g.) $W$ and $Z$ production at the
LHC at the (sub-)percent level, as well as future measurements at the
high-luminosity LHC or future high-energy colliders, demand equally precise
resummed predictions in QCD$\otimes$EW. In this paper we studied the technical
aspects of achieving this joint resummation at higher orders in generic coupled
gauge environments.

In particular, at the level of the Sudakov evolution kernel appearing in any
resummation formalism, we showed that the commonly used methods of evaluating
the associated integrals in the Sudakov exponent via analytic approximations can
cause numerical differences of the same size as the contributions coming from
moving to a higher logarithmic accuracy or including EW radiation. In other
words, systematic integration errors that are typically assumed to be subleading
cannot be overlooked when attempting percent-level precision.

To show this we first studied five methods for integrating the evolution kernels
in the case of a single gauge interaction:\ numerical, seminumerical, unexpanded
analytic, expanded analytic, and reexpanded analytic. Their main difference is
in the treatment of the $\mu$ dependence of $\alpha(\mu)$ that one has to
integrate over. Although all methods employ a priori justifiable assumptions, we
showed that the latter three analytic methods can introduce errors at and above
the percent-level compared to the exact result (obtained via a fully numerical
treatment at a sufficiently high numerical precision). One should therefore be
cautious in using them, since their approximation errors can be nonnegligible
compared to the perturbative precision one is aiming for. The reexpanded kernels,
which are often used in the literature, are particularly notable; differences
greater than $10\%$ are possible over only two decades of evolution.

On the other hand, we found that a seminumerical method, which combines an
accurate analytic approximation for the running of the couplings with a
numerical integration, yielded minimal errors (typically $\le 0.1 \%$), while
still maintaining reasonable computing times. We therefore advocate its use for
phenomenological studies at high precision and in joint resummation
environments. As an example, we have used it to obtain the complete NNLL
QCD$\otimes$QED Sudakov evolution factor.

%%%%%%%%%%%%%%%%%%%%%%%%%%%%%%%%%%%%%%%%%%%%%%%%%%%%%%%%%%%%%%%%%%%%%%%%%%%%%%%%
\acknowledgments
We would like to thank Iain Stewart, Peter Marquard, and Johannes Michel for discussions.

%%%%%%%%%%%%%%%%%%%%%%%%%%%%%%%%%%%%%%%%%%%%%%%%%%%%%%%%%%%%%%%%%%%%%%%%%%%%%%%%
\appendix
\section{Perturbative results}
\label{app:pertresults}
%%%%%%%%%%%%%%%%%%%%%%%%%%%%%%%%%%%%%%%%%%%%%%%%%%%%%%%%%%%%%%%%%%%%%%%%%%%%%%%%

In this appendix we summarize the results for the higher-order analytic RGE solutions,
and the coefficients of the QCD and QED $\beta$ functions and anomalous dimensions.

%===============================================================================
\subsection{RGE solutions}
\label{app:RGE}
%===============================================================================

The iterative solution of the $\beta$-function RGE up to N$^3$LO, $\ord{\eps^3}$, reads
%%%
\begin{align} \label{eq:alphaN3LO}
\frac{\alpha(\mu_0)}{\alpha(\mu)}
&= X + \eps\,\frac{\alpha(\mu_0)}{4\pi}\, b_1 \ln X
+ \eps^2\, \frac{\alpha(\mu_0)^2}{(4\pi)^2}\biggl(b_2\, \frac{X-1}{X} + b_1^2\, \frac{1 -X +\ln X}{X} \biggr)
\nn \\ & \quad
+\eps^3\,\frac{\alpha(\mu_0)^3}{(4\pi)^3}
\biggl[
b_3\,\frac{X^2 - 1}{2X^2} + b_2 b_1\Bigl(\frac{1-X}{X} +\frac{\ln X}{X^2}\Bigr)
+ b_1^3\,\frac{(1-X)^2-\ln^2 X}{2X^2}
\biggr]
\,,\end{align}
%%%
where $b_n = \beta_n/\beta_0$ and
$X = 1 + \frac{\alpha(\mu_0)}{2\pi}\,\beta_0\,\ln(\mu/\mu_0)$ as in \eq{alphaLO}.

The functions $K_{\Gamma}$, $\eta_{\Gamma}$, and $K_{\gamma}$ appearing in the unexpanded
analytic Sudakov exponents in \eq{evolutionkerneldecomposed} are given up to N$^{3}$LL,
$\ord{\eps^3}$, by
%%%
\begin{align} \label{eq:KGammaN3LL}
K_{\Gamma}(\mu_0,\mu)
&= -\frac{\Gamma_0}{4\beta_0^2} \biggl\{
\frac{4\pi}{\alpha(\mu_0)}\Bigl(1-\frac{1}{r}-\ln r\Bigr)
+ \eps \Bigl[(\hat{\Gamma}_1-b_1)(1-r+\ln r)+\frac{b_1}{2}\ln^2 r\Bigr]
\nn \\ & \quad
+ \eps^2\, \frac{\alpha(\mu_0)}{4\pi} \biggl[
   (b_1^2-b_2)\Bigl(\frac{1-r^2}{2} + \ln r\Bigr) + (b_1\hat{\Gamma}_1-b_1^2)(1-r+r\ln r)
\nn \\ & \qquad
- (\hat{\Gamma}_2-b_1\hat{\Gamma}_1)\frac{(1-r)^2}{2} \biggr]
% \nn \\ & \quad
+ \eps^3\, \frac{\alpha(\mu_0)^2}{(4\pi)^2}
\biggl[(b_2 - b_1^2)(\hat{\Gamma}_1-b_1)\frac{(1-r)^2(2+r)}{3}
\nn \\ & \qquad
+ (\hat{\Gamma}_3-b_3-b_1(\hat{\Gamma}_2-b_2))\Bigl(\frac{1-r^3}{3}-\frac{1-r^2}{2}\Bigr)
\nn \\ & \qquad
+ b_1 (\hat{\Gamma}_2-b_2-b_1 (\hat{\Gamma}_1-b_1 ))\Bigl(\frac{1-r^2}{4}+\frac{r^2\ln r}{2}\Bigr)
\nn \\ & \qquad
+ (-b_3+2 b_1 b_2 -b_1^3 )\Bigl(\frac{1-r^2}{4}+\frac{\ln r}{2}\Bigr)
\biggr]
\biggr\}
\,, \\[0.5ex]
\label{eq:etaGammaN3LL}
\eta_\Gamma (\mu_0,\mu)
&= -\frac{\Gamma_0}{2\beta_0} \biggl[\ln r
+ \eps\, \frac{\alpha(\mu_0)}{4\pi} (\hat{\Gamma}_1-b_1)(r-1)
+ \eps^2\, \frac{\alpha(\mu_0)^2}{(4\pi)^2} (\hat{\Gamma}_2
 -b_1\hat{\Gamma}_1+b_1^2-b_2)\frac{r^2-1}{2}
\nn \\ & \quad
+ \eps^3\, \frac{\alpha(\mu_0)^3}{(4\pi)^3} \Bigl[\hat{\Gamma}_3-b_3-b_1 (\hat{\Gamma}_2-b_2)+(b_1^2-b_2
 )(\hat{\Gamma}_1-b_1)\Bigr]\frac{r^3-1}{3}
\biggr]
\,, \\[0.5ex]
\label{eq:KgammaN3LL}
K_\gamma(\mu_0,\mu)
&= -\frac{\gamma_0}{2\beta_0} \biggl[\eps \ln r
+ \eps^2\, \frac{\alpha(\mu_0)}{4\pi} (\hat{\gamma}_1\!-\!b_1)(r\!-\!1)
% \nn \\ & \quad
+ \eps^3\, \frac{\alpha(\mu_0)^2}{(4\pi)^2} (\hat{\gamma}_2
 -b_1 \hat{\gamma}_1+b_1^2- b_2)\frac{r^2\!-\! 1}{2}
\biggr]
,\end{align}
%%%
where $b_n=\beta_n/\beta_0$, $\hat{\Gamma}_n=\Gamma_n/\Gamma_0$, $\hat{\gamma}_n=\gamma_n/\gamma_0$,
and $r = \alpha_s(\mu)/\alpha_s(\mu_0)$.

The corresponding kernels for the expanded analytic method are obtained by inserting
\eq{alphaN3LO} into the above and expanding the results to $\ord{\eps^3}$.

%===============================================================================
\subsection{QCD anomalous dimensions}
\label{app:anomdimQCD}
%===============================================================================

Here we collect the pure QCD anomalous dimensions. For clarity and to avoid any
confusion we use the two-dimensional notation everywhere.
The QCD $\beta$-function coefficients in the \MSbar scheme up to 3 loops
are~\cite{Tarasov:1980au, Larin:1993tp}
%%%
\begin{align} \label{eq:beta coeff}
\beta^s_{00} \equiv \beta_0
&= \frac{11}{3}\,C_A - \frac{4}{3}\,T_F\,n_f
\,,\nn \\
\beta^s_{10} \equiv \beta_1
&= \frac{34}{3}\,C_A^2 - 2T_F\,n_f \Bigl(\frac{10}{3}\, C_A + 2 C_F\Bigr)
\,,\\
\beta^s_{20} \equiv \beta_2
 &= \frac{2857}{54}\,C_A^3
 + 2T_F\,n_f \Bigl(- \frac{1415}{54}\,C_A^2 - \frac{205}{18}\,C_F C_A + C_F^2  \Bigr)
 + 4T_F^2\,n_f^2 \Bigl(\frac{79}{54}\, C_A  + \frac{11}{9}\, C_F \Bigr)
\nn.\end{align}
%%%
The four-loop coefficient for $N_c = 3$ is given by~\cite{vanRitbergen:1997va, Czakon:2004bu}
%%%
\begin{equation}
\beta^s_{30} \equiv \beta_3 =  \Bigl( \frac{149753}{6} + 3564 \zeta_3 \Bigr)
- \Bigl( \frac{1078361}{162} + \frac{6508}{27} \zeta_3 \Bigr) n_f
% \nn \\ &\qquad
+ \Bigl( \frac{50065}{162} + \frac{6472}{81} \zeta_3 \Bigr) n_f^2
+  \frac{1093}{729} n_f^3
.\end{equation}
%%%

The \MSbar quark cusp anomalous dimension coefficients up to three loops are~\cite{Korchemsky:1987wg, Moch:2004pa}
%%%
\begin{align} \label{eq:Gamma_n}
\Gamma^q_{(1,0)} \equiv \Gamma^q_0 &= 4 C_F
\,,\nn\\
\Gamma^q_{(2,0)} \equiv \Gamma^q_1
&= 4C_F \biggl[ C_A \Bigl( \frac{67}{9} - \frac{\pi^2}{3} \Bigr)  - \frac{20}{9}\,T_F\, n_f \biggr]
\,,\\
\Gamma^q_{(3,0)} \equiv \Gamma^q_2
&= 4 C_F \biggl\{
   C_A^2 \Bigl(\frac{245}{6} -\frac{134 \pi^2}{27} + \frac{11 \pi ^4}{45} + \frac{22 \zeta_3}{3}\Bigr)
\nn \\ & \qquad\quad
   +  2T_F\,n_f \biggl[C_A \Bigl(-\frac{209}{27} + \frac{20\pi^2}{27} - \frac{28\zeta_3}{3} \Bigr)
   +  C_F\Bigl(-\frac{55}{6} + 8\zeta_3 \Bigr) \biggr]
   - \frac{16}{27}\,T_F^2\, n_f^2
   \biggr\}
\nn.\end{align}
%%%
The four-loop coefficient entering at N$^3$LL has become available during recent years, see e.g.~\refscite{Moch:2017uml, Grozin:2018vdn, Lee:2019zop, Henn:2019rmi, Bruser:2019auj}.
We use the numerical result for $N_c = 3$ and $n_f = 5$ obtained from \refcite{Bruser:2019auj},
%%%
\begin{equation}
\Gamma^q_{(4,0)} \equiv \Gamma^q_3 = 104.93\,C_F
\,.\end{equation}

Finally, the \MSbar noncusp quark anomalous dimension coefficients for the hard function up to three loops are~\cite{Idilbi:2006dg, Becher:2006mr, Moch:2005id}
%%%
\begin{align} \label{eq:gaCq}
\gamma^q_{H\,(1,0)} \equiv \gamma^q_{H\,0} &= -6 C_F
\,,\nn\\
\gamma^q_{H\,(2,0)} \equiv \gamma^q_{H\,1}
&= - 2C_F \biggl[
   C_A \Bigl(\frac{41}{9} - 26\zeta_3\Bigr)
   + C_F \Bigl(\frac{3}{2} - 2 \pi^2 + 24 \zeta_3\Bigr)
   + \beta_0 \Bigl(\frac{65}{18} + \frac{\pi^2}{2} \Bigr) \biggr]
\,,\nn\\
\gamma^q_{H\,(3,0)} \equiv \gamma^q_{H\,2}
&= -2C_F \biggl[
   C_A^2 \Bigl(\frac{66167}{324} - \frac{686 \pi^2}{81} - \frac{302 \pi^4}{135} - \frac{782 \zeta_3}{9} + \frac{44\pi^2 \zeta_3}{9} + 136 \zeta_5\Bigr)
\nn\\ & \qquad\qquad
   + C_F C_A \Bigl(\frac{151}{4} - \frac{205 \pi^2}{9} - \frac{247 \pi^4}{135} + \frac{844 \zeta_3}{3} + \frac{8 \pi^2 \zeta_3}{3} + 120 \zeta_5\Bigr)
\nn\\ & \qquad\qquad
   + C_F^2 \Bigl(\frac{29}{2} + 3 \pi^2 + \frac{8\pi^4}{5} + 68 \zeta_3 - \frac{16\pi^2 \zeta_3}{3} - 240 \zeta_5\Bigr)
\\ & \qquad\qquad
   + C_A \beta_0 \Bigl(-\frac{10781}{108} + \frac{446 \pi^2}{81} + \frac{449 \pi^4}{270} - \frac{1166 \zeta_3}{9} \Bigr)
\nn\\ & \qquad\qquad
   + \beta_1 \Bigl(\frac{2953}{108} - \frac{13 \pi^2}{18} - \frac{7 \pi^4 }{27} + \frac{128 \zeta_3}{9}\Bigr)
   + \beta_0^2 \Bigl(-\frac{2417}{324} + \frac{5 \pi^2}{6} + \frac{2 \zeta_3}{3}\Bigr)
\biggr]
\nn\,.\end{align}
%%%
The corresponding gluon noncusp anomalous dimensions can be found e.g.\
in appendix A.2 of \refcite{Ebert:2017uel}.

%===============================================================================
\subsection[QED and mixed \texorpdfstring{QCD$\otimes$QED}{QCDxQED} anomalous dimensions]
{\boldmath QED and mixed QCD$\otimes$QED anomalous dimensions}
\label{app:anomdimQED}
%===============================================================================

Here we collect all QED and QCD$\otimes$QED coefficients that are required for
the full NNLL hard evolution, and which enter in the numerical results in
\subsec{beta2D} and \sec{2D}.
The mixed QCD $\beta$-function coefficients to three loops are
%%%
\begin{align} \label{eq:betaQCDmixed}
\beta_{01}^s
&= -4T_F\,\overline{Q^2}
\,,\nn \\
\beta_{02}^s
&= \frac{44}{9}\, T_F\,\overline{Q^2} \bigl(N_c\, \overline{Q^2} + n_\ell\,Q^2_\ell\bigr)
 + 2T_F\,\overline{Q^4}
\,,\nn \\
\beta_{11}^s
&= (4C_F-8C_A)\, T_F\,\overline{Q^2}
\,,\end{align}
%%%
while the pure and mixed QED $\beta$-function coefficients up to three loops are
%%%
\begin{align} \label{eq:betaQED}
\beta_{00}^e
&= \frac{4}{3}\bigl(N_c\, \overline{Q^2} + n_\ell\,Q^2_\ell\bigr)
\,,\nn \\
\beta_{10}^e
&= 4 \bigl(N_c\,\overline{Q^4} + n_\ell\,Q^4_\ell\bigr)
\,,\nn \\
\beta_{20}^e
&= -\frac{44}{9} \bigl(N_c\,\overline{Q^4} + n_\ell\, Q_\ell^4\bigr) \bigl(N_c\, \overline{Q^2} + n_\ell\,Q^2_\ell \bigr)
- 2\bigl(N_c\,\overline{Q^6} + n_\ell \, Q_\ell^6\bigr)
\,, \\[1ex]
\beta_{01}^e
&= 4C_F\,N_c \, \overline{Q^2}
\,,\nn \\
\beta_{02}^e
&= \Bigl(\frac{133}{18}C_A-C_F\Bigr)2C_F\,N_c\,\overline{Q^2}
   -\frac{44}{9}\, C_F\,T_F\, n_f\,N_c\,\overline{Q^2}
\,,\nn \\
\beta_{11}^e
&= -4\,C_F\,N_c\,\overline{Q^4}
\,.\end{align}
%%%
Here we defined
%%%
\begin{equation}
\overline{Q^n} = \sum_q Q_q^n
\,,\end{equation}
%%%
where the sum runs over the active quark flavors with $Q_{u,c}=2/3$ and $Q_{d,s,b}=-1/3$
the quark charges, $N_c = 3$ the number of colors, $Q_\ell=-1$ the lepton charge,
and $n_\ell = 3$ the number of charged leptons.
The extraction of the three-loop mixed coefficients is discussed in \app{betaextract}.

The QED coefficients for the cusp anomalous dimension are obtained straightforwardly
by taking the abelian QED limit of \eq{Gamma_n},
%%%
\begin{align}
\Gamma^q_{(0,1)}
&=4Q_q^2
\,,\nn \\
\Gamma^q_{(0,2)}
&=4 Q_q^2 \Bigl[-\frac{20}{9} \bigl( N_c\,\overline{Q^2} + n_\ell\, Q_\ell^2 \bigr) \Bigr]
\,,\nn \\
\Gamma^q_{(0,3)}
&=4 Q_q^2 \biggl[
   Q_q^2 \bigl(N_c\,\overline{Q^2} + n_\ell\, Q_\ell^2 \bigr) \Bigl(-\frac{55}{3} + 16 \zeta_3 \Bigr)
   - \frac{16}{27} \bigl(N_c\,\overline{Q^2} + n_\ell\, Q_\ell^2 \bigr)^2
\biggr]
\,.\end{align}
%%%
Up to three loops, the mixed coefficients vanish, as was noted in \refcite{Chiu:2008vv},
%%%
\begin{equation}
\Gamma^q_{(1,1)} = \Gamma^q_{(1,2)} = \Gamma^q_{(2,1)} = 0
\,.\end{equation}
%%%
A nonzero mixed contribution is expected to first appear at four loops.

The noncusp anomalous dimensions are
%%%
\begin{align}
\gamma^q_{H\,(0,1)}
&= -6 Q_q^2
\,,\nn \\
\gamma^q_{H\,(0,2)}
&=-2Q_q^2 \biggl[Q_q^2 \Bigl(\frac{3}{2} - 2\pi^2 + 24\zeta_3 \Bigr)
 - \bigl(N_c\,\overline{Q^2} + n_\ell\, Q^2_\ell \bigr) \Bigl(\frac{130}{27} + \frac{2\pi^2}{3}\Bigr)
\biggr]
\,,\nn \\
\gamma^q_{H(1,1)}
&= -2Q_q^2\, C_F\, (3-4\pi^2+48\zeta_3)
\,.\end{align}
%%%
The pure QED coefficients are given by the abelian QED limit of the QCD coefficients
in \eq{gaCq}. The result for the mixed two-loop coefficient, $\gamma^q_{H(1,1)}$,
is obtained by following the abelianization procedure in \refcite{deFlorian:2018wcj},
and agrees with that given in \refcite{Bacchetta:2018dcq}.

%===============================================================================
\section{\boldmath Extraction of three-loop mixed \texorpdfstring{QCD$\otimes$QED}{QCDxQED} coefficients}
\label{app:betaextract}
%===============================================================================

In this section, we discuss our extraction of the three-loop coefficients of the
mixed QCD$\otimes$QED $\beta$ functions from the results in \refcite{Mihaila:2014caa},
which are required for the complete NNLO running.

\Refcite{Mihaila:2014caa} considers the coupled $\beta$-function RGE for a generic
gauge group given as the product of $n$ simple groups, $G=G_1\otimes G_2\otimes \dots \otimes G_n$.
They explicitly calculate the case of three distinct simple groups $G=G_1\otimes G_2\otimes G_3$,
since at most three gauge bosons can propagate simultaneously at three loops.
They consider the case where each simple group $G_i$ is nonabelian, but it is straightforward
to apply their results to the abelian case by a proper modification of the Casimir invariants.

To obtain the QCD$\otimes$QED coefficients, we specify ourselves to the product group
$G = SU(3)_c\otimes U(1)_{EM} \otimes \mathbf{1}$, where $\mathbf{1}$ is the trivial
identity group. In addition, we only keep fermionic matter couplings, setting the Yukawa and scalar (quartic) couplings
and the Casimir invariants of the scalar representation $S$ to zero, as they appear in eq. (3.1) of \refcite{Mihaila:2014caa}.
We have explicitly checked that with this procedure we reproduce the pure QCD $\beta$-function coefficients.

Since we only have two couplings, namely $\alpha_i$ with $i=s,e$ for QCD and QED respectively,
the usual Casimir invariants for the $SU(3)_c$ gauge group are
%%%
\begin{align}
[T_F^A, T_F^B] &= \img f^{ABC} T_F^C\,
\,, \nn \\
\Tr T_F^A T_F^B &= \delta^{AB} \, T(F_s)
\,, \nn \\
(T_F^A)_{ab} (T_F^{A})_{bc} &= \delta_{ac}\,C(F_s)
\,, \nn \\
f^{ABC}f^{DBC} &= \delta^{AD}\,C(G_s)
\,, \nn \\
\delta^{AA} &= d(G_s)
\,,\end{align}
%%%
where $F_s$ and $G_s$ stand for the fundamental and adjoint representations, so
$C(F_s) \equiv C_F$ and $C(G_s) \equiv C_A$.
For the case of $U(1)_{EM}$, we take the abelian limit
%%%q
\begin{align}
C(F_e) \rightarrow Q_f^2
\,,\qquad
T(F_e) \rightarrow Q_f^2
\,,\qquad
C(G_e) \rightarrow 0
\,,\end{align}
%%%
with $f$ representing quarks and leptons. Note that sums over fermion species do
appear in the $\beta$-function coefficients that each gauge group involves,
i.e.\ quarks for $SU(3)_c$ and both quarks and leptons for $U(1)_{EM}$.
Also, since \refcite{Mihaila:2014caa} decomposes Dirac fermions into chiral fermions,
we have to substitute $n_f\rightarrow 2n_f$ for the number of quarks and
$n_\ell\rightarrow 2n_\ell$ for the number of leptons in their results.

All of this considered, the $\beta^e_{11}$ coefficient can be extracted from the
generic results of \refcite{Mihaila:2014caa}, finding%
\footnote{%
We found a typo in this term in eq.~(3.1) of \refcite{Mihaila:2014caa}, which is
missing the last $D(F_i)$ factor. This is confirmed by comparing this term for
the full SM gauge group with the results of \refscite{Mihaila:2012fm, Bednyakov:2012rb}.}
%%%
\begin{align}
\beta^e_{11} = \sum_F 2[2C(G_i)-C(F_i)]T(F_i)C(F_j)D(F_{ij})D(F_i)
\,,\end{align}
%%%
where $i=e$, since we are considering the $U(1)_{EM}$ $\beta$ function, and $j=s$.
These specifications then give
%%%
\begin{align} \label{eq:betaQED_11}
\beta^e_{11}
&= -2 \sum_{F=q,\ell} C(F_e)T(F_e)C(F_s)D(F_{es})D(F_e)
= -4 C_F\, \overline{Q^4}\, N_c
\,,\end{align}
%%%
where the fermion sum over the leptons does not contribute due to the presence of
$C(\ell_s)=0$ and the multiplicity factors
%%%
\begin{equation}
D(F_{es}) = \prod_{k\neq e,s}\,d(F_k) = d(\mathbf{1}) = 1
\,,\qquad
D(F_e) = \displaystyle \prod_{k\neq e}\,d(F_k) = d(F_s) \, d(\mathbf{1}) = N_c
\end{equation}
%%%
in our case correspond to the dimension of the trivial group factor and of the
fundamental $SU(3)_c$ representation.

We can also read off the $\beta^e_{02}$ coefficient%
\footnote{%
We also found a typo in these terms in eq.~(3.1) of \refcite{Mihaila:2014caa},
which are missing the last $D(F_i)$ and $D(F_{m,i})$ factors. This is confirmed by
comparing the result for $\beta^e_{02}$ with the results in \refcite{Baikov:2012zm}.}
%%%
\begin{align}
\beta^e_{02} &= \sum_{F}\Bigl(\frac{133}{18}C(G_j)-C(F_j)\Bigr)C(F_j)T(F_i)D(F_{ij})D(F_i)
\nn \\ & \quad
-\sum_{F_m,F_n}\frac{11}{9}C(F_{m,j})T(F_{n,j})T(F_{m,i})D(F_{m,ij})D(F_{n,j})D(F_{m,i})
\,,\end{align}
%%%
where again $i=e$ and $j=s$. We therefore find
%%%
\begin{align} \label{eq:betaQED_02}
\beta^e_{02} &= \Bigl(\frac{133}{18}C_A-C_F\Bigr)C_F\,2\overline{Q^2} \, d(\mathbf{1}) N_c - \sum_{F_m}\frac{11}{9}C(F_{m,s})T(F_{m,e})D(F_{m,es})D(F_{m,e})(2\,T_F\,n_f)
\nn \\
&= \Bigl(\frac{133}{18}C_A-C_F\Bigr)2C_F\,N_c\,\overline{Q^2}
   - \frac{44}{9}\, C_F\,T_F\,n_f\, N_c\,\overline{Q^2}
\,.\end{align}
%%%
Observe that again only the sum over quarks contributes, since the leptonic sums include a $C(\ell_s)=T(\ell_s)=0$.

Following the same procedure we also obtained the mixed QCD coefficients $\beta^s_{11}$ and $\beta^s_{02}$
as given in \eq{betaQCDmixed}, in agreement with the corresponding results given in \refcite{Huber:2005ig}, where
$\beta_{11}^s$ was obtained from an explicit three-loop calculation.

%%%%%%%%%%%%%%%%%%%%%%%%%%%%%%%%%%%%%%%%%%%%%%%%%%%%%%%%%%%%%%%%%%%%%%%%%%%%%%%%
\addcontentsline{toc}{section}{References}
\bibliographystyle{jhep}
\bibliography{2DSudakov}

\end{document}